\documentclass{aa}  
\usepackage{graphicx}
\usepackage{txfonts}
\usepackage[bookmarks=true,bookmarksopen=true,pdfstartview = FitV,pdfpagelayout = SinglePage,colorlinks = true,linkcolor=blue,citecolor=blue,urlcolor = blue]{hyperref} 
\usepackage{xcolor} 
\usepackage{pgf, tikz} 
\usetikzlibrary{arrows} 
\newcommand{\ur}{u_\mathrm{r}}
\newcommand{\ut}{u_\mathrm{\theta}}
\newcommand{\up}{u_\mathrm{\phi}}
\newcommand{\urr}{u_\mathrm{rr}}
\newcommand{\utt}{u_\mathrm{\theta \theta}}
\newcommand{\urt}{u_\mathrm{r \theta}}
\newcommand{\utr}{u_\mathrm{\theta r}}
\newcommand{\rsun}{R_{\odot}}
\newcommand{\msun}{M_{\odot}}
\newcommand{\progname}{LIFELINE}

\begin{document}
\title{\huge\bf \progname{}: The program for the simulation of the X-ray line profiles in massive colliding wind binaries}
\author{E.\ Mossoux\inst{1} \and G.\ Rauw\inst{1}}
\institute{Space sciences, Technologies and Astrophysics Research (STAR) Institute, Universit\'e de Li\`ege, All\'ee du 6 Ao\^ut, 19c, B\^at B5c, 4000 Li\`ege, Belgium} 
\date{}
\abstract{}
{The study of the X-ray line profiles produced by massive colliding wind binaries is a powerful tool for the characterisation of the stellar winds.
We built a self-consistent program for the computation of line profiles named \progname{}.
The resulting theoretical profiles can be compared to the line profile that will be observed with future high-resolution X-ray spectrographs to retrieve the characteristics of the stellar winds generating them.} 
{We considered a grid of 780 O-type binaries and computed, for each of them, the wind velocity distribution of each star, taking the impact of the radiation pressure and gravity force of the companion star into account.
We then computed the characteristics of the wind shock region and followed the emitted photons towards the observer to compute their absorption.
Finally, the Fe~K line profiles near 6.7\,keV were constructed from the distribution of the photons as a function of the radial velocities of their emitting region.
\progname{} can be used to compare the theoretical line profiles to the observed ones or to compute theoretical profiles for a new binary system.}
{We highlight the results for three systems.
While the line profiles created in adiabatic wind collision regions are quite simple, the line profiles arising from regions in the radiative regime, as found in short-period binaries, are more sophisticated notably because of the Coriolis effect on the shape of the shock.
The predicted differences in line morphology between systems with different wind properties are quite significant, allowing a detailed comparison between the theoretical profiles and those that will be observed with future high-resolution X-ray spectrometers.}
{}

\authorrunning{E.\ Mossoux et al.}
\titlerunning{LIFELINE}
\maketitle

\section{Introduction}
\label{intro}
The current generation of X-ray satellites has allowed researchers to observe, for the first time, details of the morphology of line profiles in the spectra of massive stars (e.g. \citealt{gudel09}).
Line profiles emitted by single massive stars have been analysed by numerous observational and theoretical studies (see e.g. \citealt{feldmeier03,owocki06, cohen10, herve12, herve13, rauw15b}).
However, only a few studies have been dedicated to the X-ray line profiles emitted by the material in the wind interaction region in massive binary systems \citep{henley03,rauw16b}.
In such systems, the powerful winds (with huge mass-loss rates of about $10^{-6}\,\mathrm{\msun{}\,yr^{-1}}$ and highly supersonic velocities of about $2000\,\mathrm{km\,s^{-1}}$) emitted by the massive stars collide to create an interaction zone where the plasma may be heated up to high temperatures thanks to the conversion of the kinetic energy normal to the shock front \citep{stevens92}.
Two hydrodynamical shocks separated by a contact discontinuity are thus created. 
Their characteristics depend on those of the involved winds.

Several theoretical studies have been conducted to determine the characteristics of the wind shock region in close binaries where the shock-heated plasma undergoes radiative cooling \citep{antokhin04} as well as in wide binaries where the plasma is in the adiabatic regime \citep[e.g.][]{stevens92}.
However, from the observational point of view, the X-ray spectra of such binary systems have been studied mostly in broadband medium-resolution spectroscopy (see \citealt{rauw16} for a review).
Only two theoretical studies have focused on high-resolution spectroscopy that allows the characterisation of the X-ray line profiles:
\citet{henley03} studied the Ly\,$\alpha$ transitions of O~{\sc viii}, Ne~{\sc x}, Mg~{\sc xii}, Si~{\sc xiv}, and S~{\sc xvi} located below 3\,keV using hydrodynamical simulations, while \citet{rauw16b} studied the Fe~K lines near 6.7\,keV in adiabatic wind shock regions.
The model from \citet{henley03} was successfully applied to fit the Si~{\sc xiv} and S~{\sc xvi} lines of the {\it Chandra} High-Energy Transmission Grating spectra of $\eta$~Car as observed at various phases around the orbit \citep{henley08}.
The advantage of the Fe~K lines emitted at higher energies is the absence of contamination by the softer intrinsic X-ray emission of the individual stellar winds. 
Indeed, the emission lines studied by \citet{henley03} are also emitted by the plasma that is distributed throughout the individual winds outside the wind shock region.
On the other hand, the emissivity of the Fe~K lines peaks around plasma temperatures of $kT = 5.4$\,keV, implying that this line forms in a very hot plasma. For massive stars, such temperatures are not reached in the intrinsic wind shocks; they can only be found in colliding wind interactions or in the winds of stars with strong magnetic fields that efficiently confine the outflow (e.g.\ \citealt{schulz00}).
Moreover, the lines emitted at high energy undergo a very low absorption by the cool unshocked wind material.
These high energy lines are thus useful tools for studying the wind interaction region.
The study of \citet{rauw16b} focused on the line profiles emitted by an adiabatic wind shock region, where the orbital separation of the system is high enough to allow the winds to reach their terminal velocities before colliding.
However, in most systems, this implies very long orbital periods.
In contrast, radiative wind shock regions prevail in short period binaries where the stellar wind has not accelerated to its terminal velocity before colliding.
Since many massive stars reside in close binaries, it is important to investigate their line profiles for the sake of comparison with future observational data.

This paper presents the \progname{} program for the simulation of the X-ray LIne proFiles in massivE coLliding wInd biNariEs.
This program performs a complete simulation of the line profiles from the computation of the velocities of the stellar winds including radiative driving and radiative inhibition \citep{stevens94}, to the characterisation of the wind shock region in the adiabatic and radiative regimes including the effects of the Coriolis deflection \citep{parkin08}, and the computation of the line profiles (Sect.~\ref{simu}).
The line profiles of the Fe~K helium-like triplet were computed for a grid of binary systems of O-type stars and several orientations of the line of sight (Sect.~\ref{grid}).
The resulting line profiles of three representative systems are presented in Sect.~\ref{discuss}.
Finally, a brief guideline on how to retrieve and use the \progname{} program is provided in Sect.~\ref{prog} before summarising our results in Sect.~\ref{summary}.

\section{Simulation}
\label{simu}
We name $d$ the distance between the stars. We define a spherical coordinate system centred on the star with the less powerful wind. 
In these coordinates, the wind velocity vector is expressed as $(\ur,\ut,\up)$.

\subsection{Radiative inhibition of the stellar winds}
\label{inhib}
\begin{figure}
\centering
\includegraphics[height=5cm]{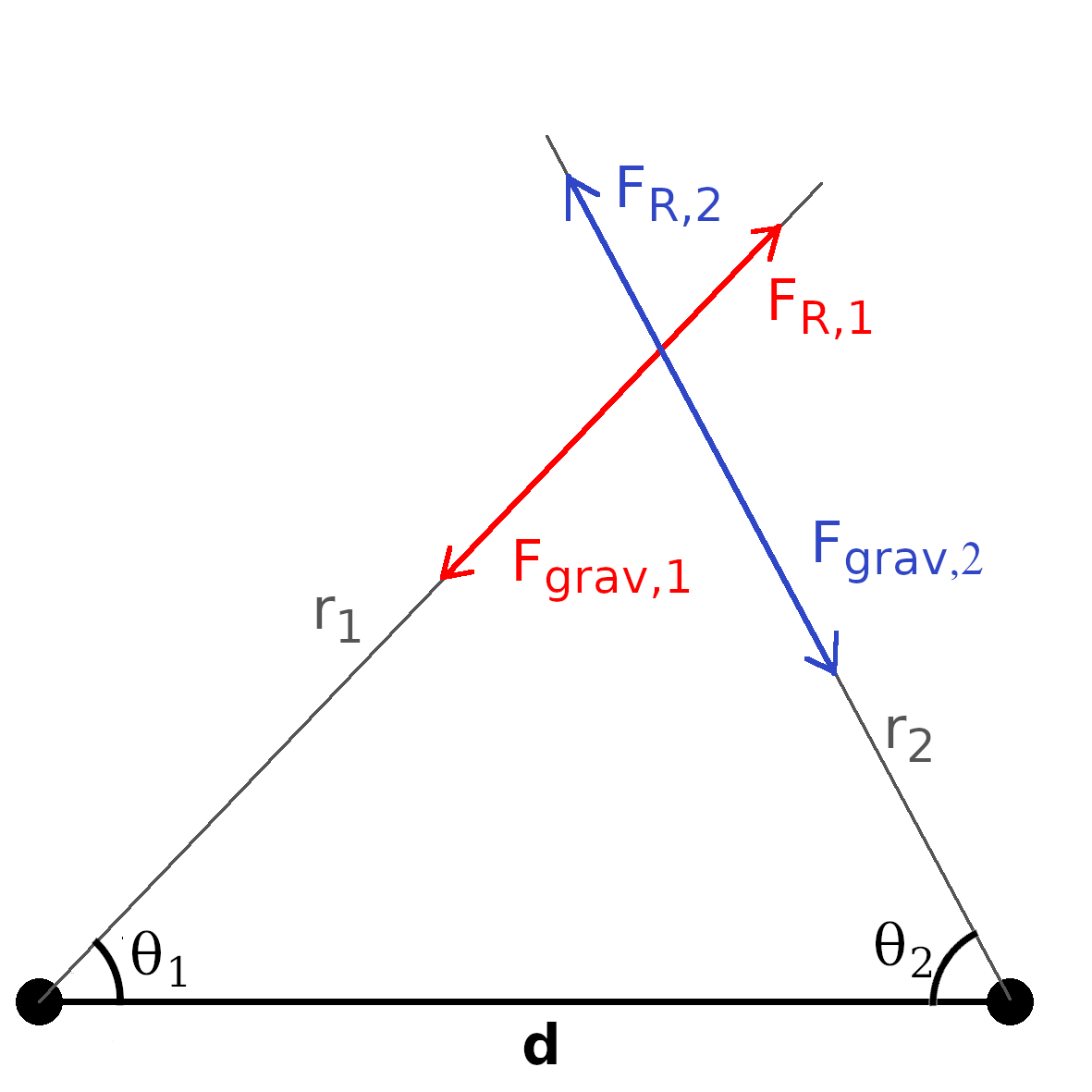}
\caption{Representation of the forces acting upon a small element of material located between the two stars.}
\label{forces}
\end{figure}

The velocity and direction of a small volume of wind material is controlled by the equilibrium of the forces acting on it.
In a binary system (see Fig.~\ref{forces}), the material is first accelerated by the line-radiation pressure and braked by the gravitation of the emitting star.
Then, approaching the companion star, the material encounters the gravitational field of the companion \citep{stevens88}, and the dynamical impact of the companion's radiation field \citep{stevens92}. 
Moreover, the line-driving may be suppressed by a change in the ionisation of the wind material \citep{stevens91}.
\citet{stevens94} computed the effects of the radiation and gravitation fields of both stars on the velocity of the wind along the line of centres between the stars.
We extend their work to compute the velocity and direction of the wind all around the emitting star.

Forces acting on a small element of wind material are the gravity ($F_\mathrm{grav}$) and the radiative acceleration ($F_\mathrm{R}$) from both stars (see Fig.~\ref{forces}).
The gravitational acceleration at a distance $r_\mathrm{i}$ of the star $i$ is $F_\mathrm{grav,i}=G\,M_\mathrm{i}(1-\Gamma_\mathrm{i})/r^2_\mathrm{i}$ with $G$ the gravitational constant, $M_\mathrm{i}$ the stellar mass, and $\Gamma_\mathrm{i}$ the Eddington ratio computed as $\sigma_\mathrm{t}\,L_\mathrm{i}/(M_\mathrm{i}\,4\pi\,G\,c\,m_\mathrm{p})$ with $\sigma_\mathrm{t}$ the Thomson cross-section, $L_\mathrm{i}$ the luminosity of the star, $c$ the speed of light, and $m_\mathrm{p}$ the proton mass.
The CAK theory \citep{castor75} allows the parametrisation of the radiative pressure with two parameters ($\alpha$ and $k$) computed by \citet{abbott82} for a set of effective temperatures and wind densities:
\begin{equation}
   g_\mathrm{rad}=\frac{\sigma^{1-\alpha}_\mathrm{e}k}{c\rho^{\alpha} V^{\alpha}_\mathrm{th}}\left|\frac{dv}{dr}\right|^{\alpha}\,,
\end{equation}
where $\sigma_\mathrm{e}=(1+H)\sigma_\mathrm{t}/(2m_\mathrm{H})$ is the scattering electron opacity with $H$ the hydrogen mass fraction, $V^2_\mathrm{th}=2k_\mathrm{B}T_\mathrm{eff}/m_\mathrm{H}$ is the ion thermal speed, $k_\mathrm{B}$ is the Boltzmann constant, and $T_\mathrm{eff}$  isthe effective temperature of the star.
The radiative acceleration is thus $F_\mathrm{R,i}=g_\mathrm{rad}F_\mathrm{i}K_\mathrm{i}$ with $F_\mathrm{i}=L_\mathrm{i}/(4\pi r^2_\mathrm{i})$ and $K_\mathrm{i}=1-(1-R_\mathrm{i}^2/r_\mathrm{i}^2)^{1+\alpha}/((1+\alpha)R_\mathrm{i}^2/r_\mathrm{i}^2)$ the finite disc correction factor, with $R_\mathrm{i}$ the radius of the considered star.

Equations of the evolution of the wind velocity presented hereafter are derived from the static Navier-Stokes equation in spherical coordinates.
Details about the equations can be found in appendix~\ref{NS}.
We assume that the wind is axisymmetric about the line of centres.
The norm of the wind velocity is thus $v^2=\ur^2+\ut^2$.
We consider an ideal gas characterised by a zero viscosity, an isothermal speed of sound $a^2=\gamma k_\mathrm{B}T/(\mu m_\mathrm{H})$ with $\mu$ the mean molecular weight, $\gamma$ the adiabatic index, and $T$ the temperature of the winds that we assume to be equal to $T_\mathrm{eff}$.

Along the line of centres, we only work with the $r$ component of the Navier-Stokes equation.
This is the configuration considered by \citet{stevens94}.
We report here the resolution for star 1.
A similar solution can be found for star 2.
For star 1, we thus have to solve:
\begin{equation}
\begin{split}
   &\left(1-\frac{a^2}{\gamma\,v^2}\right)\,v\,\frac{dv}{dr}-2\frac{a^2}{\gamma\,r_1}-\frac{G\,M_2\,(1-\Gamma_2)}{(d-r_1)^2}+\frac{G\,M_1\,(1-\Gamma_1)}{r_1^2}\\
   &-\frac{\sigma_\mathrm{e}^{1-\alpha}\,k\,(F_1\,K_1-F_2\,K_2)}{c\,V_\mathrm{th}^{\alpha}\,\rho^{\alpha}}\,\left|\frac{dv}{dr}\right|^{\alpha}=0
\end{split}
   \label{eq_loc}
\end{equation}
from the critical point $r_\mathrm{c}=1.05\,R_1$ \citep{stevens94} where the velocity is computed using Eq.~17 of \citet{stevens94}.
The stagnation mass-loss rate $d\dot{M}/d\Omega$ used to compute the density $\rho$ is given by Eq.~24 of \citet{stevens94}.

We then solve the evolution of the wind velocity outside the line of centres.
The distance between star 2 and a point with an angle $\theta_1\in\ ]0,\pi]$ and a distance $r_1$ from star 1 is $r_2=(d^2+r_1^2-2\,r_1\,d\,\cos\theta_1)^{0.5}$.
The angle between this point and the x-axis as viewed from star 2 is $\cos\theta_2=(d-r_1\,\cos\theta_1)/r_2$ (see Fig.~\ref{forces}).

We thus have to solve an equation describing the $r$ component of the Navier-Stokes equation:
\begin{equation}
\begin{aligned}
&\left(1-\frac{a^2}{\gamma v^2}\right)\,\ur\,\frac{\partial \ur}{\partial r} - \frac{\ut^2}{r_1} + \frac{\ut}{r_1}\,\frac{\partial \ur}{\partial \theta} - \frac{a^2\,\ut}{\gamma v^2}\frac{\partial \ut}{\partial r} \\
&- \frac{G\,M_2\,(1-\Gamma_2)\,\cos{(\theta_1+\theta_2)}}{r_2^2} + \frac{G\,M_1\,(1-\Gamma_1)}{r_1^2} \\
&- g_\mathrm{rad}\,(F_1\,K_1-F_2\,K_2\,\cos{(\theta_1+\theta_2)}) - 2\frac{a^2}{\gamma\,r_1}=0\,,
\end{aligned}
\end{equation}
and an equation describing the $\theta$ component of the total Navier-Stokes equation:
\begin{equation}
\begin{aligned}
&\left(1-\frac{a^2}{\gamma v^2}\right)\frac{\ut}{r_1}\,\frac{\partial \ut}{\partial \theta} + \frac{\ut\,\ur}{r_1} + \ur\frac{\partial \ut}{\partial r} - \frac{a^2\,\ur}{\gamma r_1 v^2}\frac{\partial \ur}{\partial \theta} \\
&- \frac{G\,M_2\,(1-\Gamma_2)\,\sin{(\theta_1+\theta_2)}}{r_2^2} + g_\mathrm{rad}\,F_2\,K_2\,\sin{(\theta_1+\theta_2)}=0\,.
\end{aligned}
\end{equation}
with $g_\mathrm{rad}=\frac{\sigma_\mathrm{e}^{1-\alpha}\,k}{c\,(V_\mathrm{th}\,\rho\,v)^{\alpha}}\,\left| \ur\frac{\partial \ur}{\partial r}+\ut\frac{\partial \ut}{\partial r}\right|^{\alpha}$.
These equations were solved using the finite-differences method for a mono-atomic ($\gamma=5/3$) totally ionised gas ($\mu=0.62$) with the fractional mass abundance of hydrogen equal to that of the Sun ($0.7381$).
We used a radial step of $0.03\rsun{}$ and an angular step of $1^\circ$.

\subsection{Position of the contact discontinuity}
\label{cd}
\begin{figure}
\centering
\includegraphics[width=7cm]{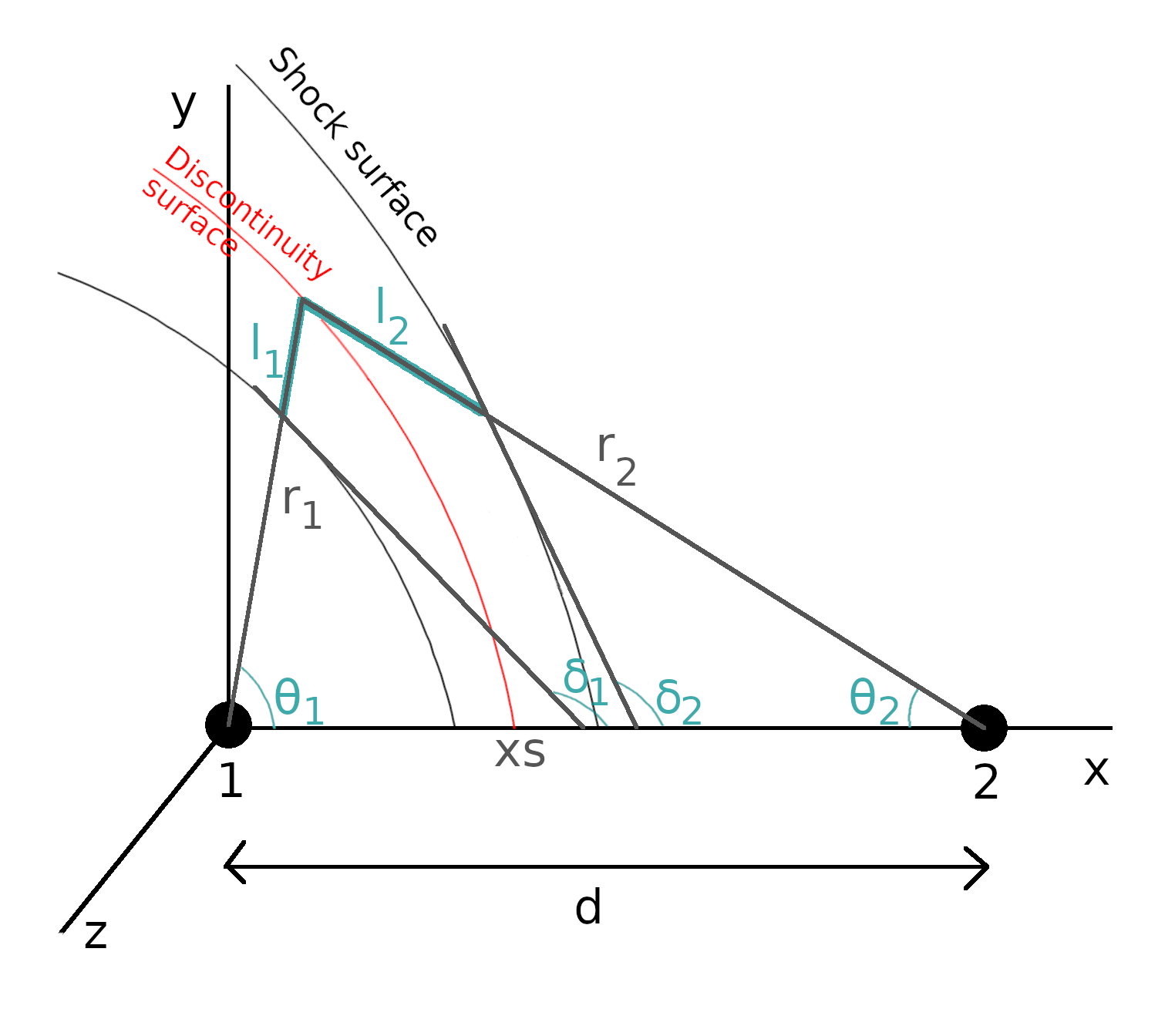}
\caption{Schematic view of the wind shock region.
Star 1 orbits star 2 in the direction of the z-axis.}
\label{sketch}
\end{figure}
The contact discontinuity is the very thin surface (thickness comparable to the mean free path of the particles) separating the post-shock regions of the two winds.
In adiabatic shocks, it is often assumed that the winds have reached their terminal velocities before they collide. 
However, because of the radiative inhibition described in Sect.~\ref{inhib}, this might not be the case, even in relatively wide binaries, especially in the region close to the line of centres.
In such cases, the analytical solution of \citet{canto96} does not provide the exact position of the contact discontinuity.

However, to get a good sampling of the shock in 2D, we use the \citet{canto96} model as a first approximation.
We consider a range of angles $\theta_1$ from 0 to the opening angle of the shock $\theta_{\infty}$ with 30 steps.
The angle $\theta_{\infty}$ is defined as \citep{canto96}
\begin{equation}
   \theta_\infty-\tan{\theta_\infty}=\frac{\pi}{1-\beta}\,,
\end{equation}
where $\beta=\dot{M}_1\,v_{\infty,1}/(\dot{M}_2\,v_{\infty,2})$ is the wind momentum ratio, considering the terminal velocities $v_\infty$ computed as 2.6 times the escape velocity \citep{kudritzki00}.
The radial distance $r_1$ from star 1 is
\begin{equation}
   r_1=d\sin{\theta_2}\csc{(\theta_1+\theta_2)}
   \label{shock_surf_ad}
\end{equation}
where $\theta_1$ and $\theta_2$ are the angles between the position of the point on the discontinuity surface and the line of centres from star 1 and 2, respectively (see Fig.~\ref{sketch}).
The angle $\theta_2$ is related to $\theta_1$ \citep{canto96}:
\begin{equation}
   \theta_2\cot{\theta_2}=1+\beta(\theta_1\cot{\theta_1}-1)\,.
\end{equation}
This allows us to determine the values of $y=r_1\,\sin{\theta_1}$ for which we will compute the corresponding values of $x$ for the position of the contact discontinuity.

We use the formalism of \citet{antokhin04} to compute the exact solution of the position of the contact discontinuity considering that the winds have not reached their terminal velocities before colliding.
The stagnation point $xs$ of the contact discontinuity is the result of the non-linear equation $x=\eta(x,0)\,d/(1+\eta(x,0))$ with the wind momentum ratio varying with the distance from star 1:
\begin{equation}
   \eta(x,y)=\sqrt{\frac{\dot{M}_1\,v_1(x,y)}{\dot{M}_2\,v_2(x,y)}}\,.
\end{equation}
The positions of the discontinuity surface are the solutions of the differential equation:
\begin{equation}
   \frac{dx}{dy}=\frac{1}{y}\left(x-\frac{d\,r^2_1(x,y)}{r^2_1(x,y)+r^2_2(x,y)\eta(x,y)}\right)\,,
   \label{shock_surf}
\end{equation}
where $r_1(x,y)$ is the distance to star 1 and $r_2(x,y)$ is the distance to star 2 (see Fig.~\ref{sketch}).

\subsection{Characteristics of the shocked gas}
The evolution of the characteristics of the gas in the post-shock region depends on the efficiency of radiative cooling.
\citet{stevens92} established a criterion to determine whether the gas is in an adiabatic or in a radiative regime: The shock is considered as adiabatic if $v^4xs/\dot{M}>1$, with $\dot{M}$ is the mass-loss rate in $10^{-7}\,\mathrm{\msun{}\,yr^{-1}}$ and $v$ is the pre-shock velocity in $1000\,\mathrm{km\,s^{-1}}$ at the distance of the stagnation point $xs$ (in $10^7\,$km).

\subsubsection{Adiabatic cooling}
\label{eq:ad}
The width of an adiabatic wind interaction zone is computed as the ratio between the surface density and the volume density at the discontinuity surface. The surface density is computed by means of the formalism of \citet{canto96}:
\small
\begin{equation}
   \sigma=\frac{\sigma_0\sin{(\theta_1+\theta_2)}\csc{\theta_1}\csc{\theta_2}(\beta(1-\cos{\theta_1})+\frac{v_2(x,y)}{v_1(x,y)}(1-\cos{\theta_2}))^2}{\sqrt{(\beta(\theta_1-\sin{\theta_1}\cos{\theta_1})+(\theta_2-\sin{\theta_2}\cos{\theta_2}))^2+(\beta\sin^2{\theta_1}-\sin^2{\theta_2})^2}}\,,
   \label{width_ad}
\end{equation}
\normalsize
where $\sigma_0=\dot{M}_1/(2\pi\beta d v_1(x,y))$.
The volume density is the average density of the mixed gas inside the interaction region:
\begin{equation}
   \rho=0.5\left(\frac{\dot{M}_1}{v_1(x,y)\pi r^2_1(x,y)}+\frac{\dot{M}_2}{v_2(x,y)\pi r^2_2(x,y)}\right)\,.
\end{equation}
The shock width is discretised using 20 linearly spaced steps on each side of the discontinuity surface.

We can thus compute the temperature at the shock surface as
\begin{equation}
   kT_\mathrm{i}=\frac{3m_\mathrm{p}v^2_\mathrm{p,i}}{16k_\mathrm{b}}\,,
\end{equation}
where $v_\mathrm{p,1}=v_1(x,y)\sin(\delta_1-\theta_1)$ and $v_\mathrm{p,2}=v_2(x,y)\sin(\pi-\delta_2-\theta_2)$ is the component of the velocities that is normal to the shock surface with $\delta_\mathrm{i}$ the slope of the shock surface.
We assume here that the normal component of the velocity is fully thermalised and that the electrons and ions are characterised by the same temperature.
These approximations may not be valid in all cases, as shown by several authors \citep{usov92,zhekov00, zhekov07,pollock05}.
But they mainly affect the strengths of the lines, not their shape.
Since we are dealing with normalised line profiles, such effects should be less important for our purpose.
For adiabatic shocks, we also assume that the shocked plasma of the two winds are mixed and have the same temperature for a given position along the contact discontinuity.

\subsection{Radiative cooling}
\label{eq:rad}
The width of a radiative shock on the side of star $i$ is computed as in Eq.\,19 of \citet{antokhin04}:
\begin{equation}
   l_\mathrm{i}=\frac{15\,\bar{\mu}^2m_\mathrm{p}^2}{512\sum(X_zZX_H)}\frac{(v_\mathrm{i}(x,y)\sin \delta_\mathrm{i})^3}{\rho_0\lambda(T_0)}\,,
   \label{width_rad}
\end{equation}
where $\bar{\mu}=1.3$, $\sum(X_zZX_H)=0.99$, $\rho_0=4\dot{M}_\mathrm{i}/(v_\mathrm{i}(x,y)4\pi\,r_\mathrm{i}^2)$ is the post-shock density and $\lambda(T_0)$ is the cooling function for $T_0=1.21(\mu/0.62)(v_\mathrm{i}(x,y)\sin{\delta_\mathrm{i}})^2$ the post-shock temperature. 
The shock width is again discretised using 20 linearly spaced steps on each side of the discontinuity surface.
As for adiabatic shocks, we assume that the normal component of the velocity is totally thermalised and that the electrons and ions are characterised by the same temperature.

From the post-shock temperature, we can also compute the evolution of the temperature across the interaction zone with Eq.\,16 of \citet{antokhin04}:
\begin{equation}
   \frac{dT}{dl}=-C(T)\frac{\lambda(T)}{T^2}\,,
   \label{eq:rad_temp}
\end{equation}
where $C(T)=9\sum(X_zZX_H)\mu^3m_\mathrm{p}\rho_0\,(v_\mathrm{i}(x,y)\sin{\delta_\mathrm{i}})^3/(40\bar{\mu}^2k^3_\mathrm{B})$ and $v_\mathrm{i}(x,y)$ is the velocity at the shock surface.
The temperature is thus decreasing towards the discontinuity surface.

In the same manner, considering an isobaric gas, the variation of the ideal gas law over the shock width is:
\begin{equation}
   \frac{k}{\mu\,m_\mathrm{p}}\left(\rho\frac{dT}{dl}+T\frac{d\rho}{dl}\right)=0\,.
\end{equation}
Considering Eq.~\ref{eq:rad_temp}, we have:
\begin{equation}
   \frac{d\rho}{dl}=C(T)\rho\frac{\lambda(T)}{T^3}\,.
   \label{eq:rad_rho}
\end{equation}
The density is thus increasing towards the discontinuity surface.

The computation of the surface density depends on the off-axis angle from the line of centres.
Close to the line of centres ($y<0.2xs$), we use Eq.~A2 or A3 of \citet{antokhin04} depending on the side of the interaction region to compute the parameter $\sigma_0$. 
The surface density is then:
\begin{equation}
   \sigma=\sigma_0\left(1-\frac{y^2(1+2xs\,z_0)}{6xs^2}\right)\,,
   \label{eq:rad_sigma1}
\end{equation}
with
\begin{equation}
   z_0=\frac{\frac{4\sqrt{1-\eta}}{xs\,\,\,\eta}+\frac{xs(c_1-c_2)\eta}{1+\eta}}{6-xs^2\eta\frac{c_1+c_2\eta}{1+\eta}}\,,
\end{equation}
where $c_1=\beta_1/(xs^2((xs/R_1)-1))$, $c_2=\beta_2/((d-xs)^2((d-xs/R_2)-1))$, and $\beta_\mathrm{i}$ is the parameter of the beta-law of wind velocities.

Far from the line of centres ($y>0.2xs$), we solve the differential equation A7 of \citet{antokhin04} to compute the auxiliary function $\zeta_\mathrm{i}$ from the value of $\sigma$ when $y=0.2xs$:
\begin{equation}
   \frac{d\zeta_\mathrm{i}}{dy}=\frac{\dot{M}_\mathrm{i}\,\cos{\theta_\mathrm{i}}\,y}{4\pi\,r^2_\mathrm{i}\,\sin{\delta_\mathrm{i}}}\,.
\end{equation}
The surface density is:
\begin{equation}
   \sigma=\frac{\zeta_\mathrm{i}}{y\,v_\mathrm{i}(x,y)\cos{\delta_\mathrm{i}}}\,.
   \label{eq:rad_sigma2}
\end{equation}

\citet{vishniac94} showed that non-linear thin-shell instabilities can arise in narrow radiative shocks.
This instability leads to the reduction of the X-ray emission.
Using a 2D study of the effect, \citet{kee14} computed a reduction by a factor of about 50 of the emission for a collision between two equal winds. These authors stressed that further investigations are needed to assess how this reduction factor depends on the physical parameters of the shocks. This instability is not taken into account in \progname{}. 
To first order, we expect that this effect mostly affects the flux of the lines as well as the overall level of X-ray emission. For the shape of the line profiles, we expect that these instabilities will probably lead to a broadening of the lines due to an increased turbulence in the shock region.

\subsection{Coriolis deflection}
\label{eq:cor}
The methodology used to compute the Coriolis deflection is based on the work of \citet{parkin08} (see their Figure 7).
The shock is first divided into the shock cap and the shock tail.
We define the shock cap as the region where the tangential velocity is lower than 85\% of the lowest terminal velocity.
The Coriolis force has two effects on the shape of the shock: the skewing of the entire shock and the curvature of the shock tail.

The skewing angle is:
\begin{equation}
   s=\arctan{\left(\frac{v_\mathrm{orb}}{v_1(xs,0)}\right)}\,,
   \label{eq:skew}
\end{equation}
where $v_1(xs,0)$ is the wind velocity of star 1 at the stagnation point and $v^2_\mathrm{orb}=G\,(M_1+M_2)\,(2/d-1/a)$ is the orbital velocity of the star with $a$ the semi-major axis of the orbit.

The deflection is then computed going back along the orbit of the stars. 
Indeed, at a time $T_1=(E-e\sin{(E)})P/2\pi$ with $P$ the orbital period, $e$ the eccentricity, and $E$ the eccentric anomaly, the particles at the end of the shock cap move freely along a rectilinear trajectory whose direction and velocity depend on the component of the post-shock wind tangential to the shock considering the stars at rest.
At a time $T_2>T_1$, the stars have moved while the particles have continued on a linear trajectory.
At this time, a new shock cap is created.
Particles at the end of the shock cap are also ejected along a rectilinear trajectory.
The stellar motion continues leading to the escaping particles creating a tail of the shock, which is curved behind star 1.

The tail of the shock is thus created by considering several times in the past, computing the shock cap, and deducing the rectilinear trajectory of the particles at the end of the shock cap. 
The distance travelled by the particles is thus proportional to the velocity they had as they left the shock cap and to the time elapsed between the past times and the present time.

We computed the Coriolis deflection over 20 positions of the binary with a step of $0.003$ times the orbital period using ballistic motion.
After computing the linear trajectory of the particles and the centre of mass of the system $CM=d\,M_2/(M_1+M_2)$ for each position, we rotate the particles from the oldest to the closest position with:
\small
\begin{equation}
\left( \begin{array}{c} x'\\y'\\z'\end{array} \right)=\left( \begin{array}{ccc} \cos{(\phi_1-\phi_2)} & 0 & -\sin{(\phi_1-\phi_2)}\\
0 & 1 & 0 \\
\sin{(\phi_1-\phi_2)} & 0 & \cos{(\phi_1-\phi_2)}\end{array} \right)
\left( \begin{array}{c} x-CM\\y\\z\end{array} \right)
+\\
\left( \begin{array}{c} CM\\0\\0\end{array} \right)\,.
\end{equation}
\normalsize
The phases $\phi_1$ and $\phi_2$ are those of two consecutive positions with $\phi_1<\phi_2$.
At each loop, we rotate the particles ejected at the positions occurring before the time we are interested in (corresponding to phase $\phi_2$).
As a result, at the end of the 20 loops, the oldest position (farthest from the stagnation point) was rotated 20 times while the closest position was rotated only once.

\subsection{Computation of the line profile}
\label{rt}
The line profile is built as an histogram of the flux received by the observer as a function of the radial velocities of the cells of material that emit the received photons.
The radial velocity is the velocity of the cell along the line of sight.
The velocity vector of the cell is assumed to be tangential to the shock surface.
The observed flux at a given energy is a combination of emission and absorption.

We now consider a cell inside the wind interaction zone.
This cell is characterised by a temperature $T$, a volume $V$, and a volume density $\rho$.
Assuming ionisation equilibrium, the emissivity (or its decimal logarithm, $q$) for a certain line of a given ion is set by the temperature of the cell.
As for the equality between the electron and the ion temperature, the ionisation equilibrium is a strong assumption but only affects the overall strength of the line.
The volume density defines the electron ($n_\mathrm{e}=1.17n_\mathrm{H}$) and particle densities ($n_\mathrm{H}$) considering solar abundance with:
\begin{equation}
   n_\mathrm{H}=\frac{\rho}{1.34m_\mathrm{H}}\,.
\end{equation}

The absorption depends on the medium crossed by the photon.
If the photon crosses a cell inside the shock, the cross-section of the hot plasma inside the crossed cell depends on the temperature of this cell and the energy of the photon. 
The cross-sections are taken from the AtomDB database \citep{foster12} as a function of plasma temperature and photon energy.
The optical depth of the crossed cell is 
\begin{equation}
   \tau_\mathrm{shock}=\frac{\kappa\,\sigma}{\vec{P}\cdot \vec{N}}\,,
   \label{tau_shock}
\end{equation}
where $\sigma$ is the surface density, $\vec{P}$ is the position vector of the centre of the cell, and $\vec{N}$ is the direction of the observer.
The opacity $\kappa$ is computed as the cross-section divided by $1.3m_\mathrm{p}$.

\begin{figure}
\centering
\includegraphics[width=7cm]{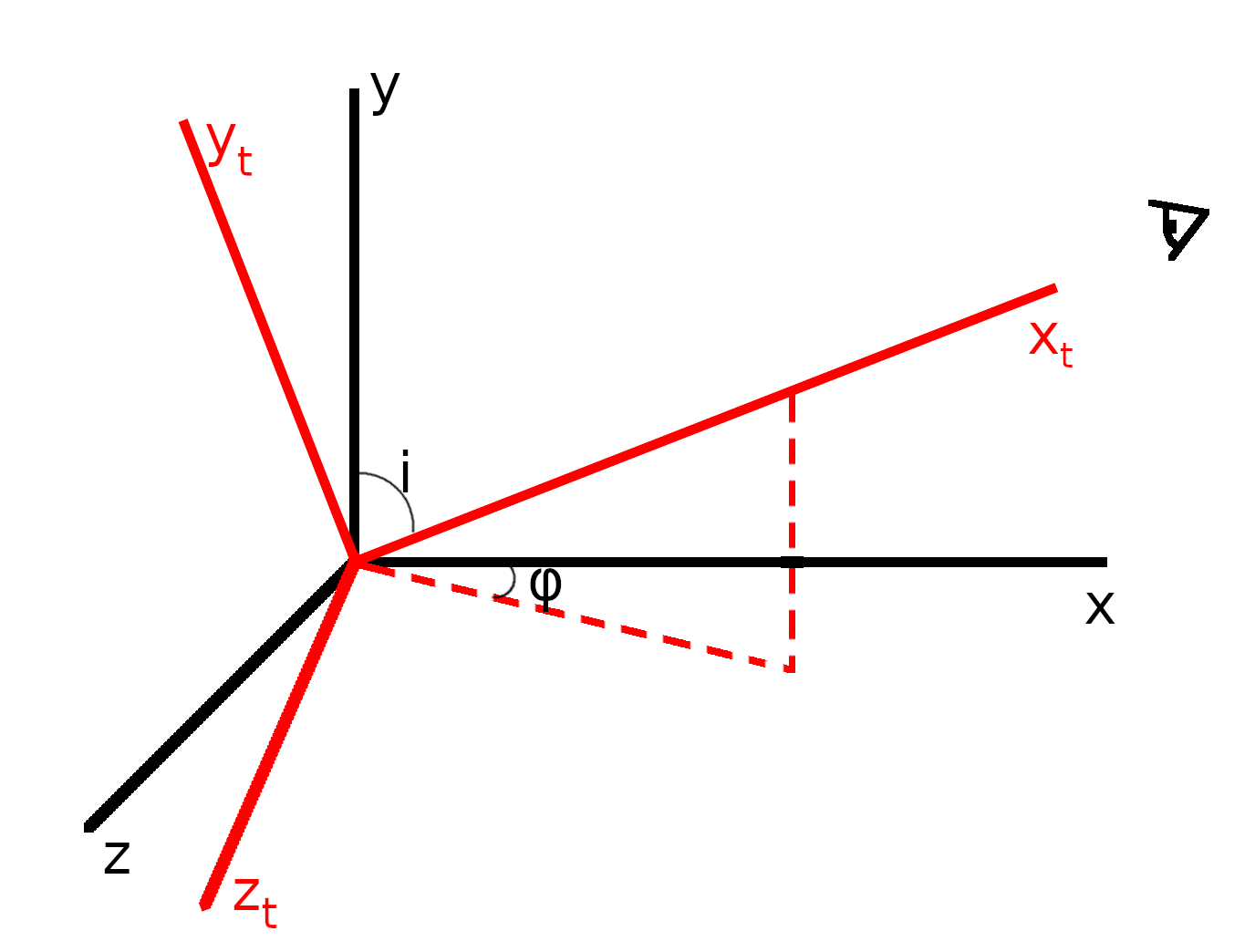}
\caption{Coordinate system used to compute the absorption by the cool stellar winds.}
\label{pz}
\end{figure}

The optical depth of the cool, unshocked, wind material along the line of sight is computed as follows \citep{rauw07}, assuming a simplified wind velocity law $v(r) = v_{\infty}\,(1 - R/r)$.
We first create a new coordinate system centred on the star that is located in front at the considered orbital phase. 
The $x_\mathrm{t}$ axis points towards the observer (characterised by an inclination $i$ and a phase $\phi$) and the axis $y_\mathrm{t}$ and $z_\mathrm{t}$ are perpendicular to $x_\mathrm{t}$ (see Fig.~\ref{pz}).
The coordinate $p_\mathrm{t}$ is defined as $(y_\mathrm{t}^2+z_\mathrm{t}^2)^{0.5}$.
The optical depth is thus:
\begin{equation}
\tau_\mathrm{wind}= \left\{
  \begin{aligned}
   \infty &\mathrm{\ \ \ for\ }p_\mathrm{t}<R\\
   \frac{2\tau_0}{\alpha-1} &\mathrm{\ \ \ for\ }p_\mathrm{t}=R\\
   \frac{2\tau_0}{\sqrt{p_\mathrm{t}^2/R^2-1}}\left(\frac{\pi}{2}-\arctan\left(\frac{\alpha\,p_\mathrm{t}/R-1}{\sqrt{p_\mathrm{t}^2/R^2-1}}\right)\right) &\mathrm{\ \ \ for\ }p_\mathrm{t}>R\,,
  \end{aligned}
\right.
   \label{tau_wind}
\end{equation}
where $R$ is the radius of the star that is located in front, and where
\begin{equation}
\alpha= \left\{
  \begin{aligned}
   \tan\left(\frac{\arctan{(p_\mathrm{t}/x_\mathrm{t})}}{2}\right) &\mathrm{\ \ \ for\ }x_\mathrm{t}<0\\
   \tan\left(\frac{\pi}{4}\right) &\mathrm{\ \ \ for\ }x_\mathrm{t}=0\\
   \tan\left(\frac{\pi-\arctan{(p_\mathrm{t}/x_\mathrm{t})}}{2}\right) &\mathrm{\ \ \ for\ }x_\mathrm{t}>0
  \end{aligned}
\right.
\end{equation}
and $\tau_0=\kappa\dot{M}_\mathrm{i}/(4\pi\,v_\mathrm{i}\,R_\mathrm{i})$ with $v_\mathrm{i}$ the velocity of the wind at the considered position.

We also test the visibility of each cell. Indeed, if the cell is hidden by one of the stars, its emission will not be recorded (which is equivalent to setting the wind optical depth to $\infty$). 
If the emitting cell is visible and has a radial velocity $v_\mathrm{rad}$ with respect to the observer, the absorbed emission by the cell of volume $V$, computed as $10^q\,n_\mathrm{H}\,n_\mathrm{e}\,V\,e^{-\sum(\tau_\mathrm{shock}+\tau_\mathrm{wind})}$, is added to the histogram of the observed fluxes at a Doppler shift corresponding to $v_\mathrm{rad}$.

\section{Simulation grid of O-type stars}
\label{grid}
We have computed the wind distribution, the shock characteristics and the iron line profiles for a long set of binary systems composed of O-type stars.
We considered each combination of two stars characterised by the parameters reported in Table~\ref{stellar_par}.
We adopted the temperatures, masses, and radii from Table 1--3 of \citet{martins05}, and the mass-loss rates from \citet{muijres12}.
For each couple of stars, we considered ten values of the orbital separation.
Six separations were linearly spaced from 1.05 times the sum of the stellar radii to 0.9 times the critical distance satisfying the criterion $v^4xs/\dot{M}=1$ leading to the construction of a radiative shock region.
The four other separations are linearly spaced from 1.1 times the critical distance to 2000$\rsun{}$ leading to the construction of an adiabatic shock region.
This leads to 780 configurations with star 1 having the less powerful wind. 
The line profiles were simulated for five orbital phases (from 0.0 to 0.8 in steps of 0.2) and five inclinations (from 18 to 90$^\circ$).
In our grid, we consider only circular orbits. Non-zero eccentricities are implemented in the code, but were not used here as they would imply a much heavier grid.
The phase $\phi=0$ is defined at the conjunction (i.e. when the radial velocities of the stars are equal to zero) with the star with the most powerful wind in front.
We used version 3.0.9 of the AtomDB atomic database in our computations.

\begin{table}
\caption{Table of parameters of the precomputed systems.}
\label{stellar_par}
\centering
\begin{tabular}{ccccc}
\hline
Spectral type & $T_\mathrm{eff}$ & $R$ & $M$ & $\log{(\dot{M})}$ \\
 & ($K$) & ($\rsun{}$) & ($\msun{}$) & ($\msun{}\,yr^{-1}$) \\
\hline
\hline
O3V&44616& 13.84& 58.34& -5.641\\
O5V&41540& 11.08& 37.28& -5.969\\
O7V&35531& 9.37& 26.52& -7.340\\
O9V&31524& 7.73& 18.03& -7.818\\
O3III&42942& 16.57& 58.62& -5.445\\
O5III&39507& 15.26& 41.48& -5.630\\
O7III&34638& 14.51& 31.17& -6.804\\
O9III&30737& 13.69& 23.07& -6.812\\
O3I&42551& 18.47& 66.89& -5.347\\
O5I&38520& 19.48& 50.87& -5.561\\
O7I&33326& 21.14& 40.91& -5.995\\
O9I&29569& 22.60& 31.95& -6.385\\
\hline
\end{tabular}
\end{table}

\section{Results}
\label{discuss}
We present here the results of the simulation of the Fe~K helium-like triplet for three representative systems.

\subsection{Fe~K helium-like triplet}
\label{iron_line}
As mentioned in Sect~\ref{intro}, the iron line has the advantage to be emitted at high energy, where the contamination by the intrinsic emission from the stellar winds themselves and the absorption by the cool unshocked plasma are negligible.
High plasma temperatures are required to reach such a high degree of ionisation leading to the emission of the Fe~K helium-like triplet at energies around 6.7~keV.
Indeed, the emissivity of this line peaks near $T\sim 10^{7.8}\,$MK (5.4\,keV).
Such temperatures can be reached inside the wind-shock region considering the very high wind velocities of massive stars.

The Fe~K helium-like triplet is composed of a resonance line (6.7004\,keV), an intercombination doublet (6.6823 and 6.6676\,keV), and a forbidden line (6.6366\,keV) as taken from the AtomDB database \citep{foster12}.
As shown by \citet{rauw16b}, we can neglect the impact of photospheric radiation on the relative strengths of the components of the triplet and the suppression of the forbidden line by collisional excitation.
Indeed, the impact of these phenomena is, at maximum (i.e. at the stellar surface) a reduction of 8\% of the ratio between the forbidden and intercombination lines.
The relative strength then rapidly reaches the ratio computed in the absence of photospheric radiation and collisional excitation as the distance from the stellar surface increases.

In the range of energy where the Fe~K lines are emitted, satellite lines from Fe~{\sc xxiv} created by dielectric transitions $1{\rm s}^2 nl{-}1{\rm s} 2{\rm p} nl$ ($n\geqslant 2$) are expected \citep{gabriel72}. 
Diagnostics inferred from high-resolution X-ray spectroscopy of hot collisional plasmas must account for these lines \citep{hitomi}.  
To discuss the contamination by these satellite lines, we included, in Appendix~\ref{total_lp}, the $n=2$ \textit{j} and \textit{k} (6.6445 and 6.6541~keV) as well as the $n\leqslant3$ \textit{q} and \textit{r} (6.6644 and 6.6533~keV) transitions in the calculation of the total Fe~K line complex.

\subsection{Analysis of three representative systems}
\label{three_sys}
\begin{figure*}
\centering
\includegraphics[trim= 2cm 1cm 2cm 1cm,clip, width=14cm]{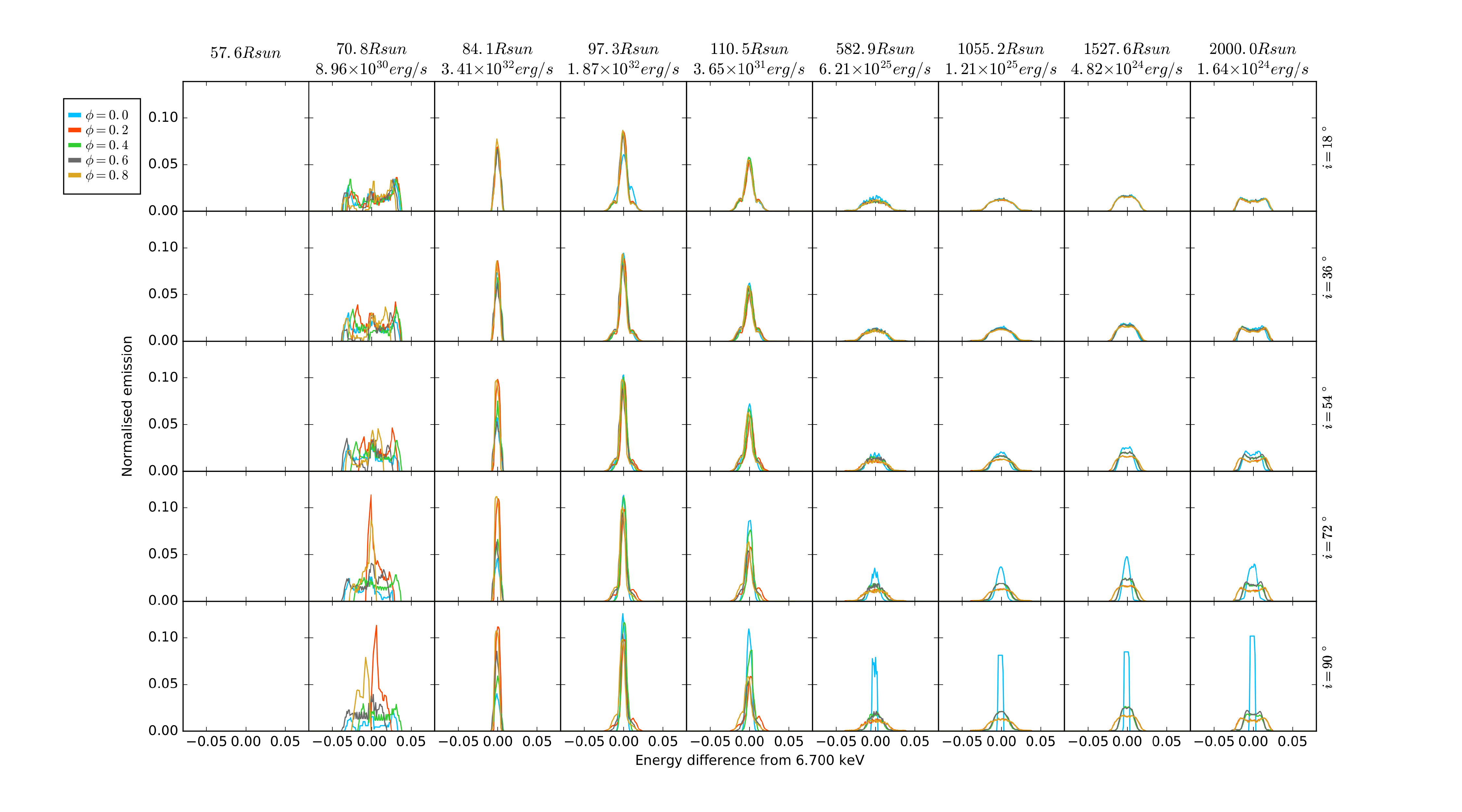}\\
\includegraphics[trim= 1cm 1cm 2cm 1cm,clip, width=14cm]{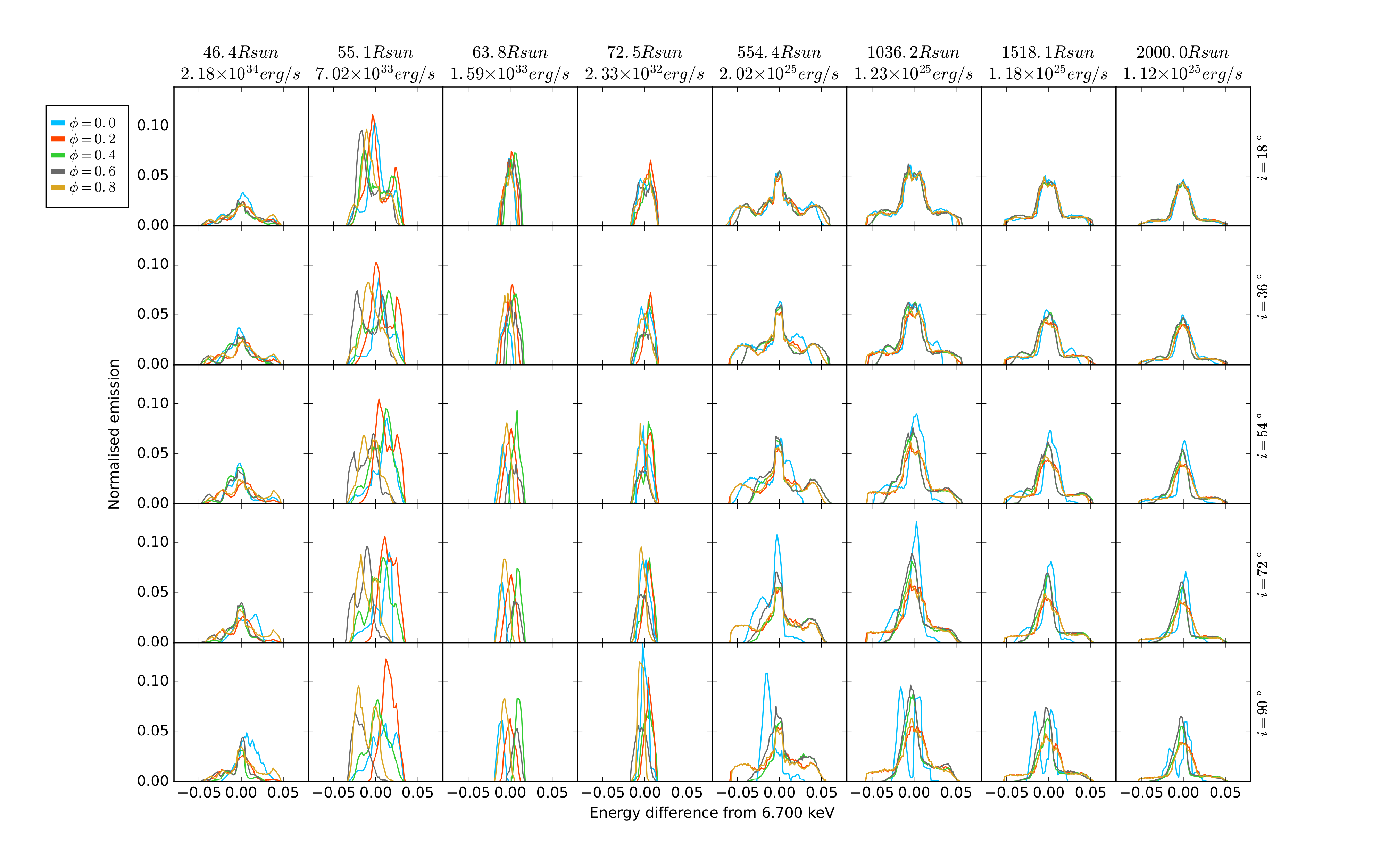}\\
\includegraphics[trim= 2cm 1cm 2cm 1cm,clip, width=14cm]{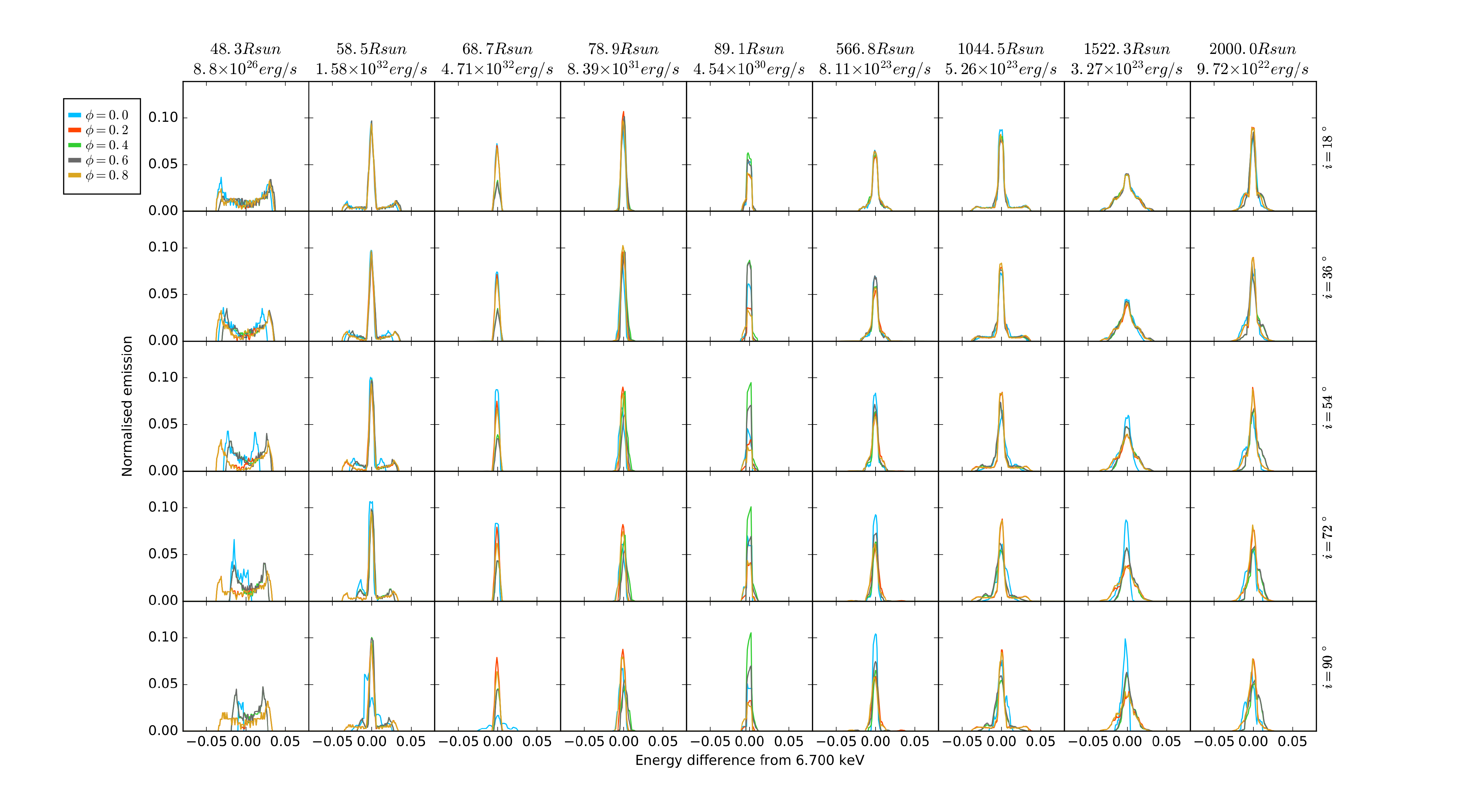}
\caption{Line profiles computed for three systems: O7V+O7V (top panel), O5I+O3III (middle panel), and O9III+O9V (bottom panel).
See text for details.}
\label{line_prof}
\end{figure*}

Since the detailed analysis of the overall 780 configurations is beyond the scope of the paper, we decided instead to highlight three systems representative for the variety of O-type binaries that exist.
We selected systems having very different values of wind momentum ratio $\beta$ and different spectral types.
We choose to present the results of the O7V+O7V ($\beta=1$), O5I+O3III ($\beta=0.66$), and O9III+O9V ($\beta=0.11$) binaries.
For each valid configuration (i.e. the dominant wind non crashing onto the companion's photosphere), Fig.~\ref{line_prof} shows the line profiles of the Fe~K resonance line for the three systems considering the ten distances and the 25 couples of inclinations and phases whose distributions were explained in Sect.~\ref{grid}.
For clarity, we only show the line profiles of the resonance line that has the highest emissivity of all components. 
Plots of the overall Fe~K helium-like triplet can be found in Appendix~\ref{total_lp}.
Each system is represented by one panel.
In each panel, the distances are represented by the columns while the five inclinations are represented by the rows.
In each subplot, the five curves illustrate the profiles at each phase as a function of the radial velocity of the emitting cell.
Since the integrated flux does not vary much with the inclination and phase, the line profiles at each separation are normalised by the highest integrated flux of the considered column.
The value of this highest flux is given at the top of each column.
The values of the normalised emission are given on the left side of the first subplots column.
If the normalised emission of a subplot is higher than that y-axis range, a new y-axis range specific to this subplot is defined (e.g. for the O7V+O7V binary in adiabatic regime with an inclination of $90^\circ{}$ in the figures of Appendix~\ref{total_lp}).

\begin{figure*}
\begin{picture}(600,700)
\put(0,600){\includegraphics[trim= 0cm 0cm 0cm 1.2cm,clip, width=5.cm]{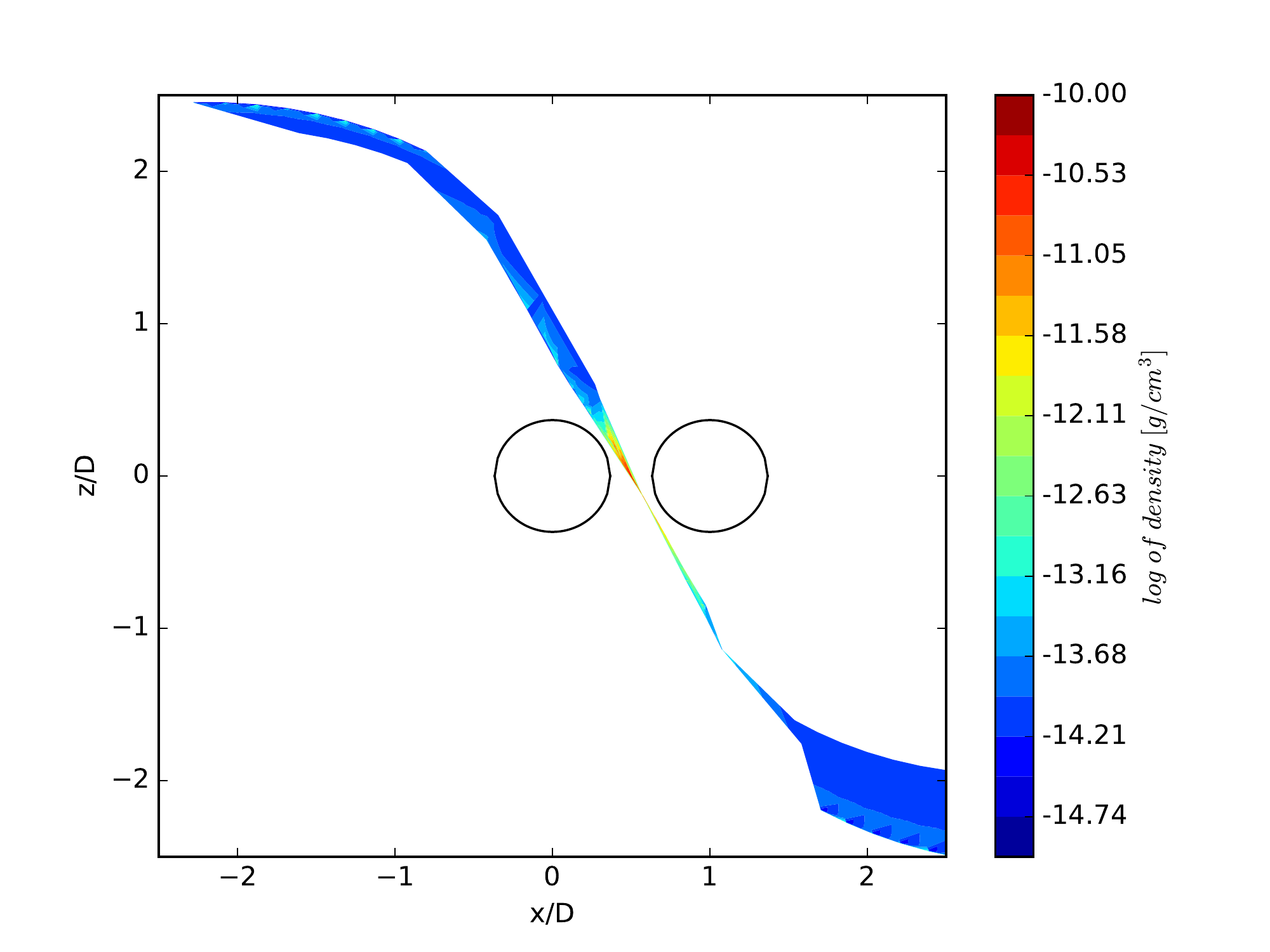}}
\put(133,600){\includegraphics[trim= 0cm 0cm 0cm 1.2cm,clip, width=5.cm]{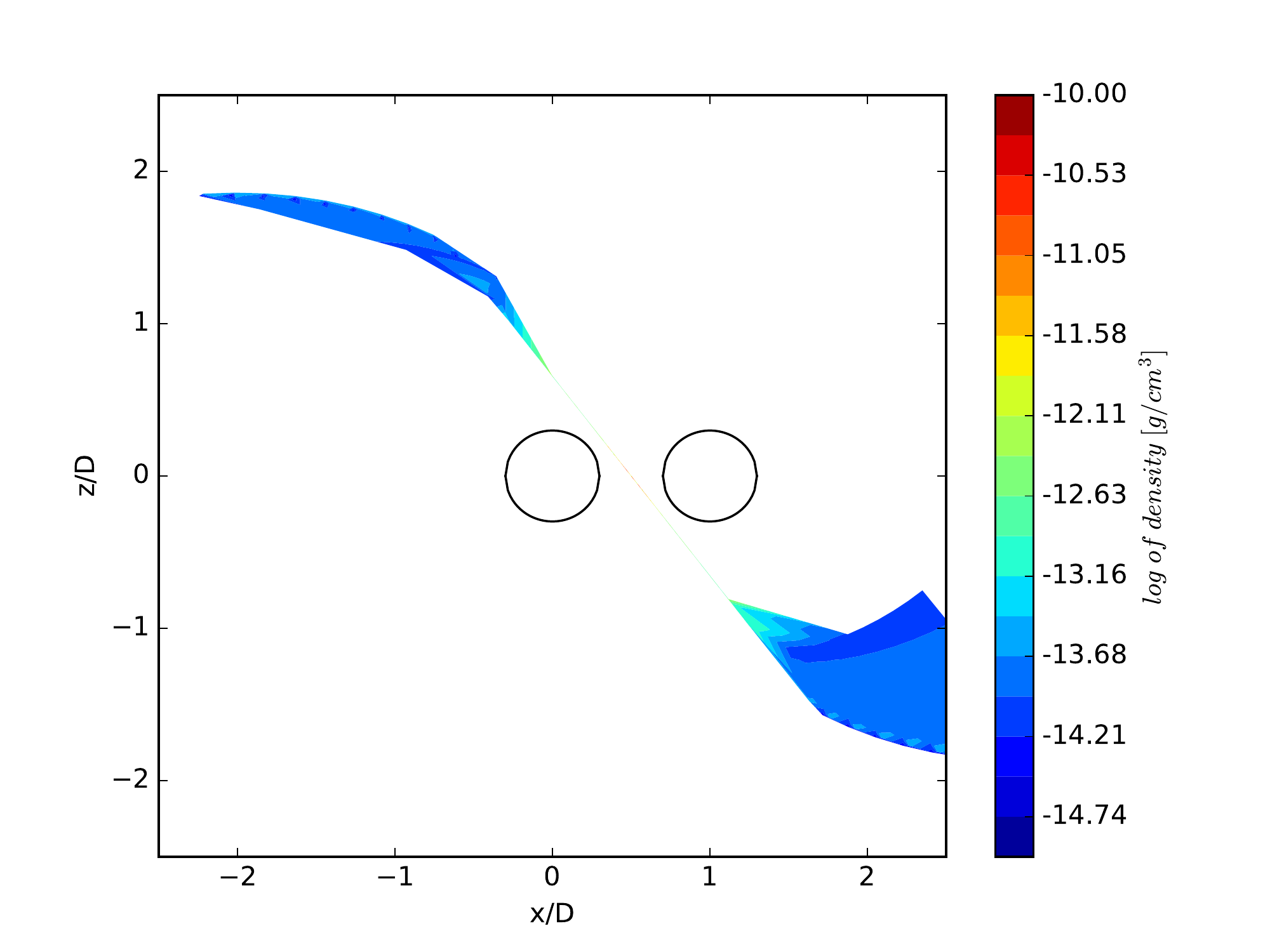}}
\put(266,600){\includegraphics[trim= 0cm 0cm 0cm 1.2cm,clip, width=5.cm]{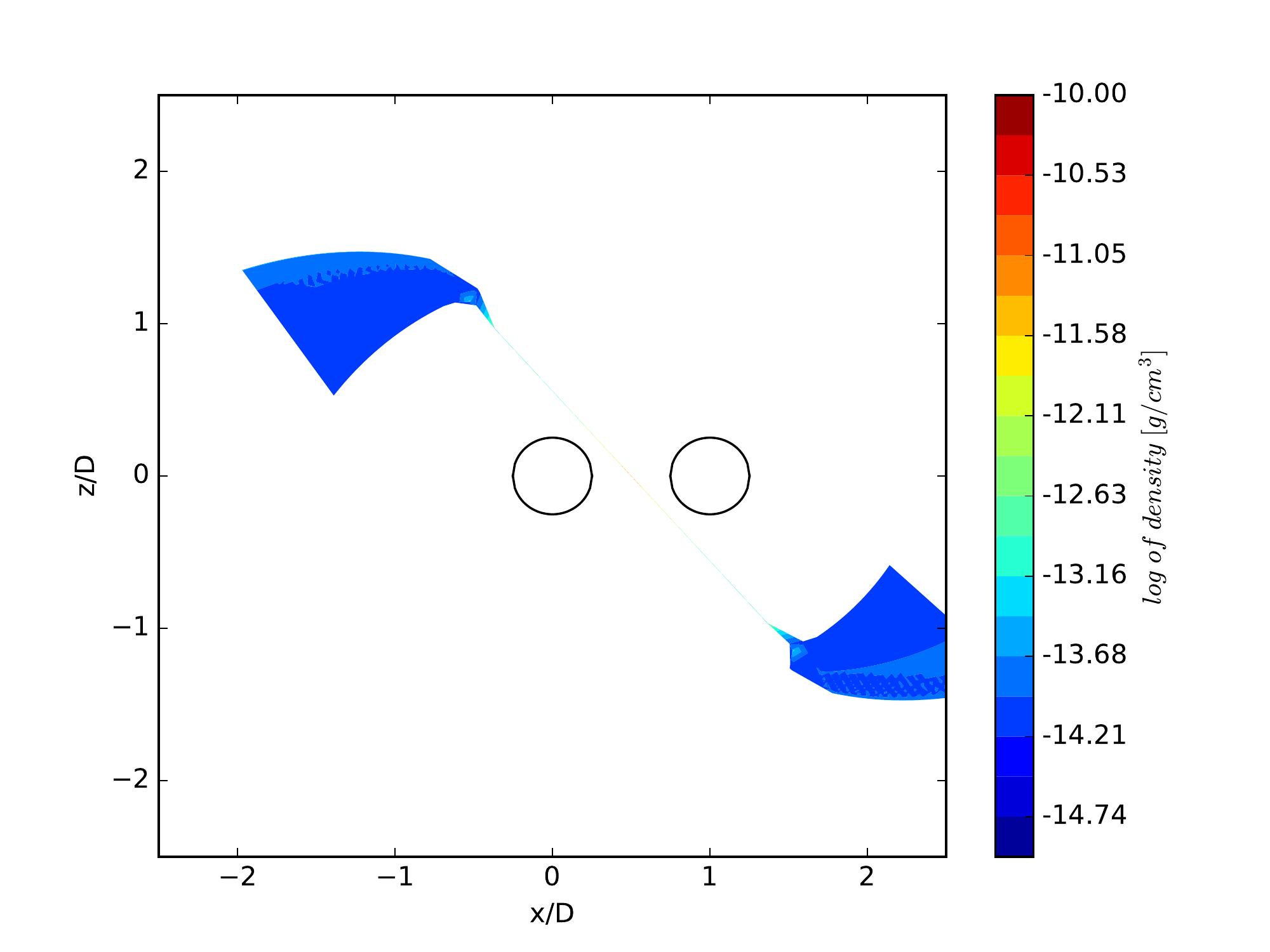}}
\put(400,600){\includegraphics[trim= 0cm 0cm 0cm 1.2cm,clip, width=5.cm]{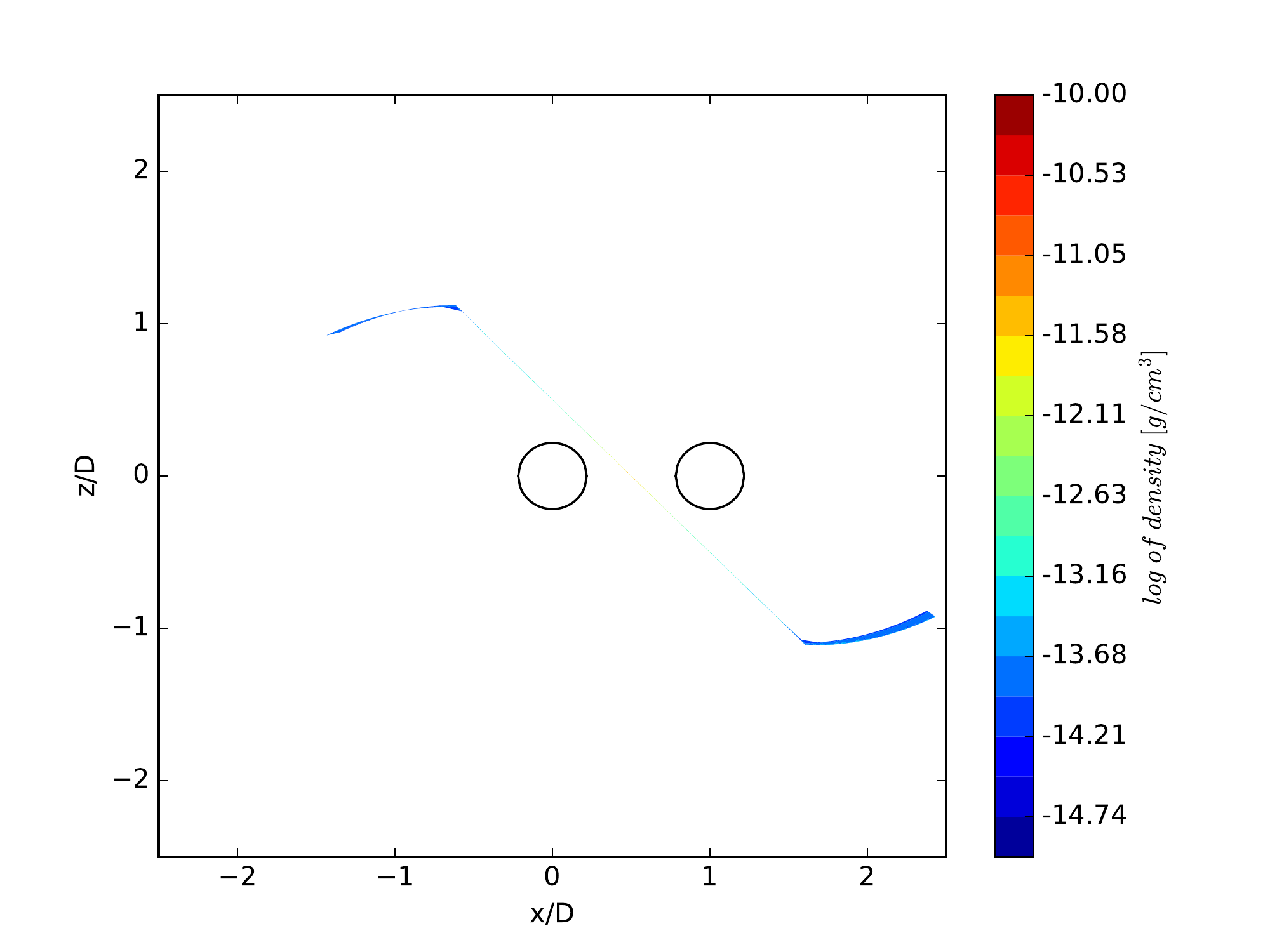}}

\put(0,500){\includegraphics[trim= 0cm 0cm 0cm 1.2cm,clip, width=5.cm]{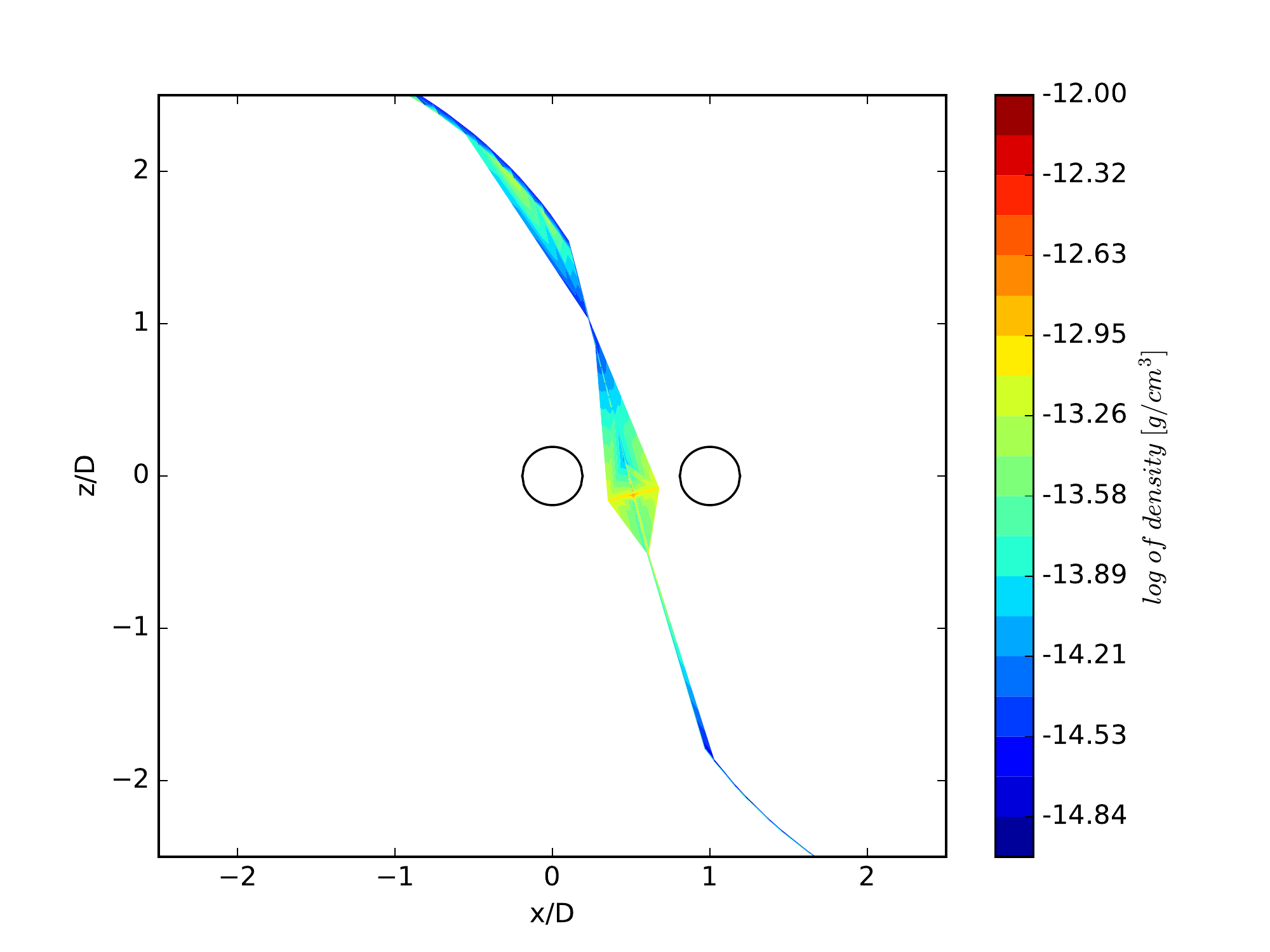}}
\put(133,500){\includegraphics[trim= 0cm 0cm 0cm 1.2cm,clip, width=5.cm]{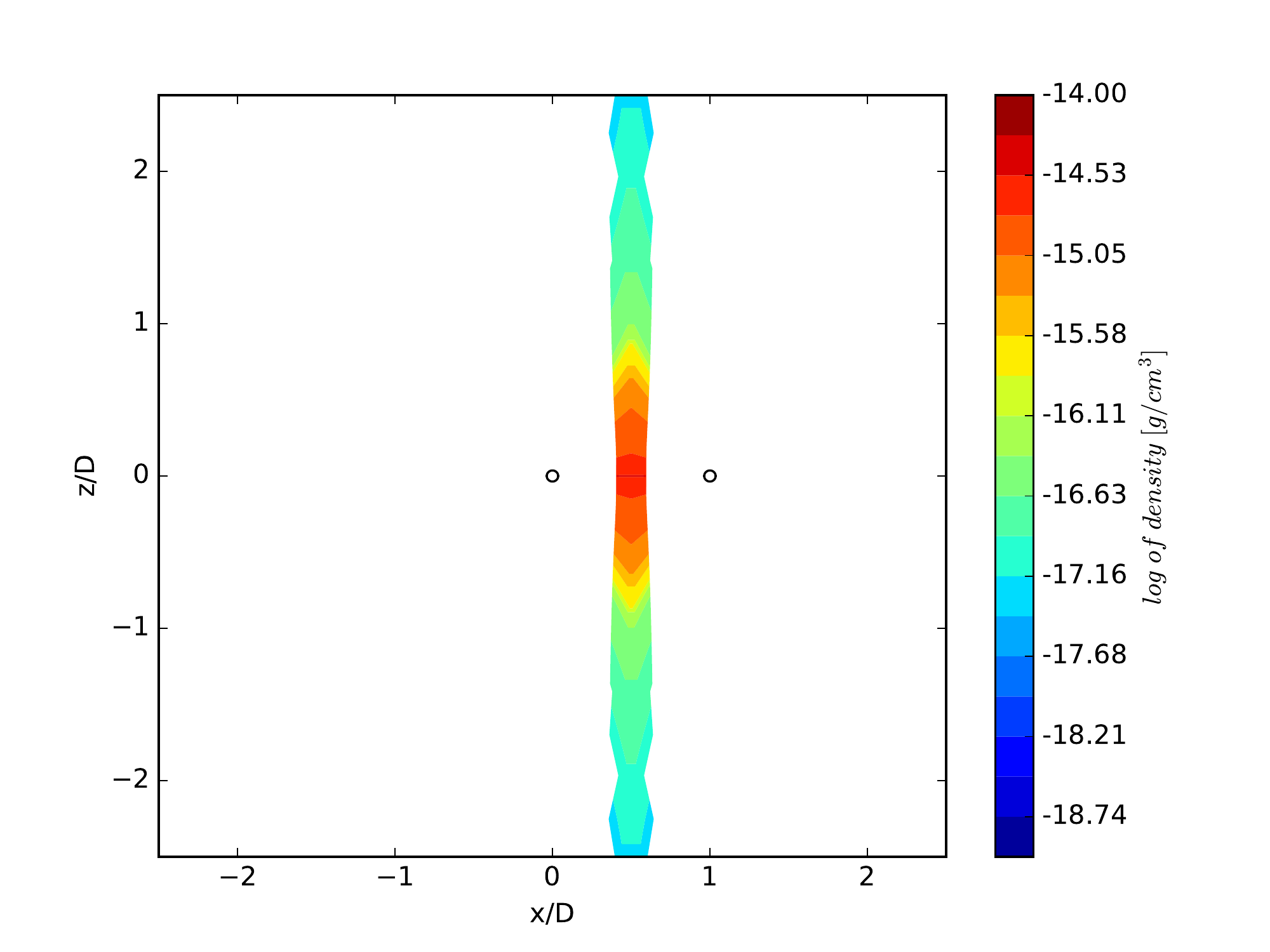}}
\put(266,500){\includegraphics[trim= 0cm 0cm 0cm 1.2cm,clip, width=5.cm]{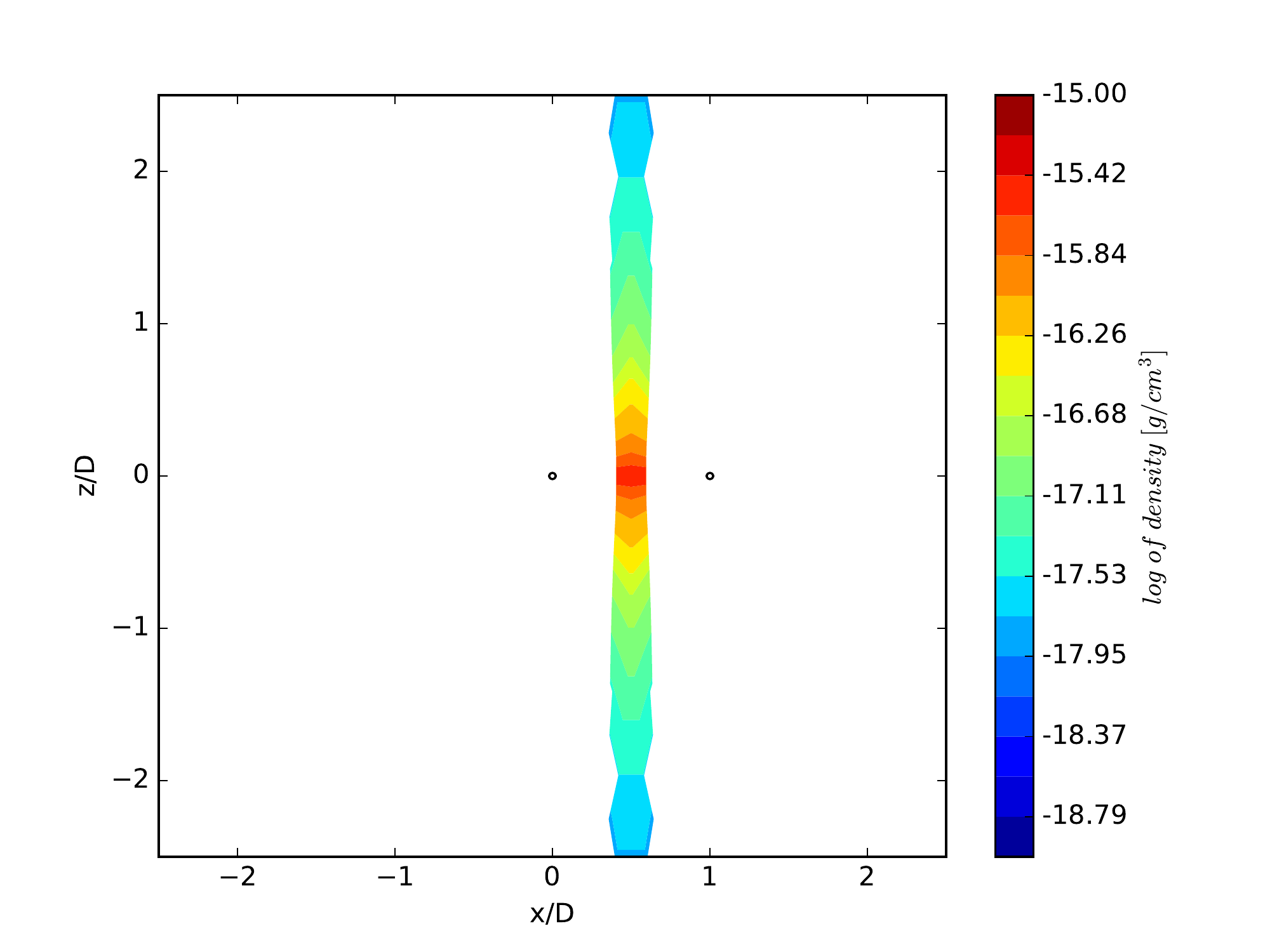}}
\put(400,500){\includegraphics[trim= 0cm 0cm 0cm 1.2cm,clip, width=5.cm]{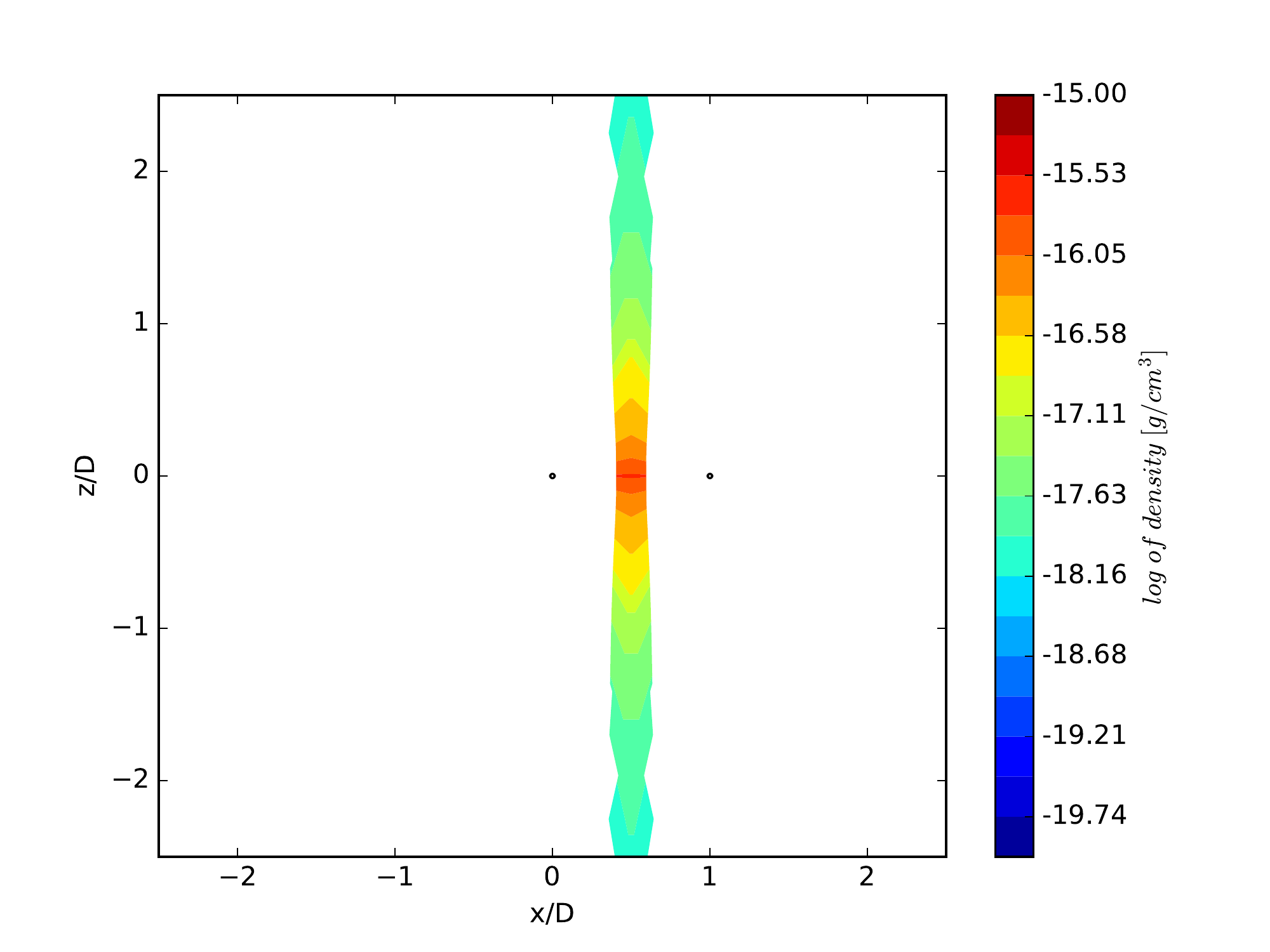}}

\put(0,400){\includegraphics[trim= 0cm 0cm 0cm 1.2cm,clip, width=5.cm]{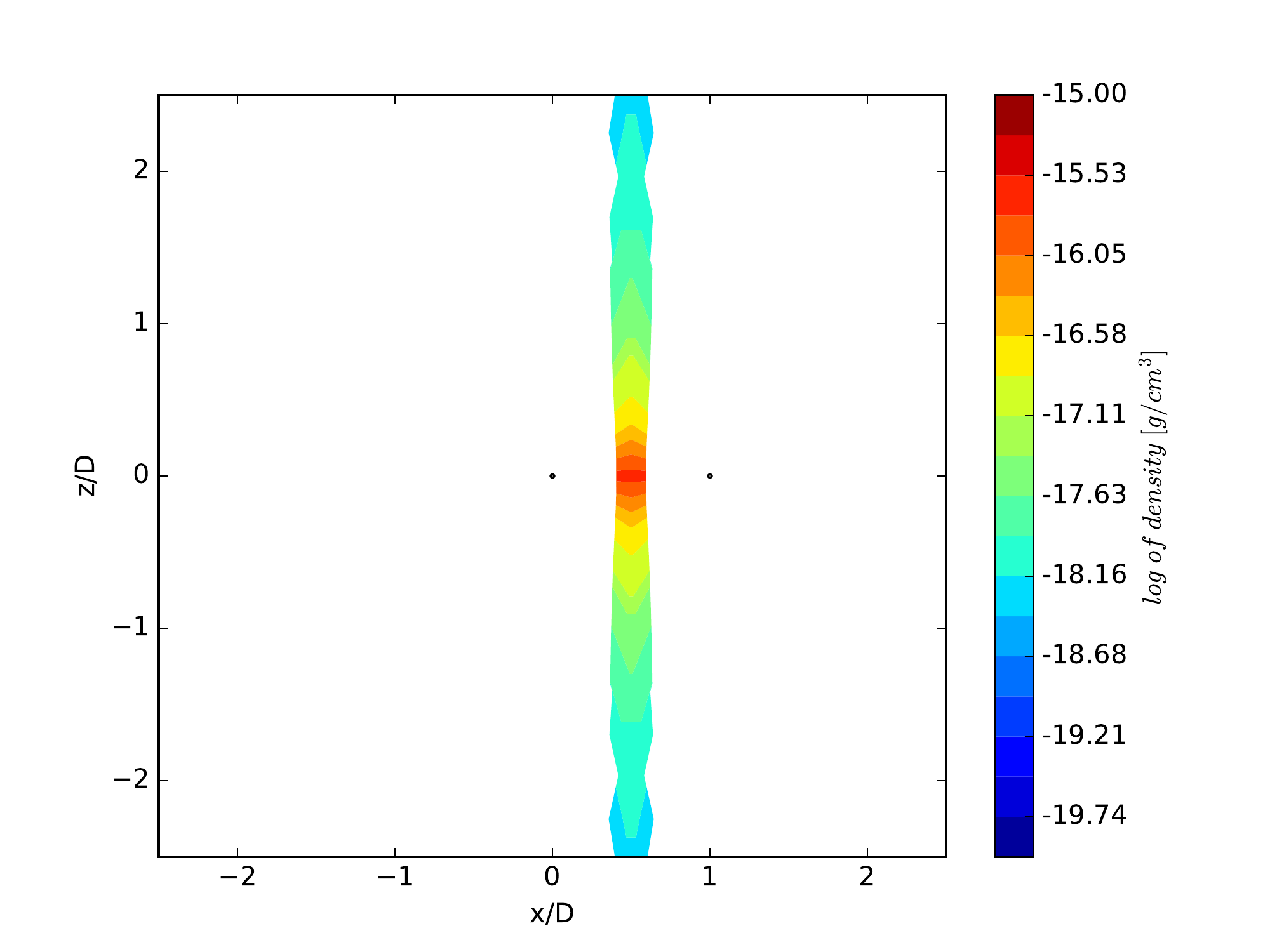}}
\put(133,400){\includegraphics[trim= 0cm 0cm 0cm 1.2cm,clip, width=5.cm]{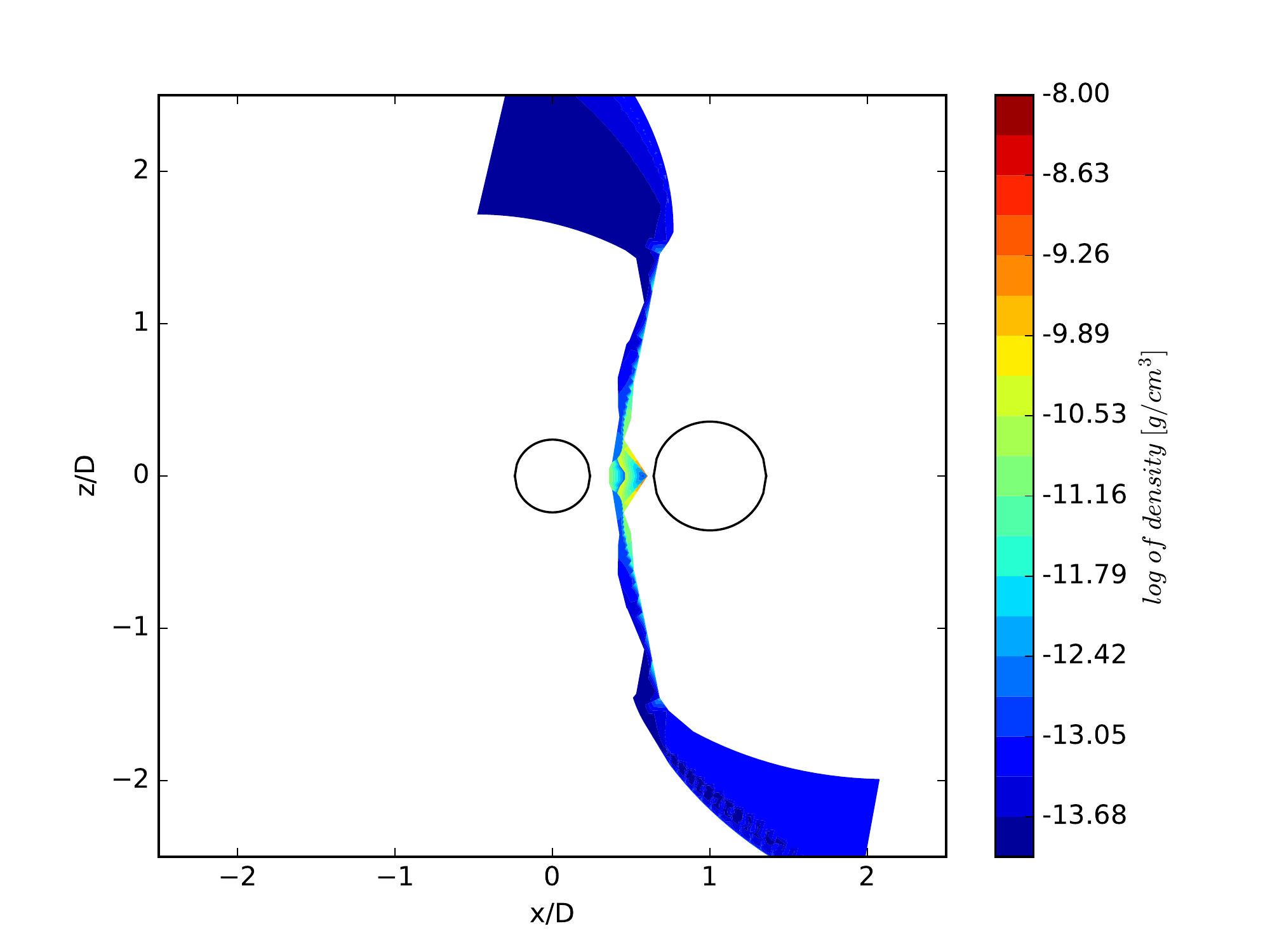}}
\put(266,400){\includegraphics[trim= 0cm 0cm 0cm 1.2cm,clip, width=5.cm]{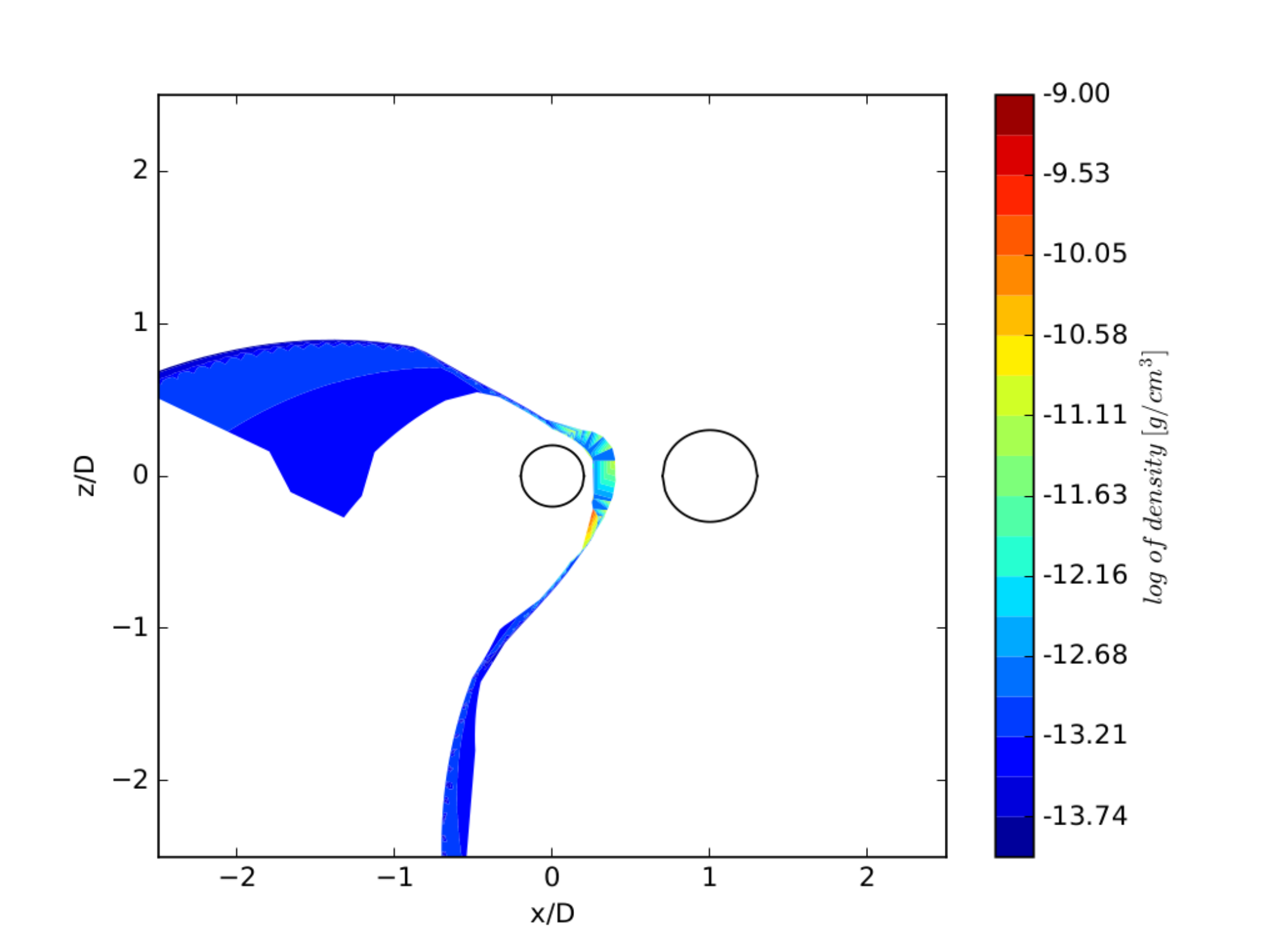}}
\put(400,400){\includegraphics[trim= 0cm 0cm 0cm 1.2cm,clip, width=5.cm]{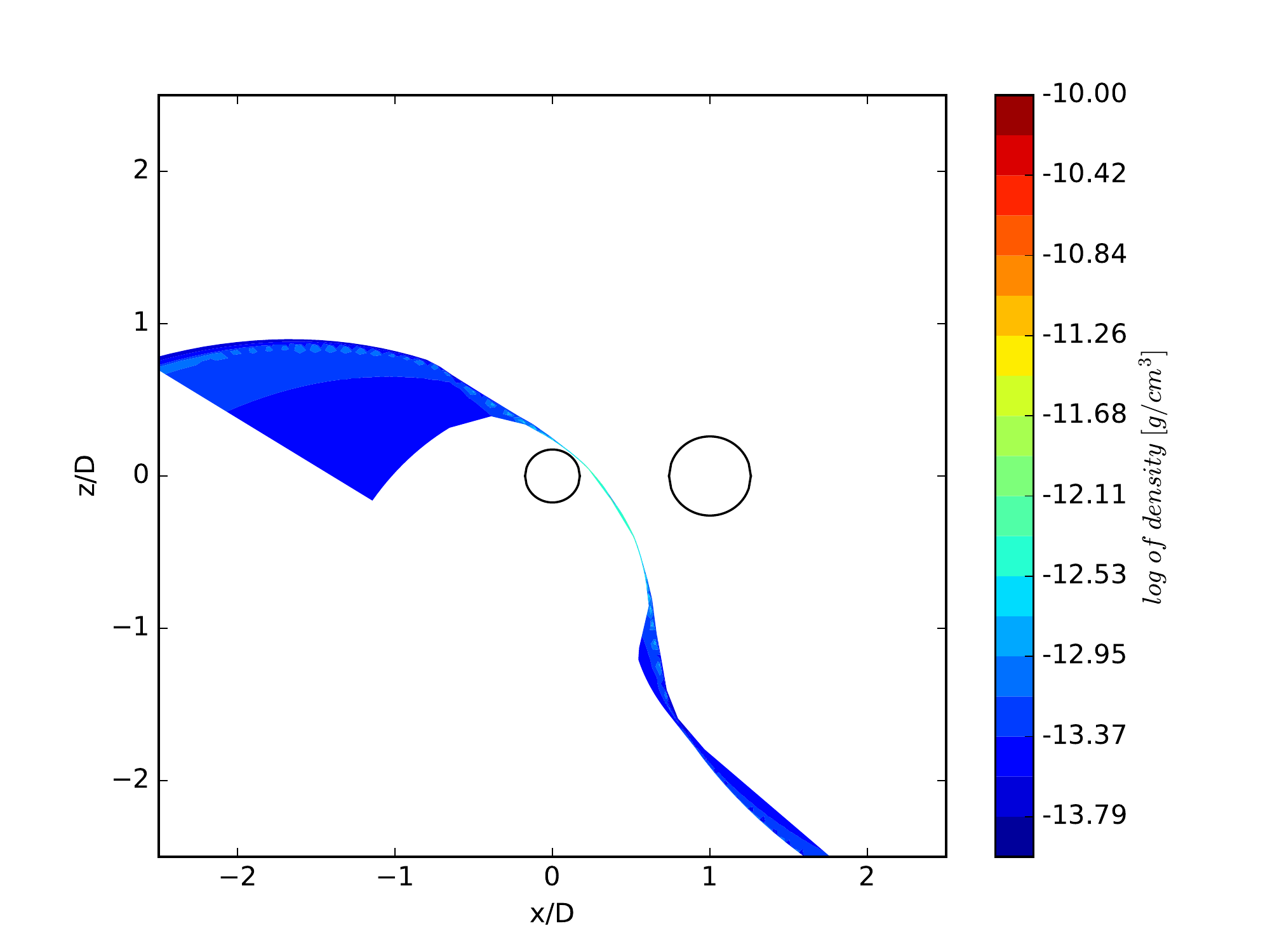}}

\put(0,300){\includegraphics[trim= 0cm 0cm 0cm 1.2cm,clip, width=5.cm]{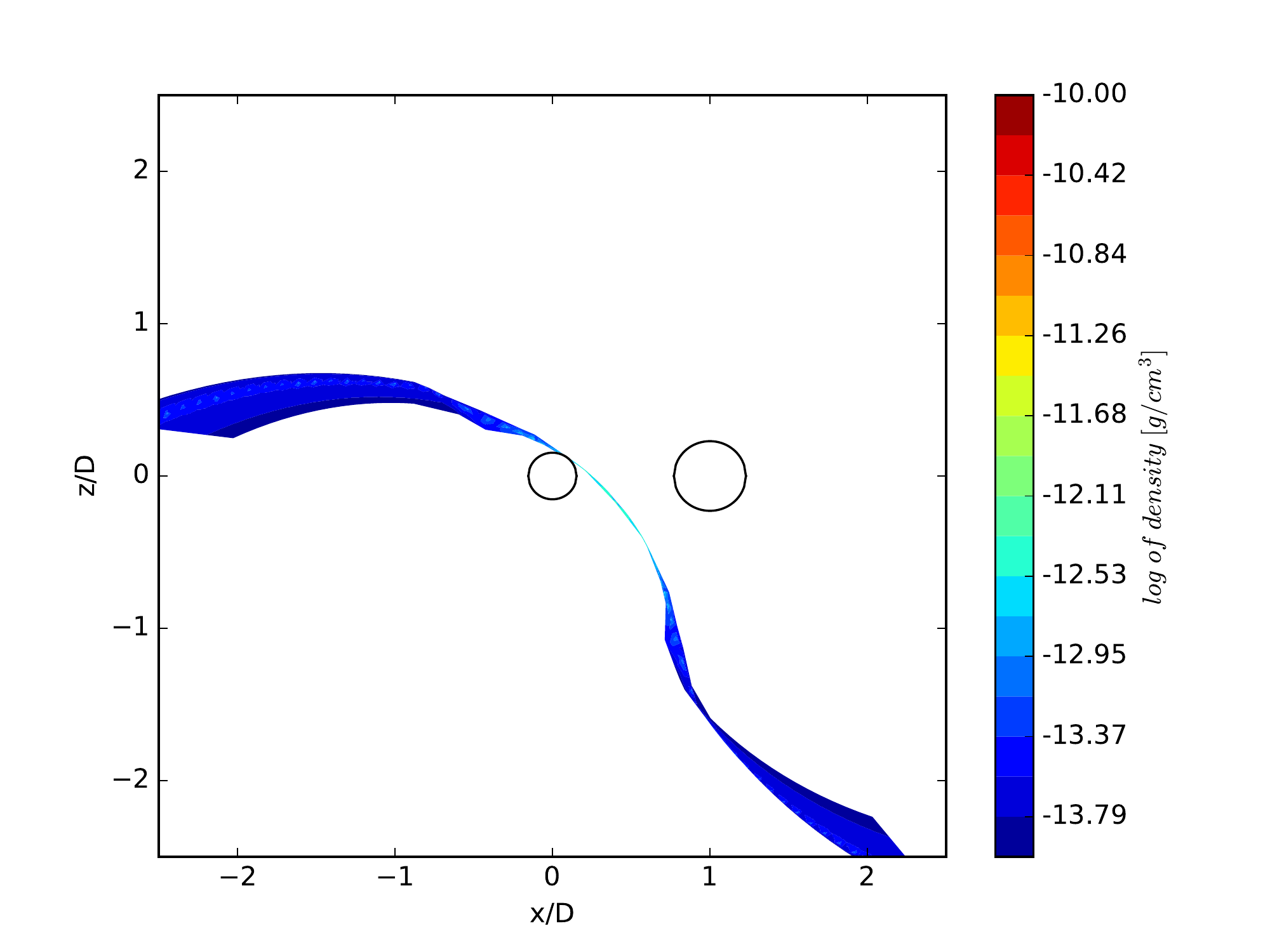}}
\put(133,300){\includegraphics[trim= 0cm 0cm 0cm 1.2cm,clip, width=5.cm]{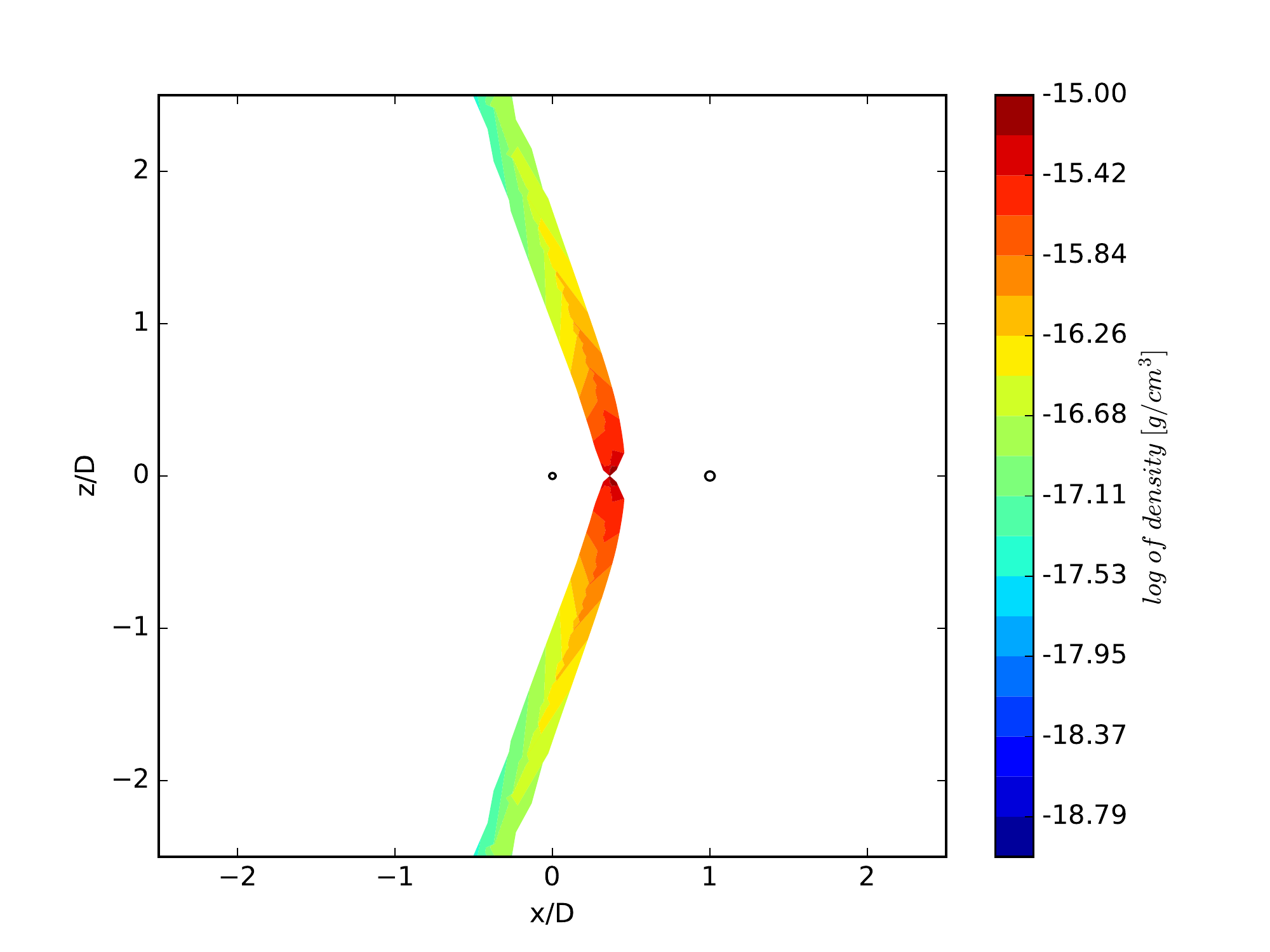}}
\put(266,300){\includegraphics[trim= 0cm 0cm 0cm 1.2cm,clip, width=5.cm]{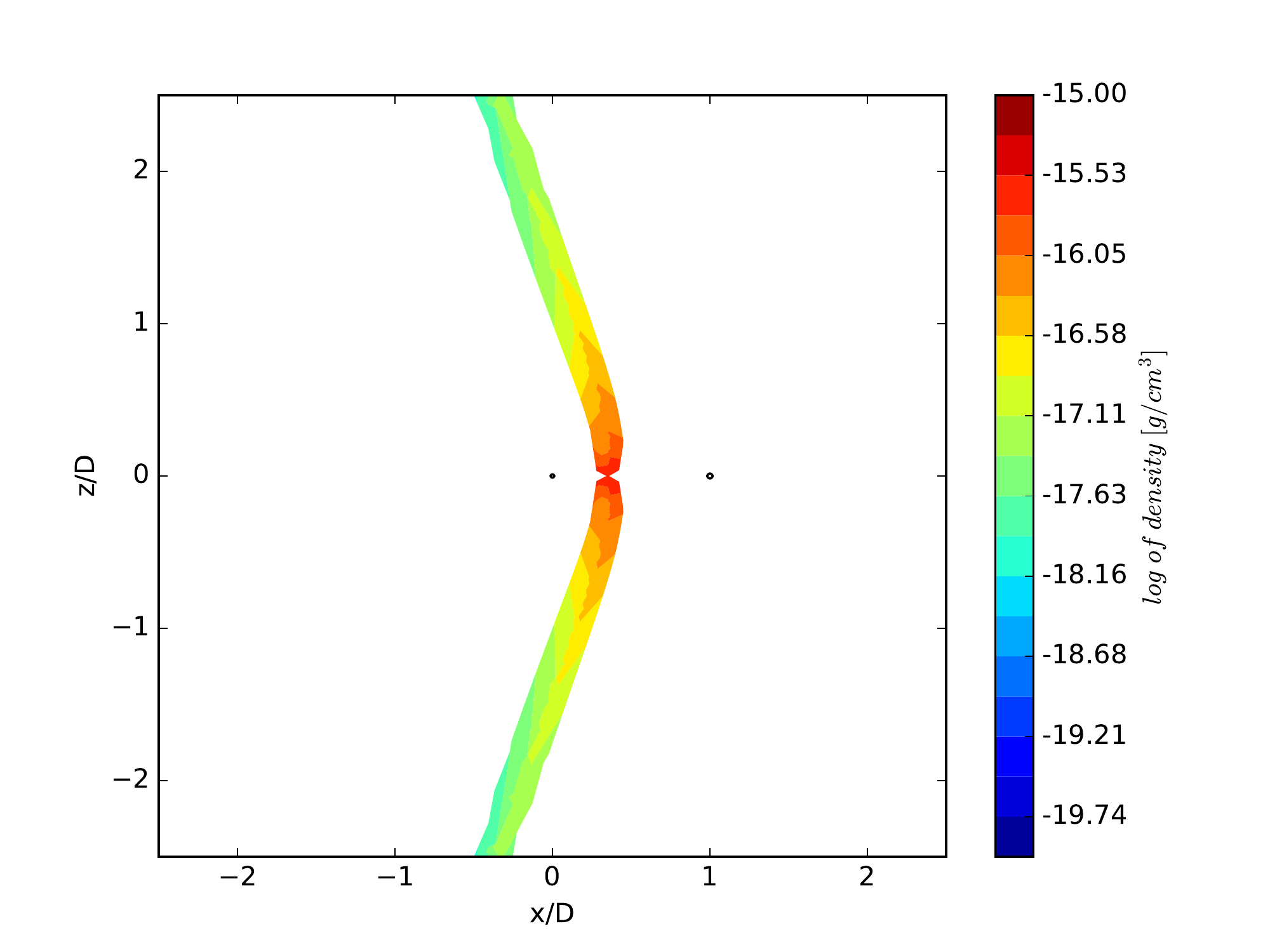}}
\put(400,300){\includegraphics[trim= 0cm 0cm 0cm 1.2cm,clip, width=5.cm]{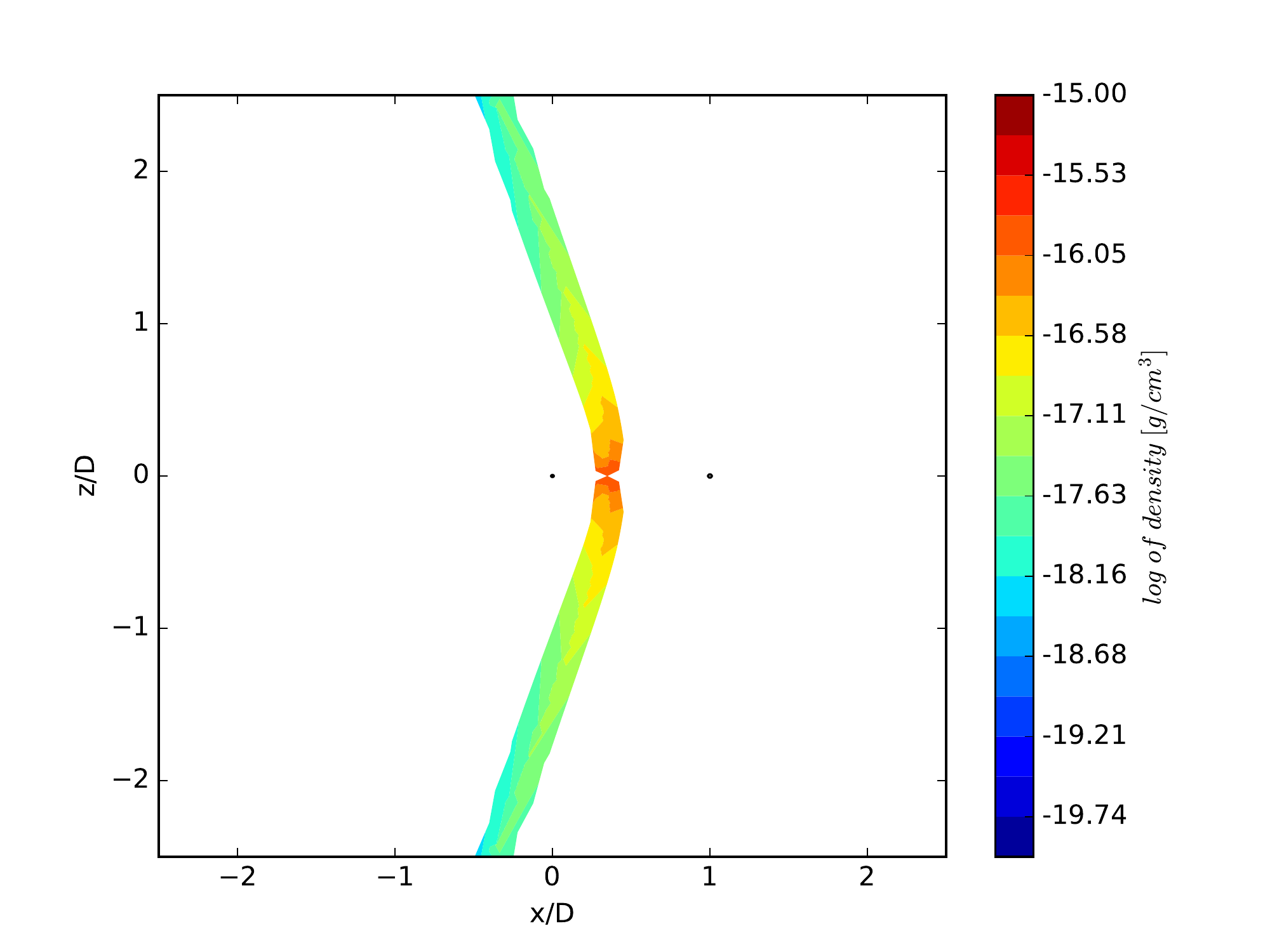}}

\put(0,200){\includegraphics[trim= 0cm 0cm 0cm 1.2cm,clip, width=5.cm]{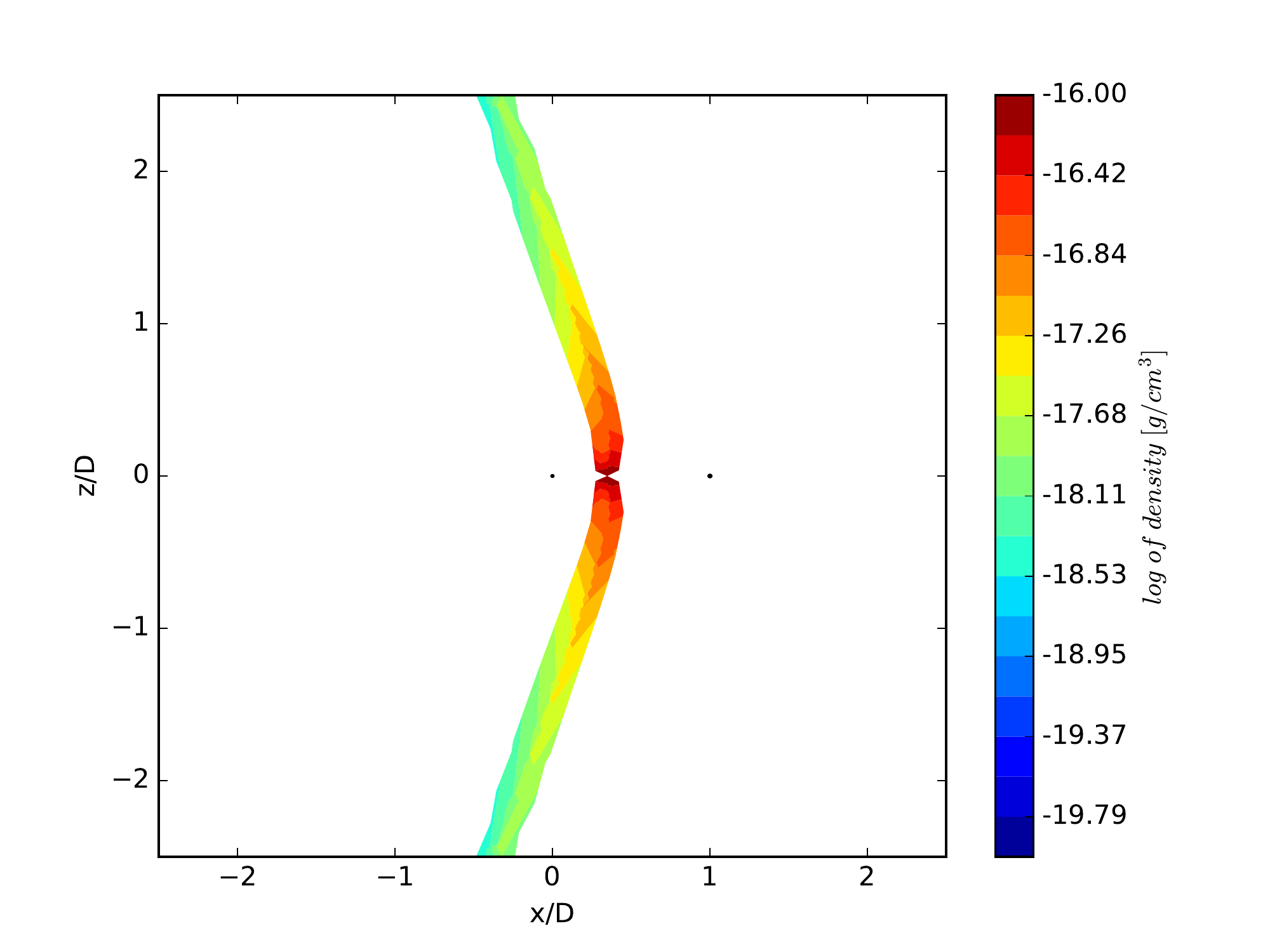}}
\put(133,200){\includegraphics[trim= 0cm 0cm 0cm 1.2cm,clip, width=5.cm]{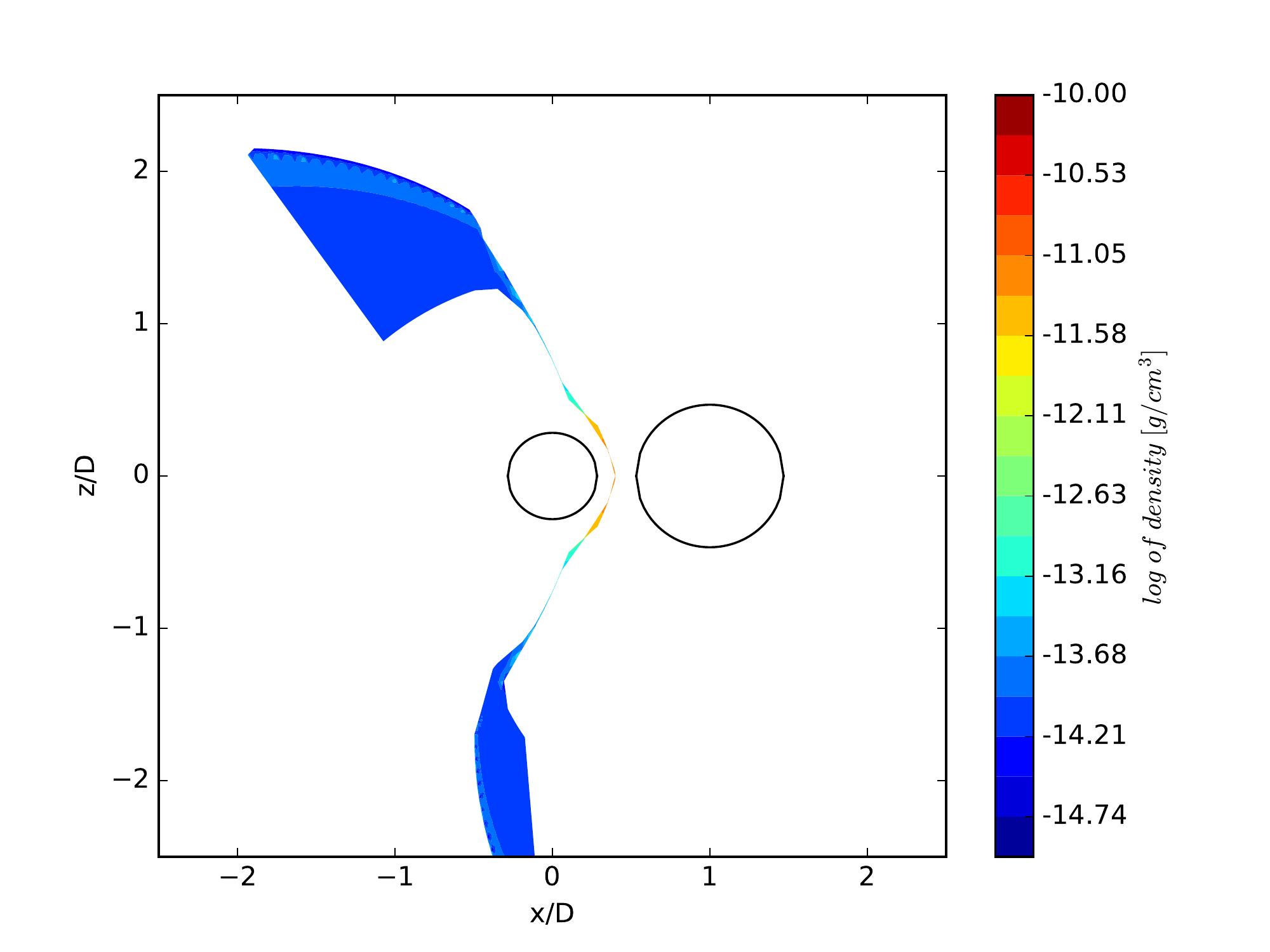}}
\put(266,200){\includegraphics[trim= 0cm 0cm 0cm 1.2cm,clip, width=5.cm]{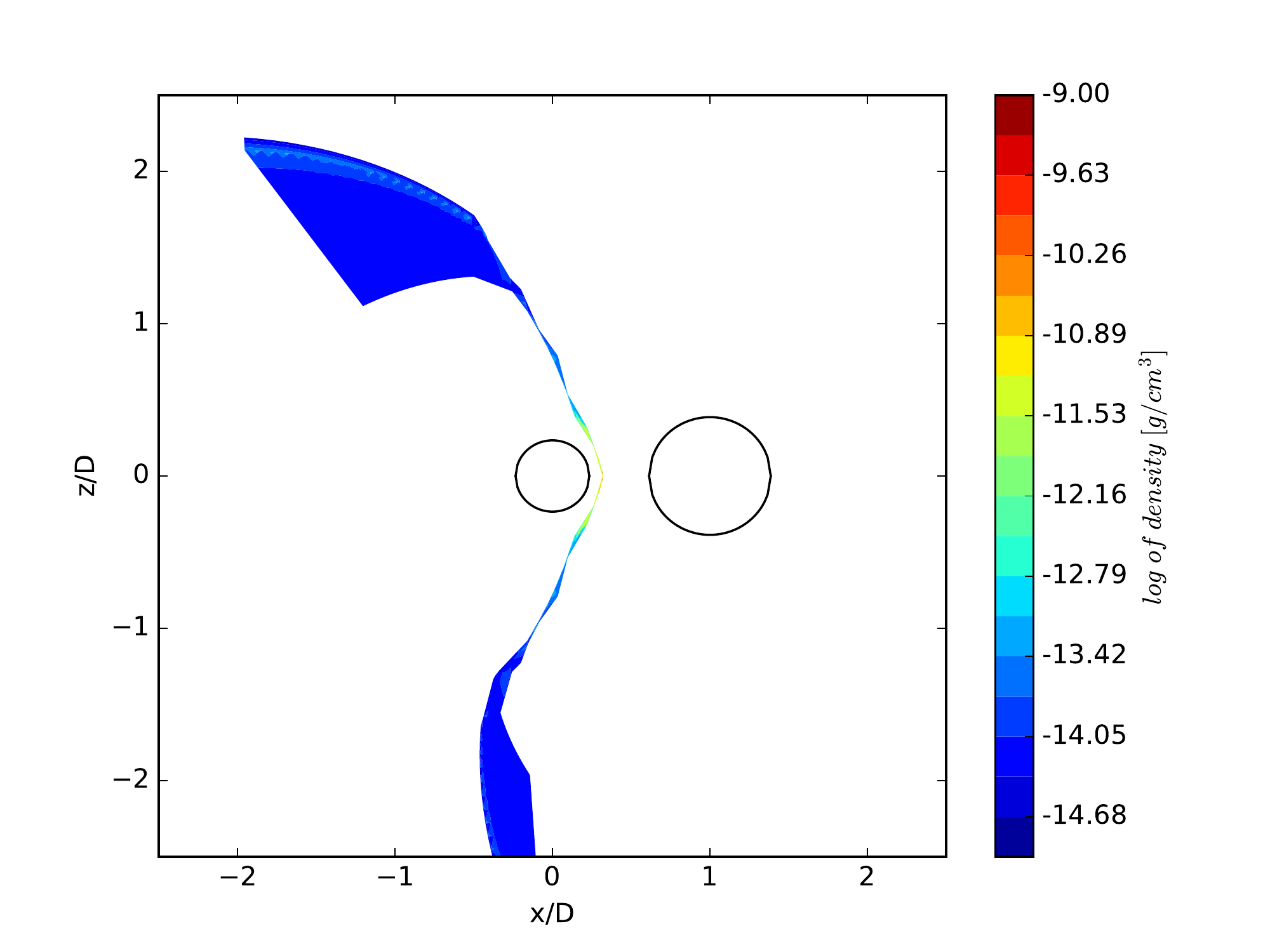}}
\put(400,200){\includegraphics[trim= 0cm 0cm 0cm 1.2cm,clip, width=5.cm]{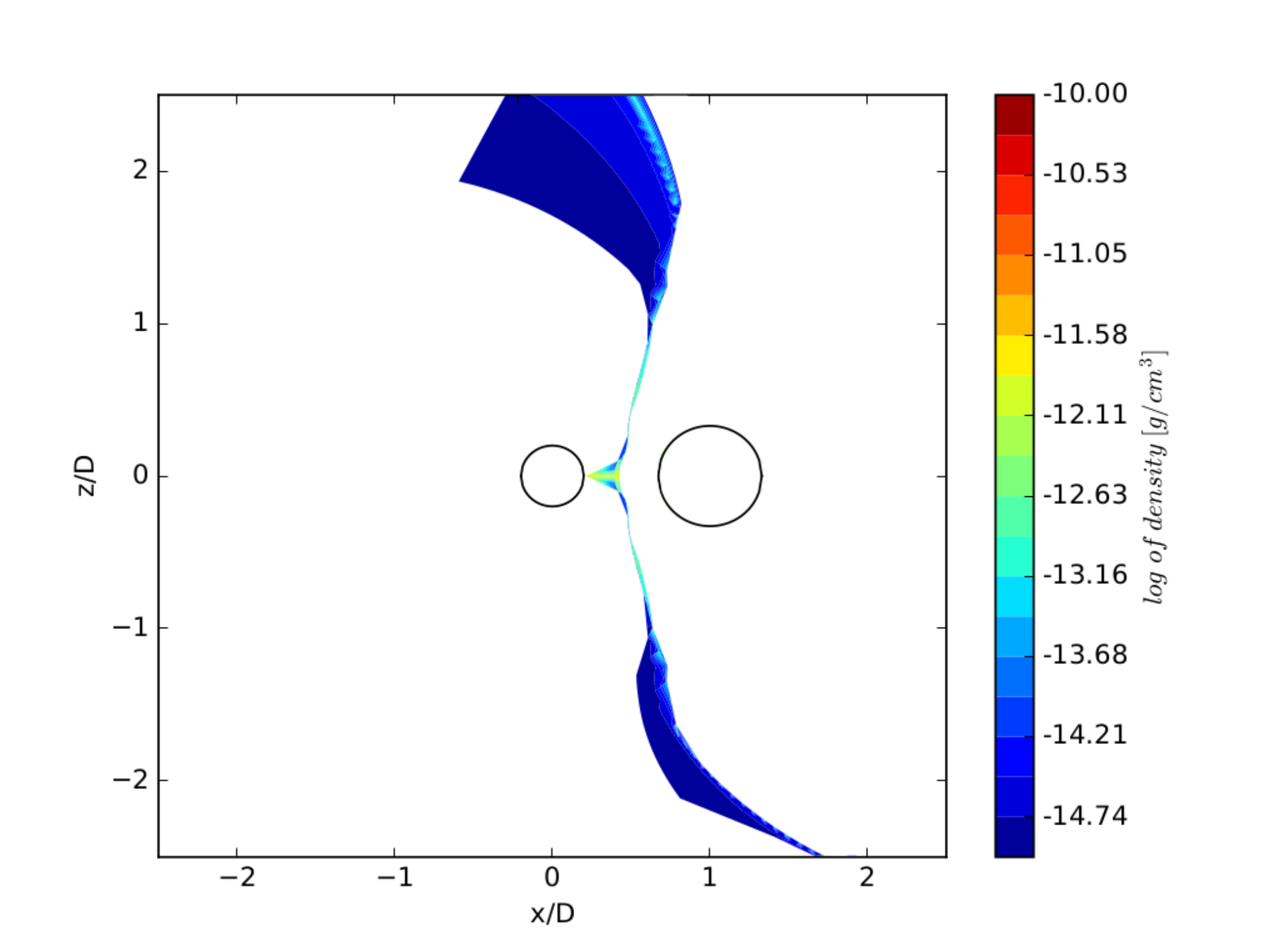}}

\put(0,100){\includegraphics[trim= 0cm 0cm 0cm 1.2cm,clip, width=5.cm]{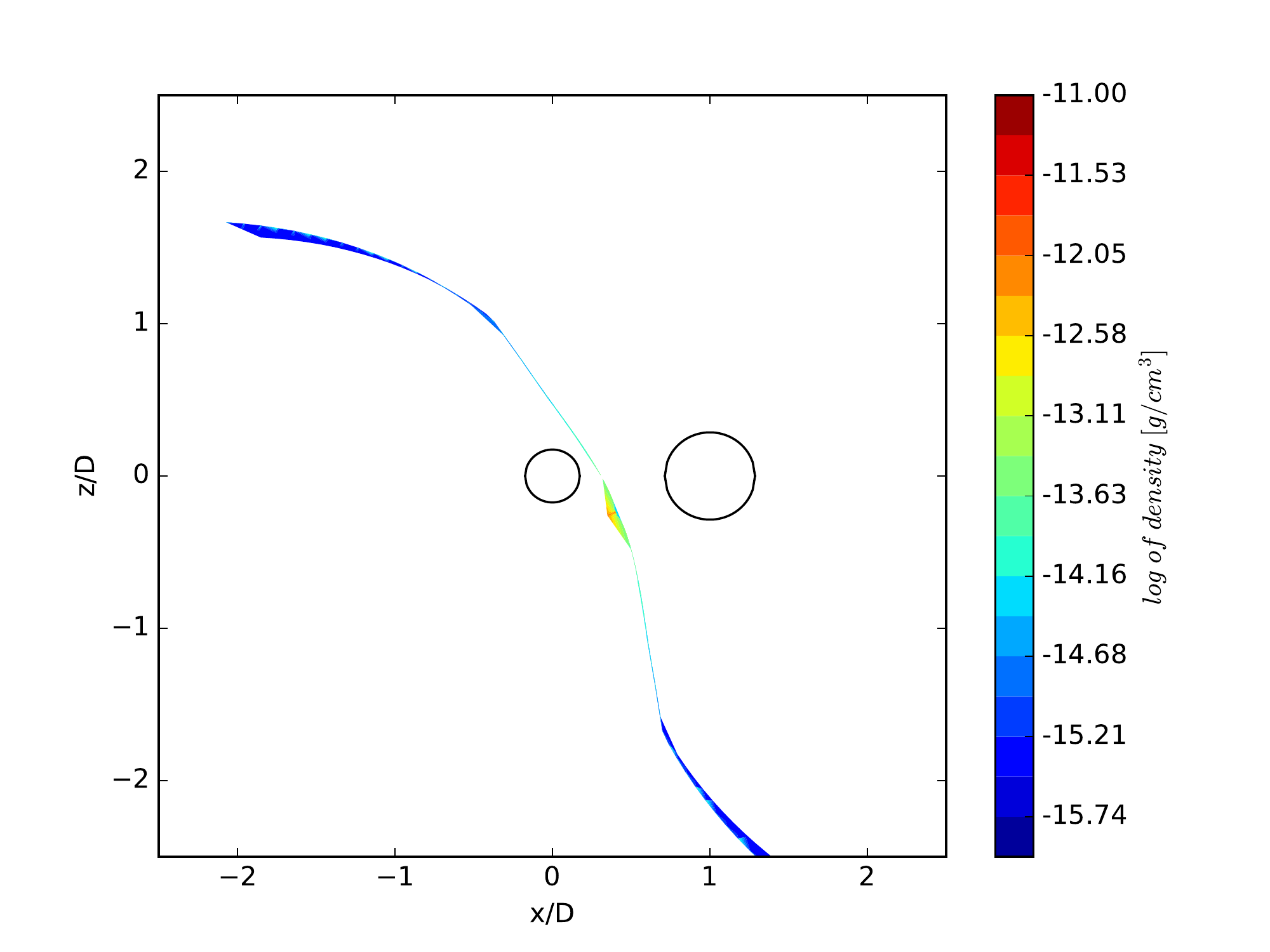}}
\put(133,100){\includegraphics[trim= 0cm 0cm 0cm 1.2cm,clip, width=5.cm]{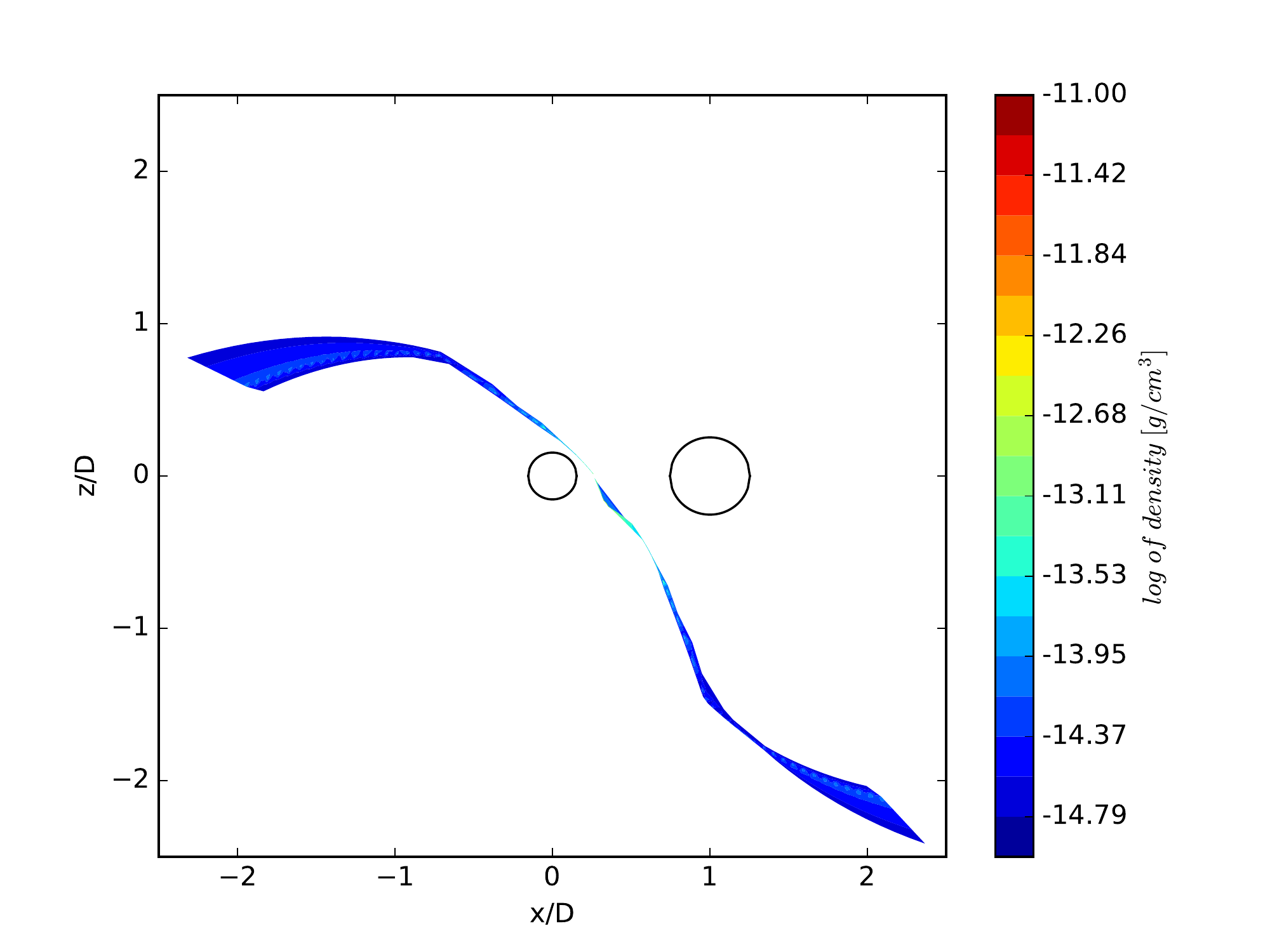}}
\put(266,100){\includegraphics[trim= 0cm 0cm 0cm 1.2cm,clip, width=5.cm]{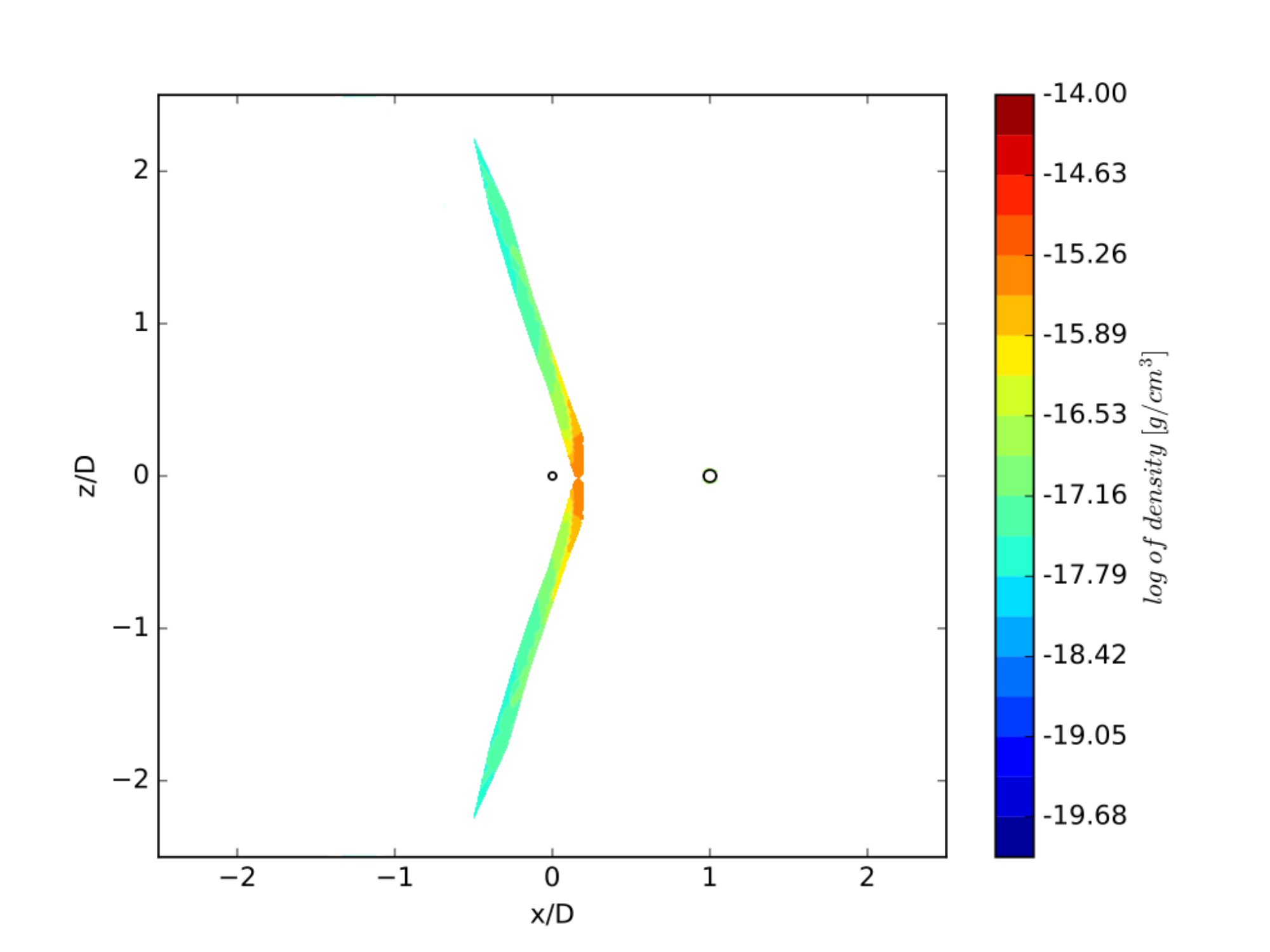}}
\put(400,100){\includegraphics[trim= 0cm 0cm 0cm 1.2cm,clip, width=5.cm]{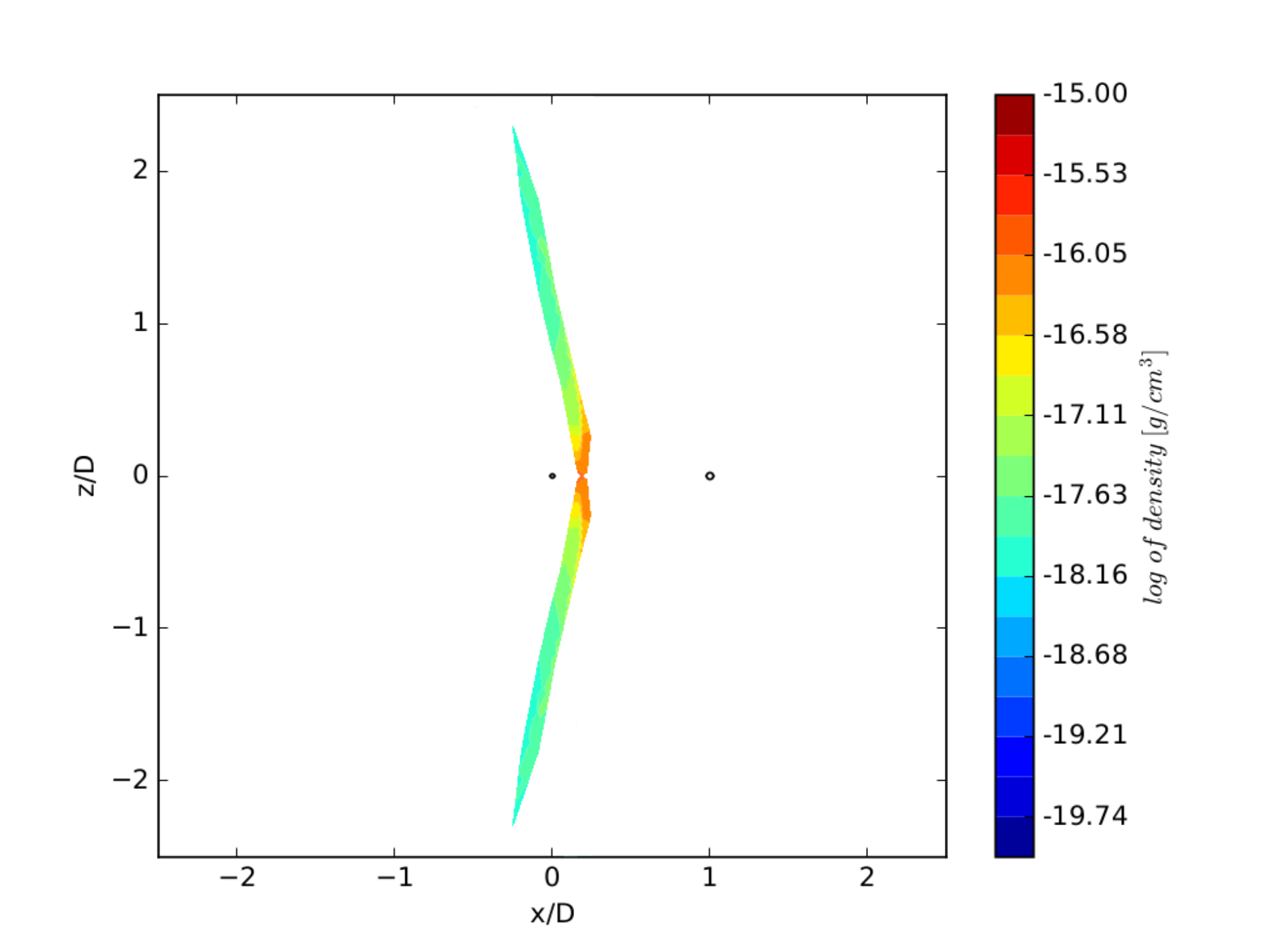}}

\put(0,0){\includegraphics[trim= 0cm 0cm 0cm 1.2cm,clip, width=5.cm]{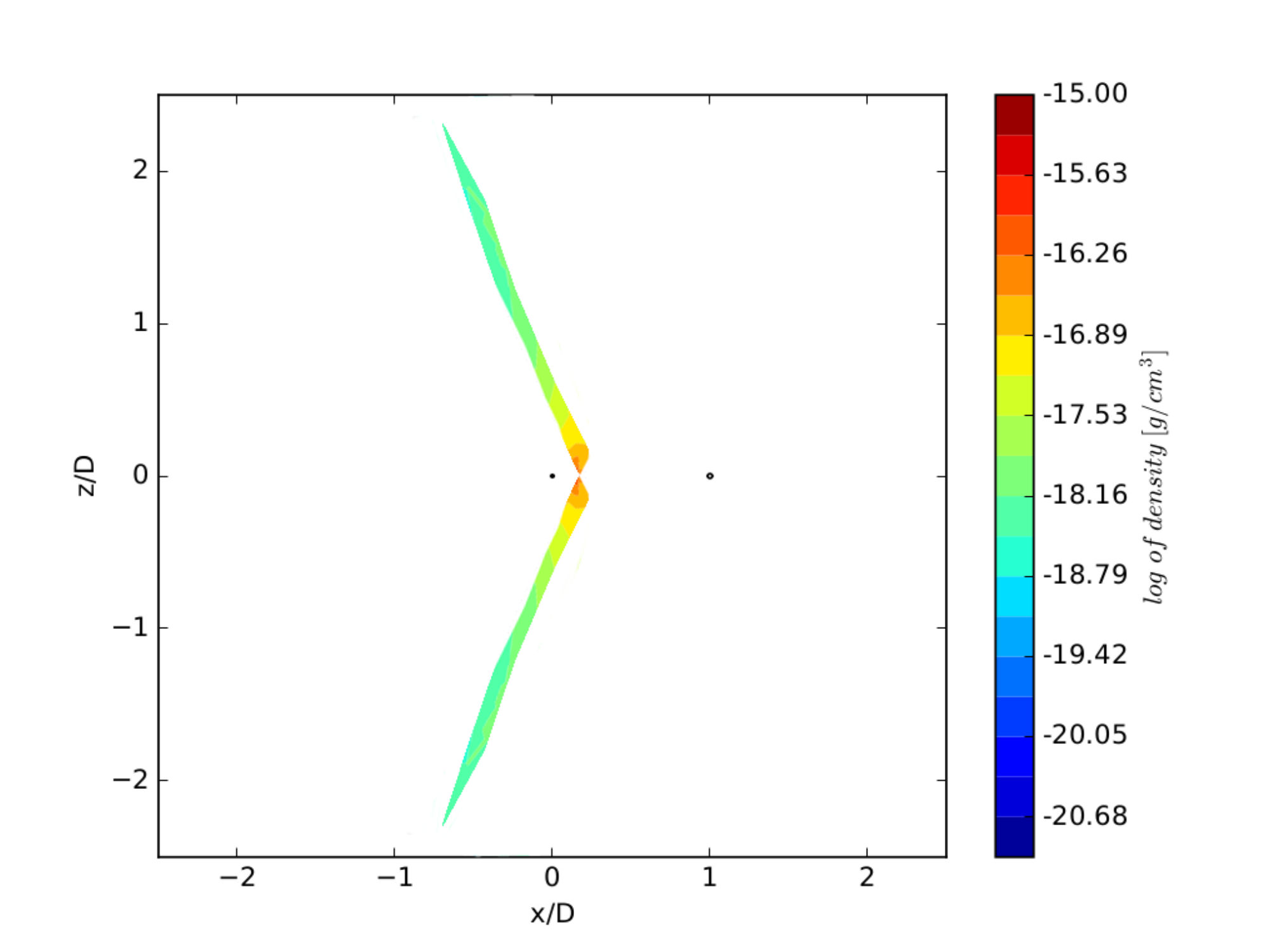}}
\put(133,0){\includegraphics[trim= 0cm 0cm 0cm 1.2cm,clip, width=5.cm]{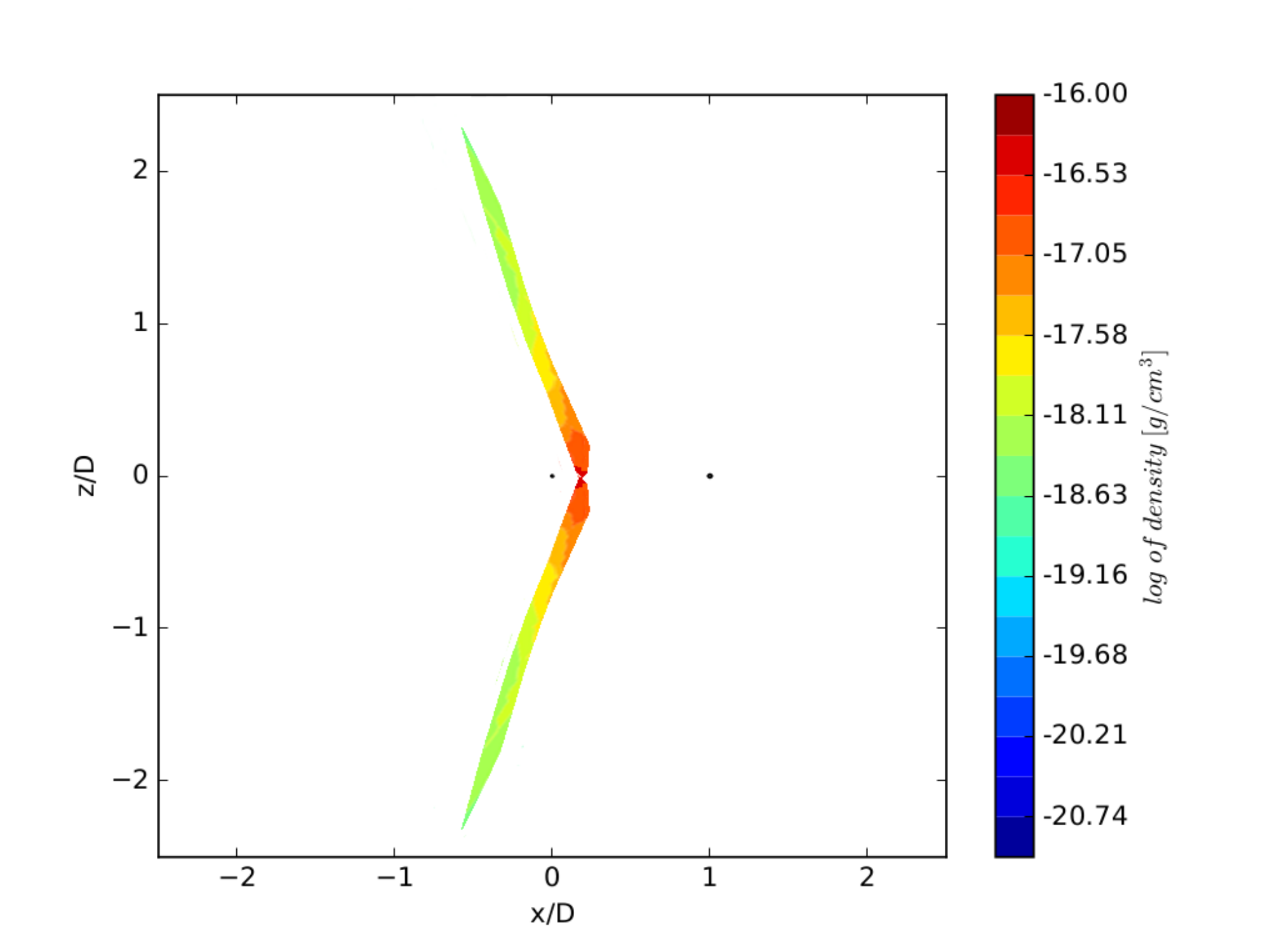}}

\put(22,690){\scriptsize{O7V+O7V}}
\put(22,682){\scriptsize{$d=57.6\,\rsun{}$}}
\put(155,690){\scriptsize{O7V+O7V}}
\put(155,682){\scriptsize{$d=70.8\,\rsun{}$}}
\put(288,690){\scriptsize{O7V+O7V}}
\put(288,682){\scriptsize{$d=84.1\,\rsun{}$}}
\put(422,690){\scriptsize{O7V+O7V}}
\put(422,682){\scriptsize{$d=97.3\,\rsun{}$}}

\put(22,590){\scriptsize{O7V+O7V}}
\put(22,582){\scriptsize{$d=110.5\,\rsun{}$}}
\put(155,590){\scriptsize{O7V+O7V}}
\put(155,582){\scriptsize{$d=582.9\,\rsun{}$}}
\put(288,590){\scriptsize{O7V+O7V}}
\put(288,582){\scriptsize{$d=1055.2\,\rsun{}$}}
\put(422,590){\scriptsize{O7V+O7V}}
\put(422,582){\scriptsize{$d=1527.6\,\rsun{}$}}

\put(22,490){\scriptsize{O7V+O7V}}
\put(22,482){\scriptsize{$d=2000.0\,\rsun{}$}}
\put(155,490){\scriptsize{O5I+O3III}}
\put(155,482){\scriptsize{$d=46.4\,\rsun{}$}}
\put(288,490){\scriptsize{O5I+O3III}}
\put(288,482){\scriptsize{$d=55.1\,\rsun{}$}}
\put(422,490){\scriptsize{O5I+O3III}}
\put(422,482){\scriptsize{$d=63.8\,\rsun{}$}}

\put(22,390){\scriptsize{O5I+O3III}}
\put(22,382){\scriptsize{$d=72.5\,\rsun{}$}}
\put(155,390){\scriptsize{O5I+O3III}}
\put(155,382){\scriptsize{$d=554.4\,\rsun{}$}}
\put(288,390){\scriptsize{O5I+O3III}}
\put(288,382){\scriptsize{$d=1036.2\,\rsun{}$}}
\put(422,390){\scriptsize{O5I+O3III}}
\put(422,382){\scriptsize{$d=1518.1\,\rsun{}$}}

\put(22,290){\scriptsize{O5I+O3III}}
\put(22,282){\scriptsize{$d=2000.0\,\rsun{}$}}
\put(155,290){\scriptsize{O9III+O9V}}
\put(155,282){\scriptsize{$d=48.3\,\rsun{}$}}
\put(288,290){\scriptsize{O9III+O9V}}
\put(288,282){\scriptsize{$d=58.5\,\rsun{}$}}
\put(422,290){\scriptsize{O9III+O9V}}
\put(422,282){\scriptsize{$d=68.7\,\rsun{}$}}

\put(22,190){\scriptsize{O9III+O9V}}
\put(22,182){\scriptsize{$d=78.9\,\rsun{}$}}
\put(155,190){\scriptsize{O9III+O9V}}
\put(155,182){\scriptsize{$d=89.1\,\rsun{}$}}
\put(288,190){\scriptsize{O9III+O9V}}
\put(288,182){\scriptsize{$d=566.8\,\rsun{}$}}
\put(422,190){\scriptsize{O9III+O9V}}
\put(422,182){\scriptsize{$d=1044.5\,\rsun{}$}}

\put(22,90){\scriptsize{O9III+O9V}}
\put(22,82){\scriptsize{$d=1522.3\,\rsun{}$}}
\put(155,90){\scriptsize{O9III+O9V}}
\put(155,82){\scriptsize{$d=2000.0\,\rsun{}$}}
\end{picture} 
\caption{Density distribution inside the wind shock region for the O5I+O3III, O7V+O7V, and O9III+O9V systems.}
\label{density}
\end{figure*}

\begin{figure*}
\begin{picture}(600,700)
\put(0,600){\includegraphics[trim= 0cm 0cm 0cm 1.2cm,clip, width=5.2cm]{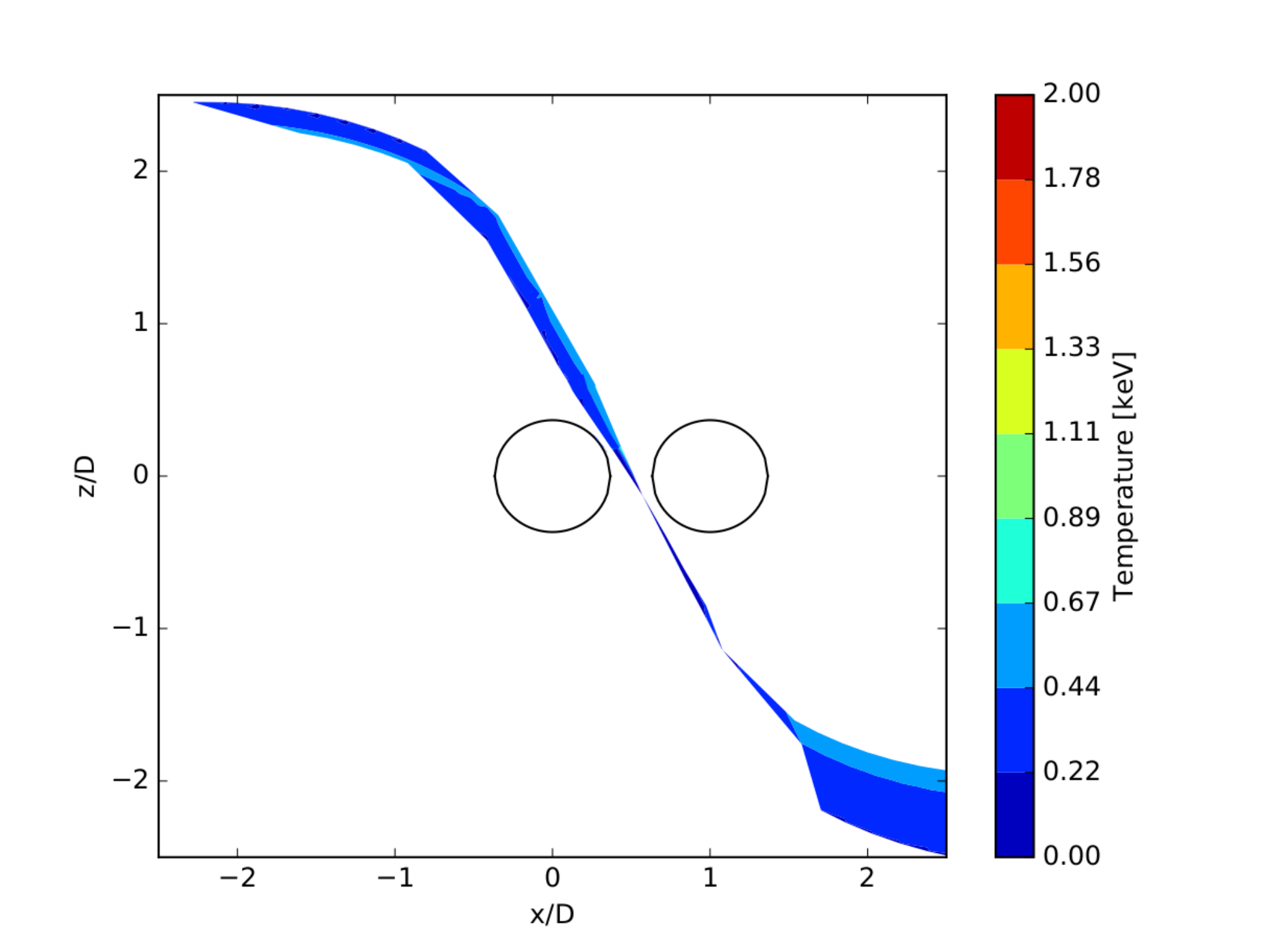}}
\put(133,600){\includegraphics[trim= 0cm 0cm 0cm 1.2cm,clip, width=5.2cm]{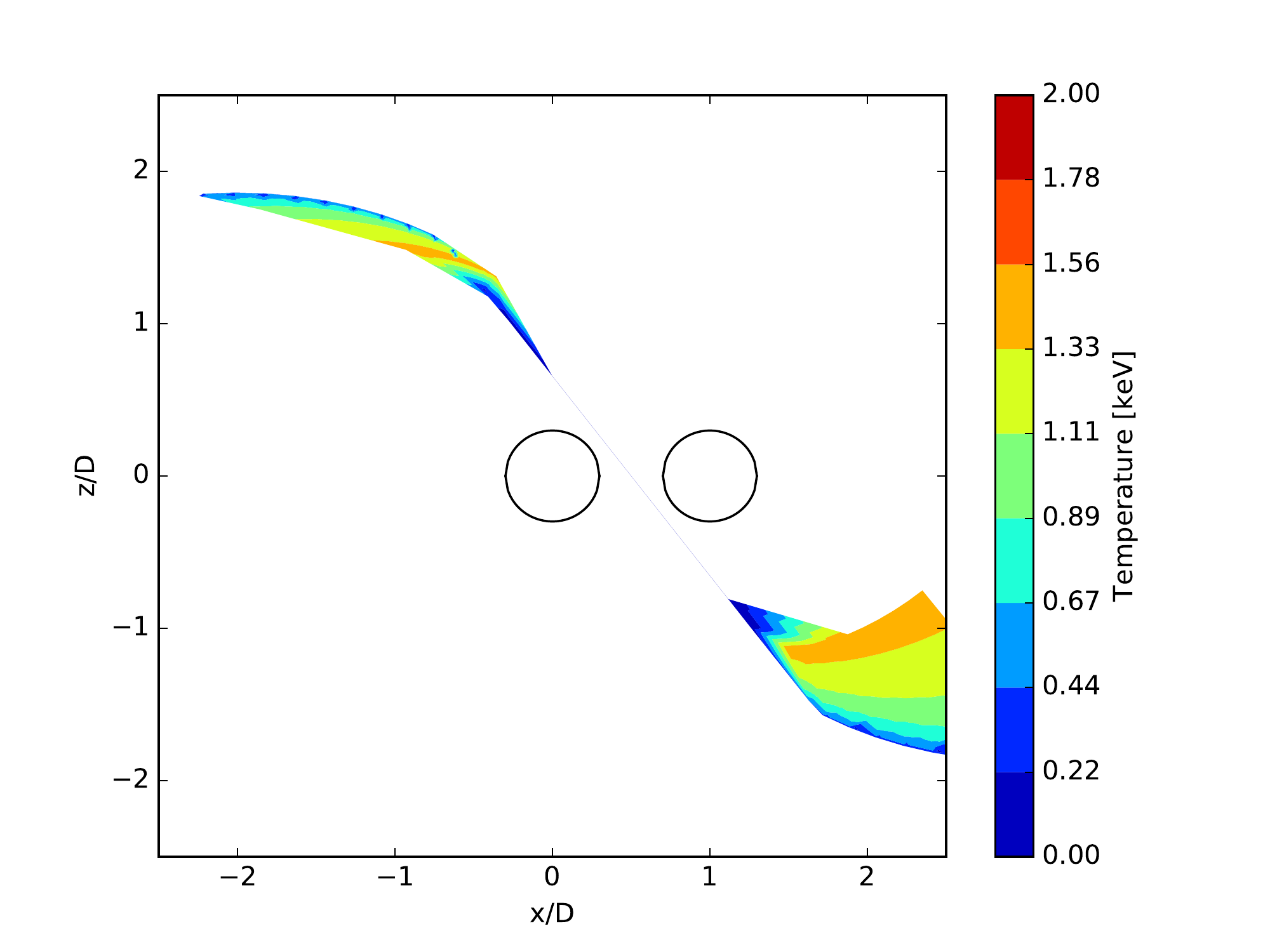}}
\put(266,600){\includegraphics[trim= 0cm 0cm 0cm 1.2cm,clip, width=5.2cm]{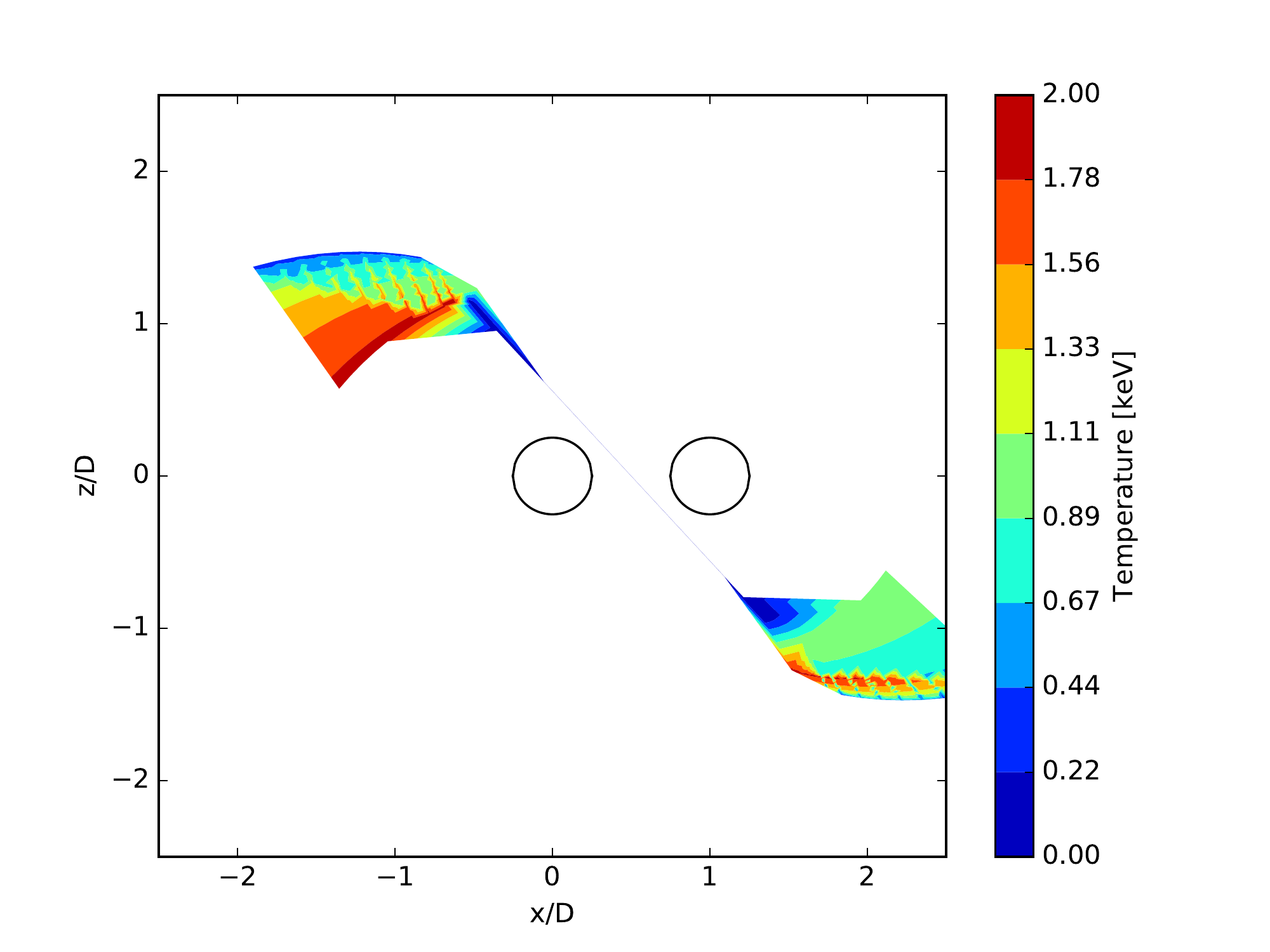}}
\put(400,600){\includegraphics[trim= 0cm 0cm 0cm 1.2cm,clip, width=5.2cm]{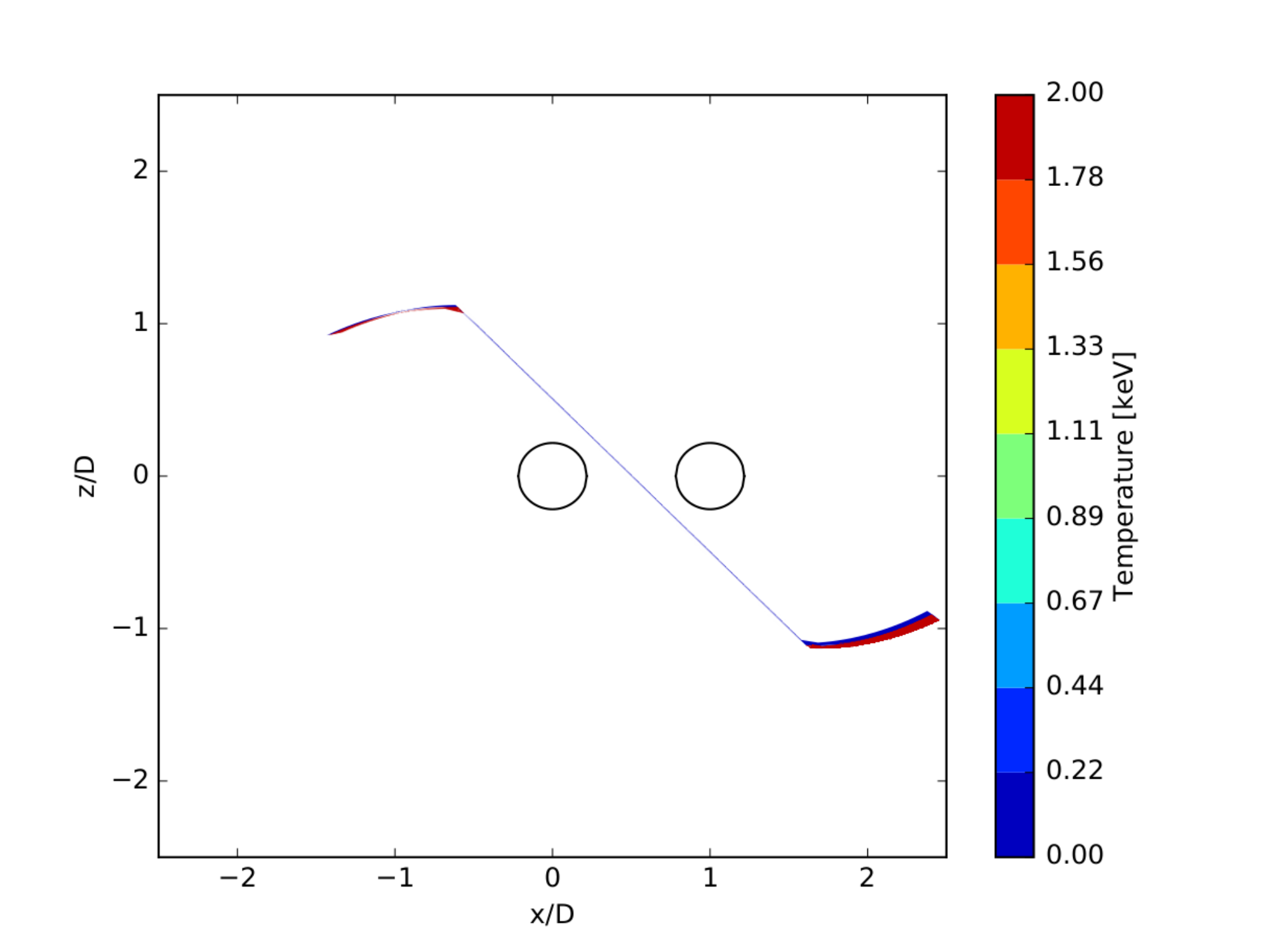}}

\put(0,500){\includegraphics[trim= 0cm 0cm 0cm 1.2cm,clip, width=5.2cm]{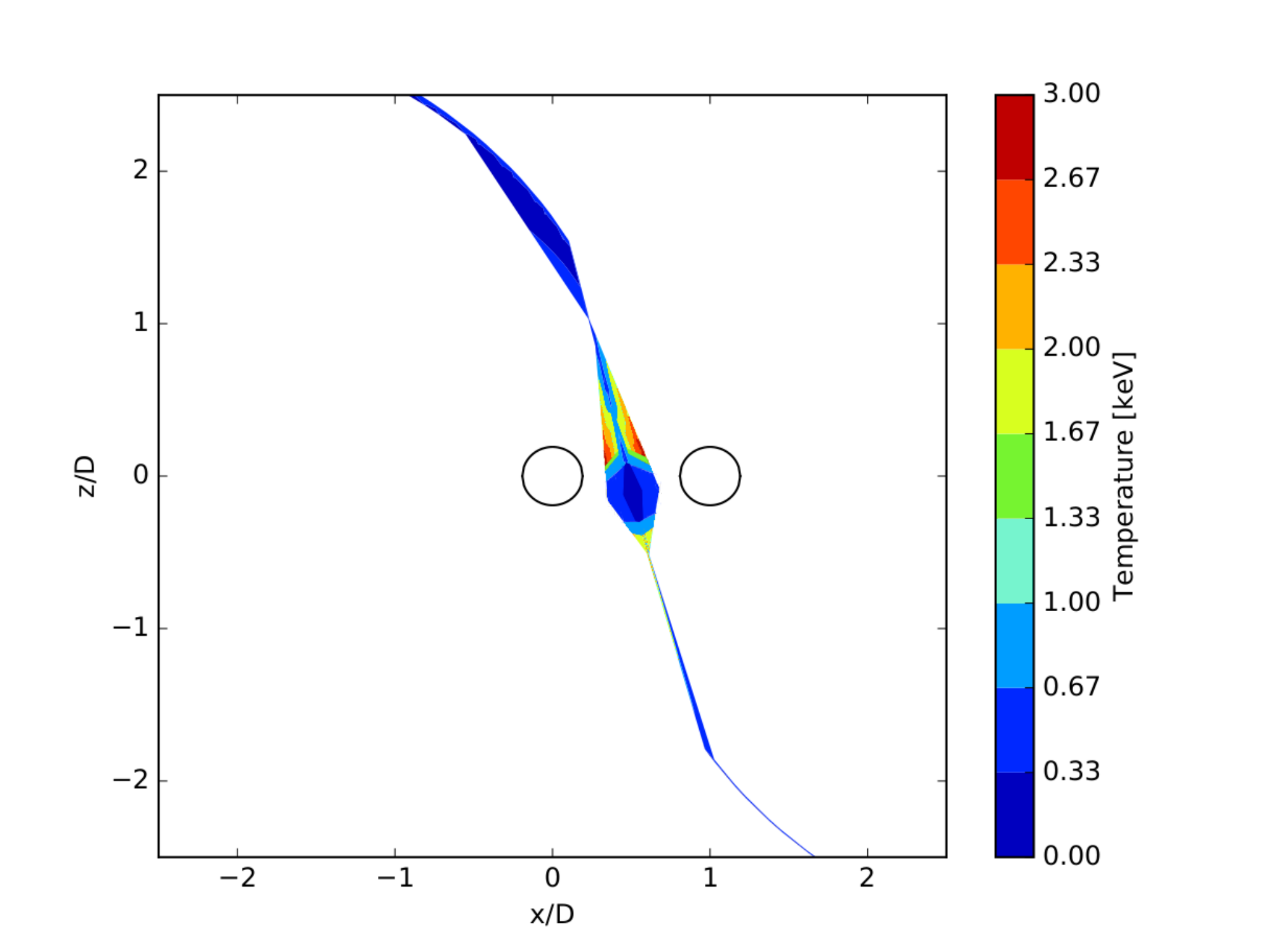}}
\put(133,500){\includegraphics[trim= 0cm 0cm 0cm 1.2cm,clip, width=5.2cm]{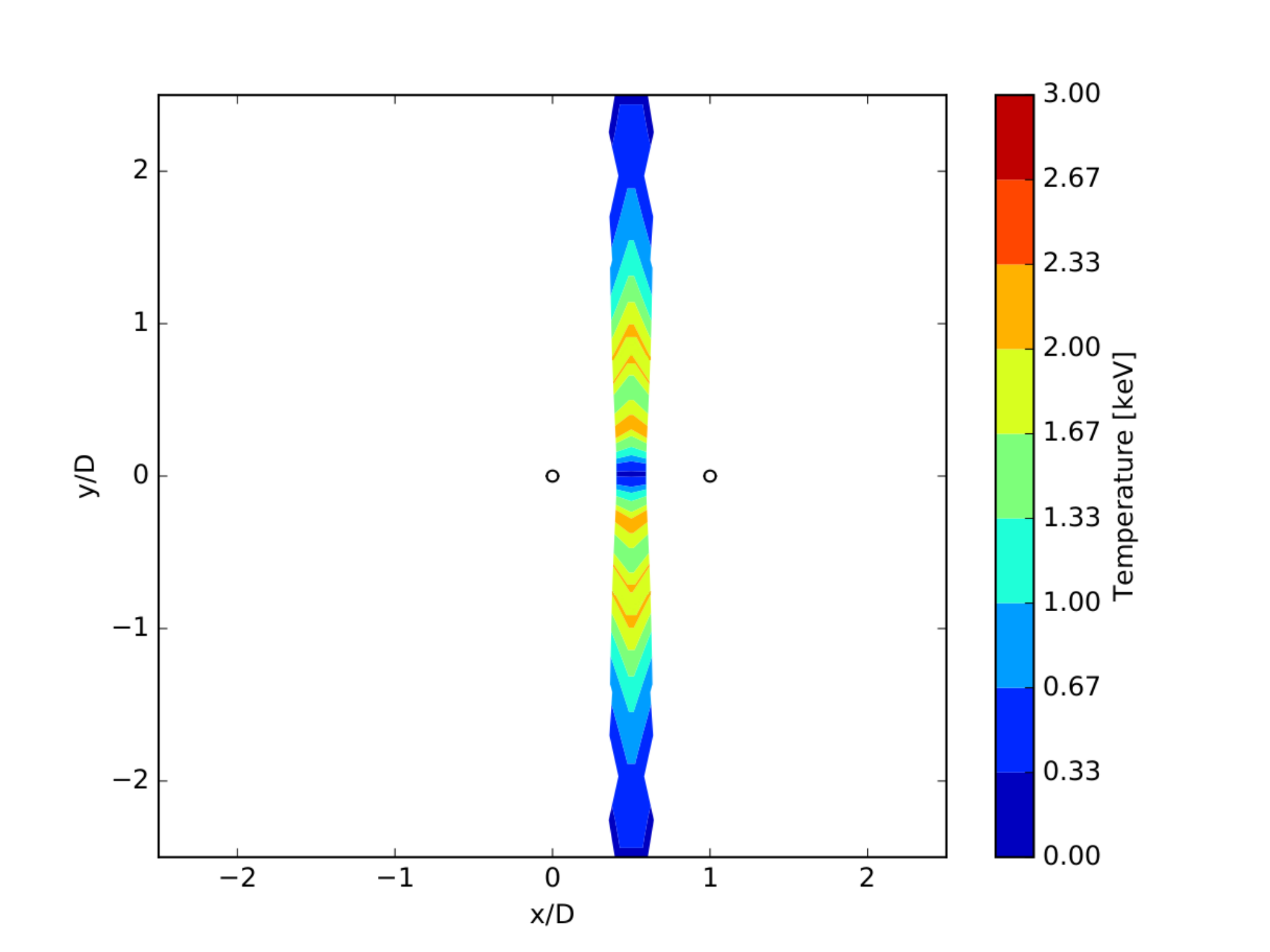}}
\put(266,500){\includegraphics[trim= 0cm 0cm 0cm 1.2cm,clip, width=5.2cm]{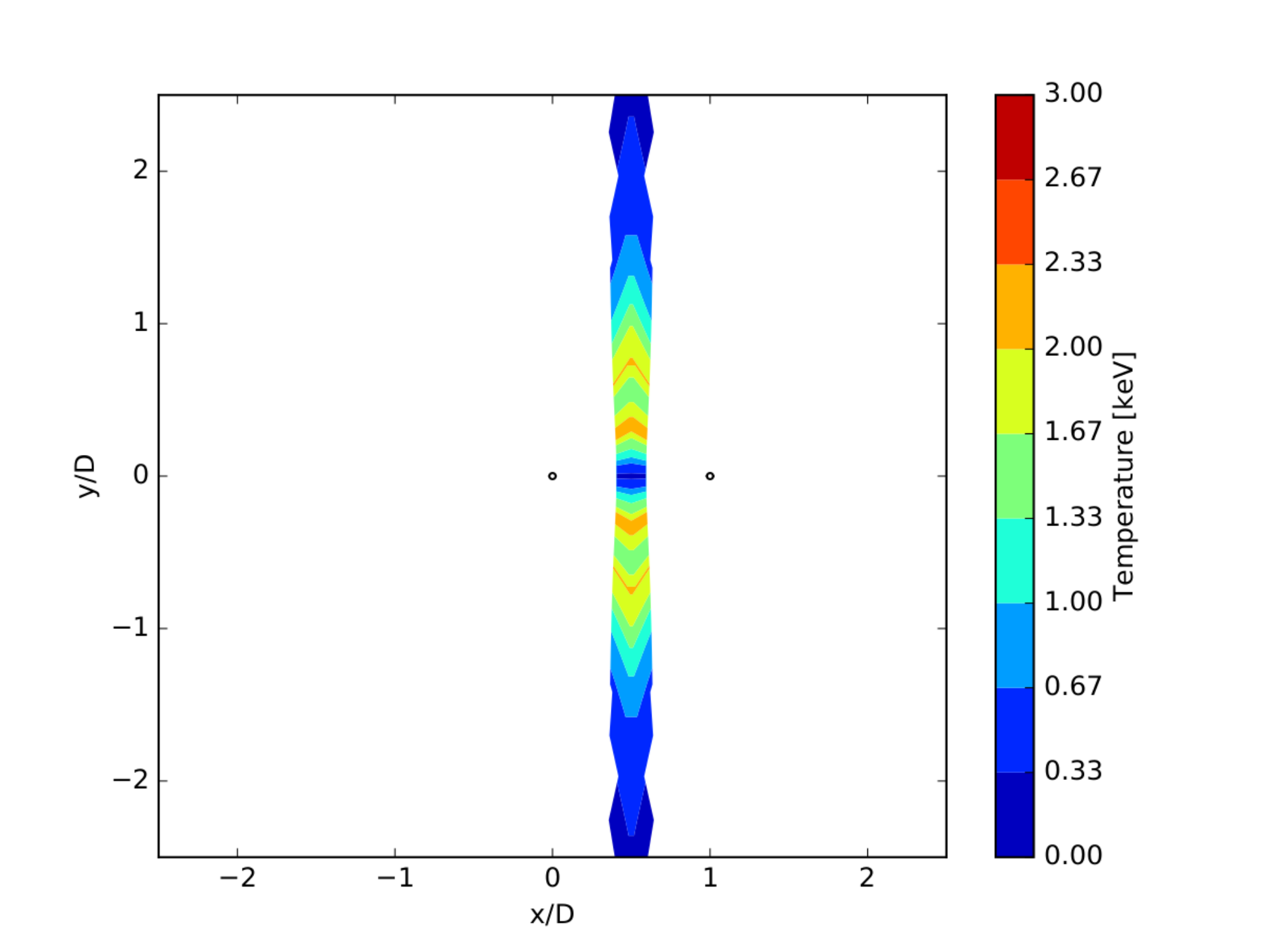}}
\put(400,500){\includegraphics[trim= 0cm 0cm 0cm 1.2cm,clip, width=5.2cm]{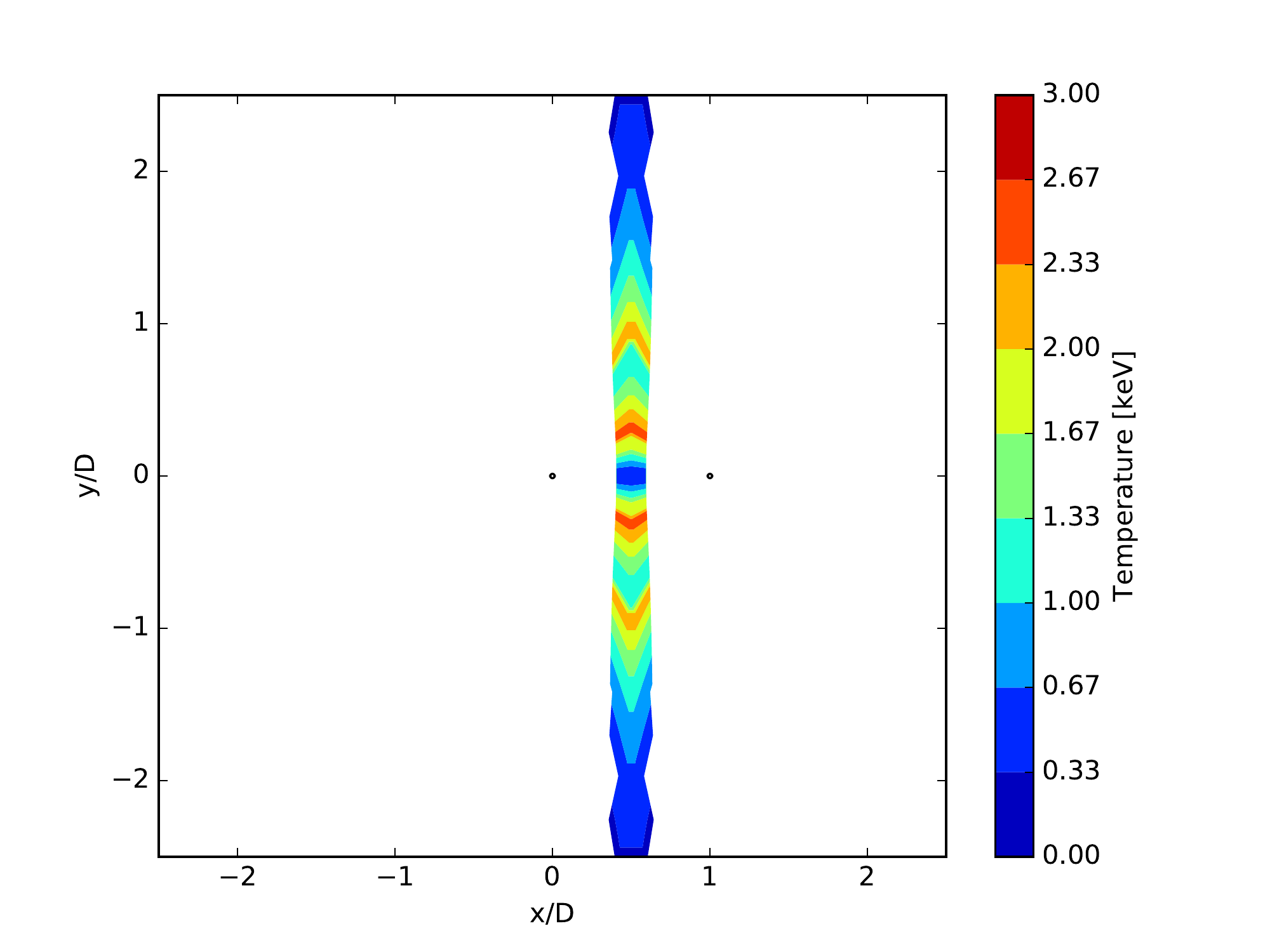}}

\put(0,400){\includegraphics[trim= 0cm 0cm 0cm 1.2cm,clip, width=5.2cm]{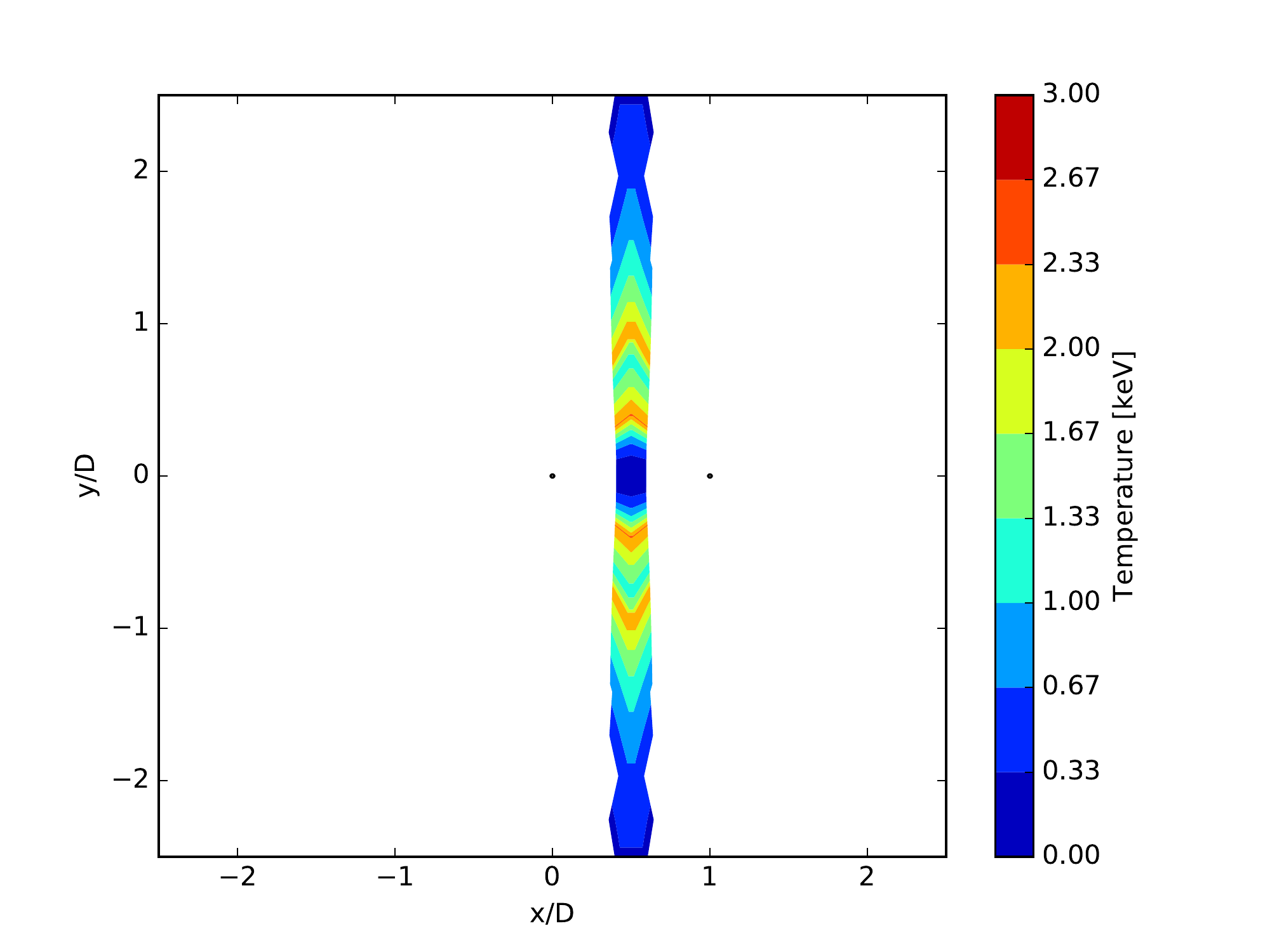}}
\put(133,400){\includegraphics[trim= 0cm 0cm 0cm 1.2cm,clip, width=5.2cm]{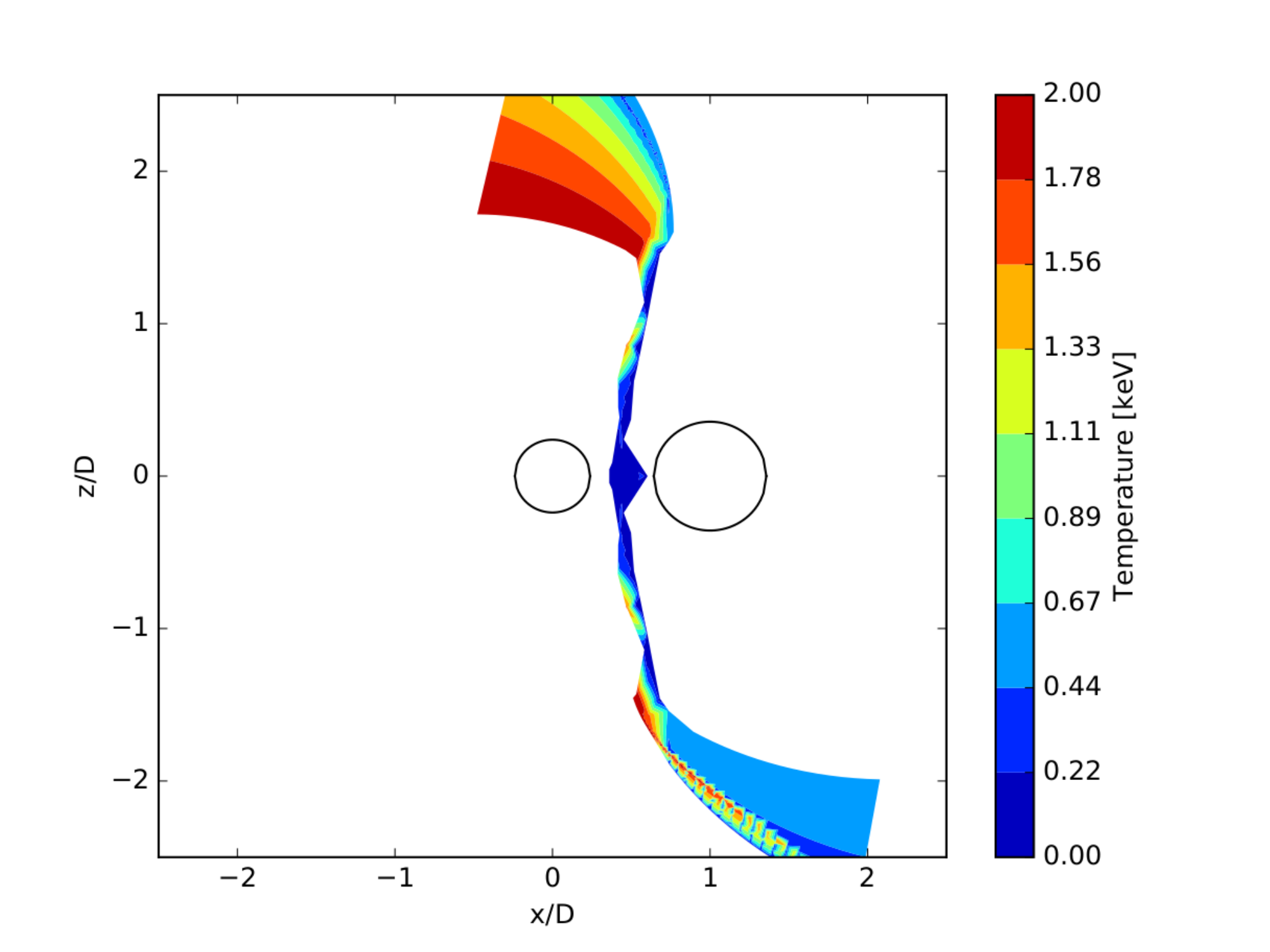}}
\put(266,400){\includegraphics[trim= 0cm 0cm 0cm 1.2cm,clip, width=5.2cm]{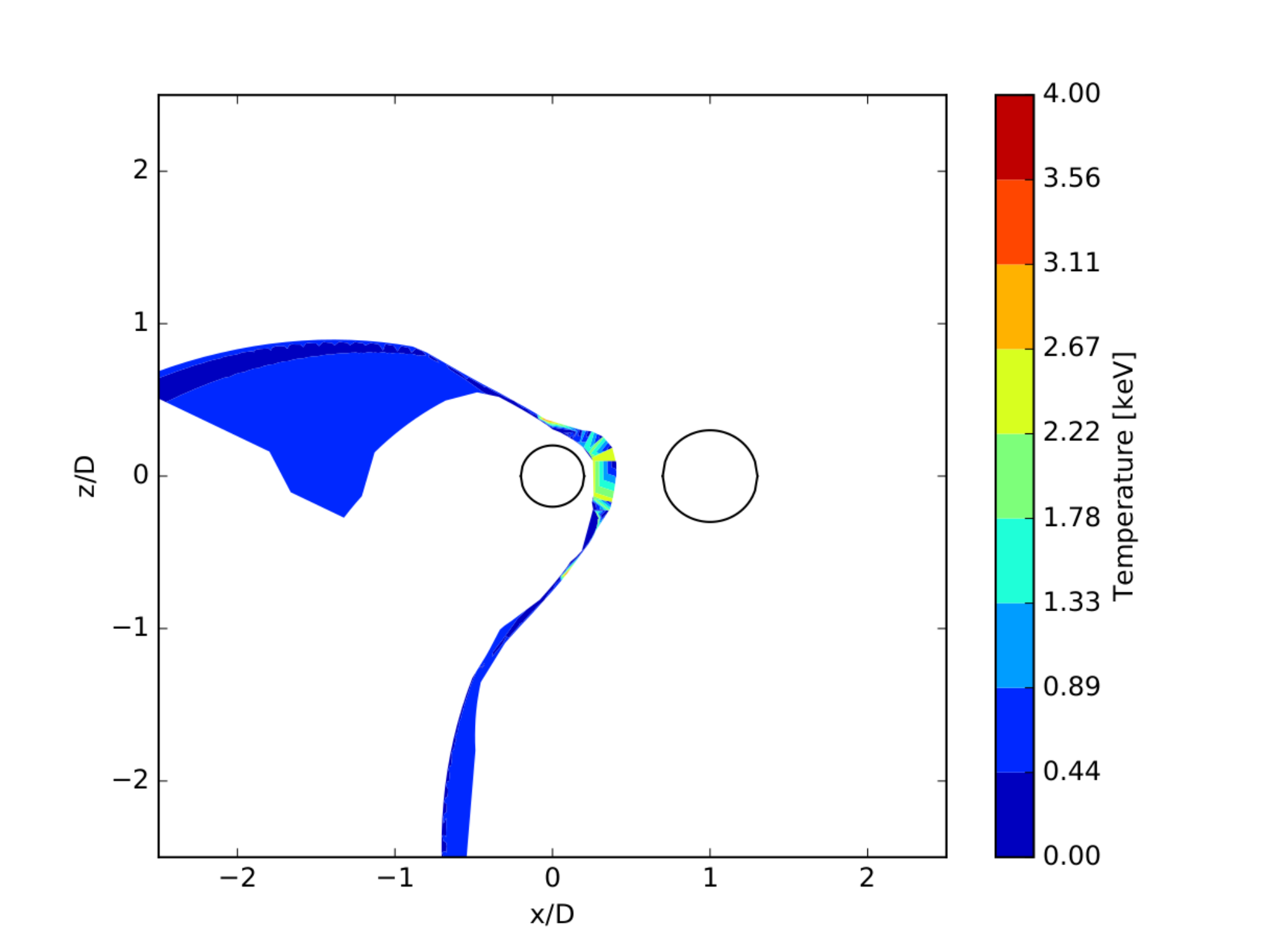}}
\put(400,400){\includegraphics[trim= 0cm 0cm 0cm 1.2cm,clip, width=5.2cm]{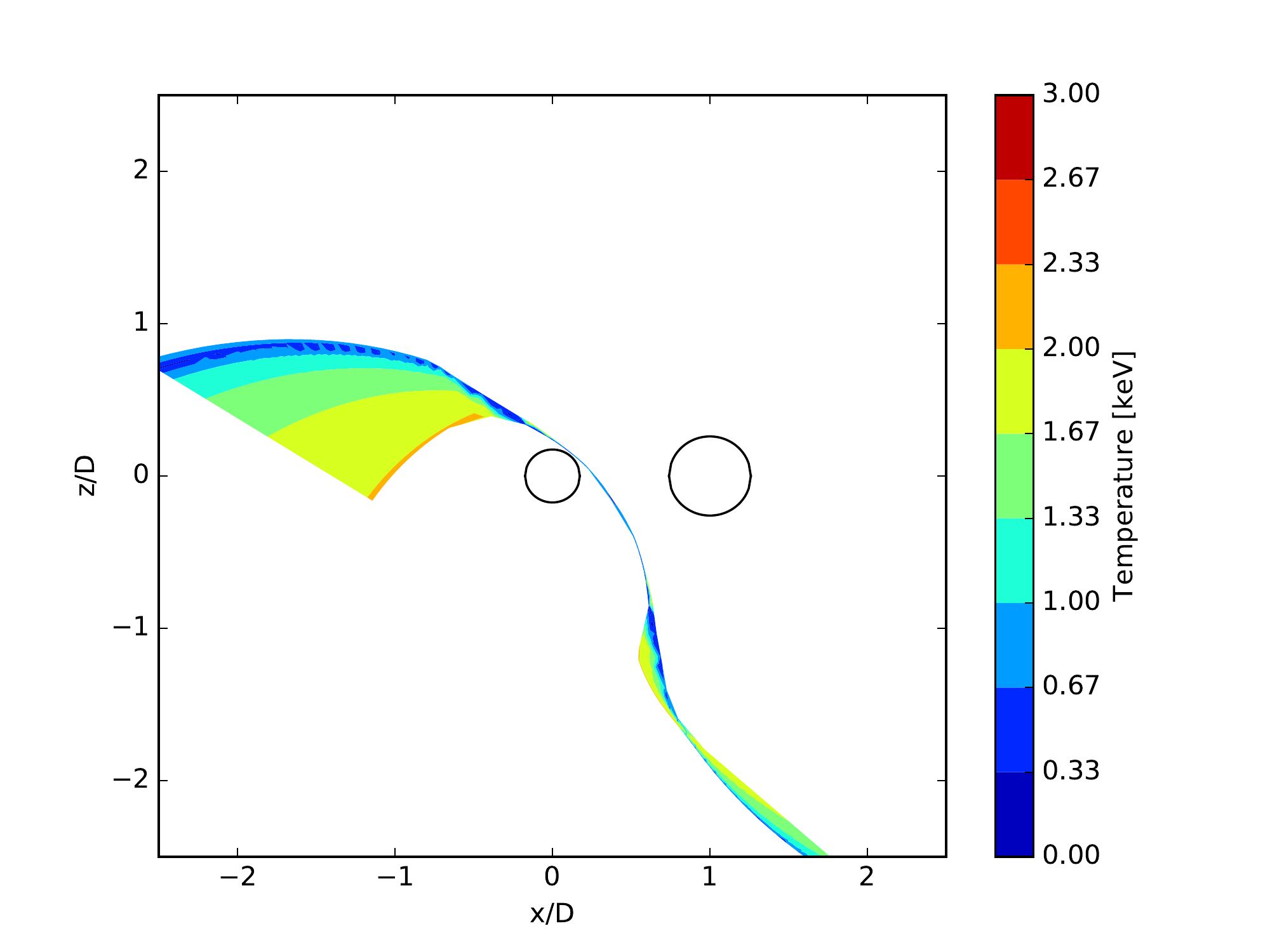}}

\put(0,300){\includegraphics[trim= 0cm 0cm 0cm 1.2cm,clip, width=5.2cm]{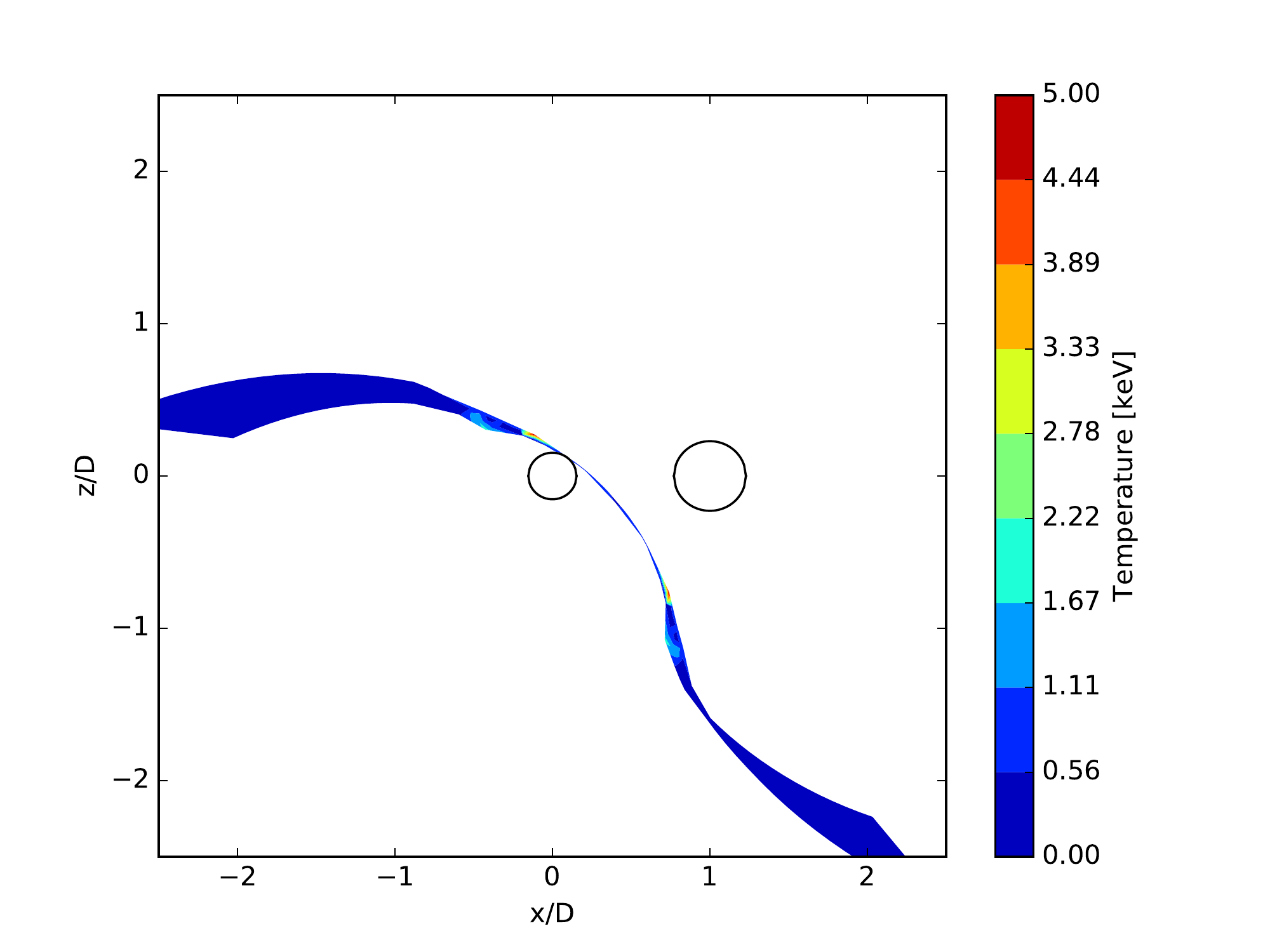}}
\put(133,300){\includegraphics[trim= 0cm 0cm 0cm 1.2cm,clip, width=5.2cm]{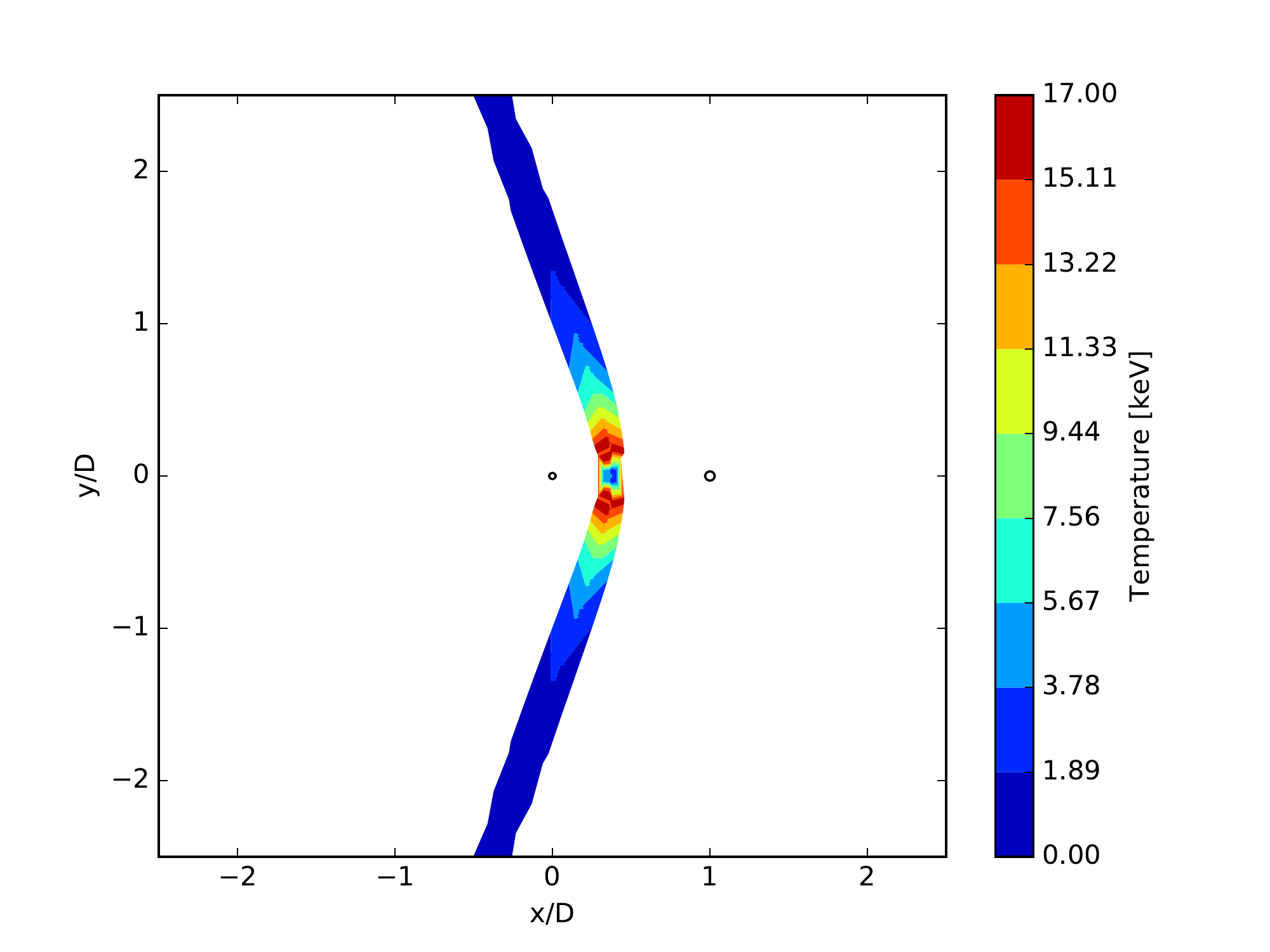}}
\put(266,300){\includegraphics[trim= 0cm 0cm 0cm 1.2cm,clip, width=5.2cm]{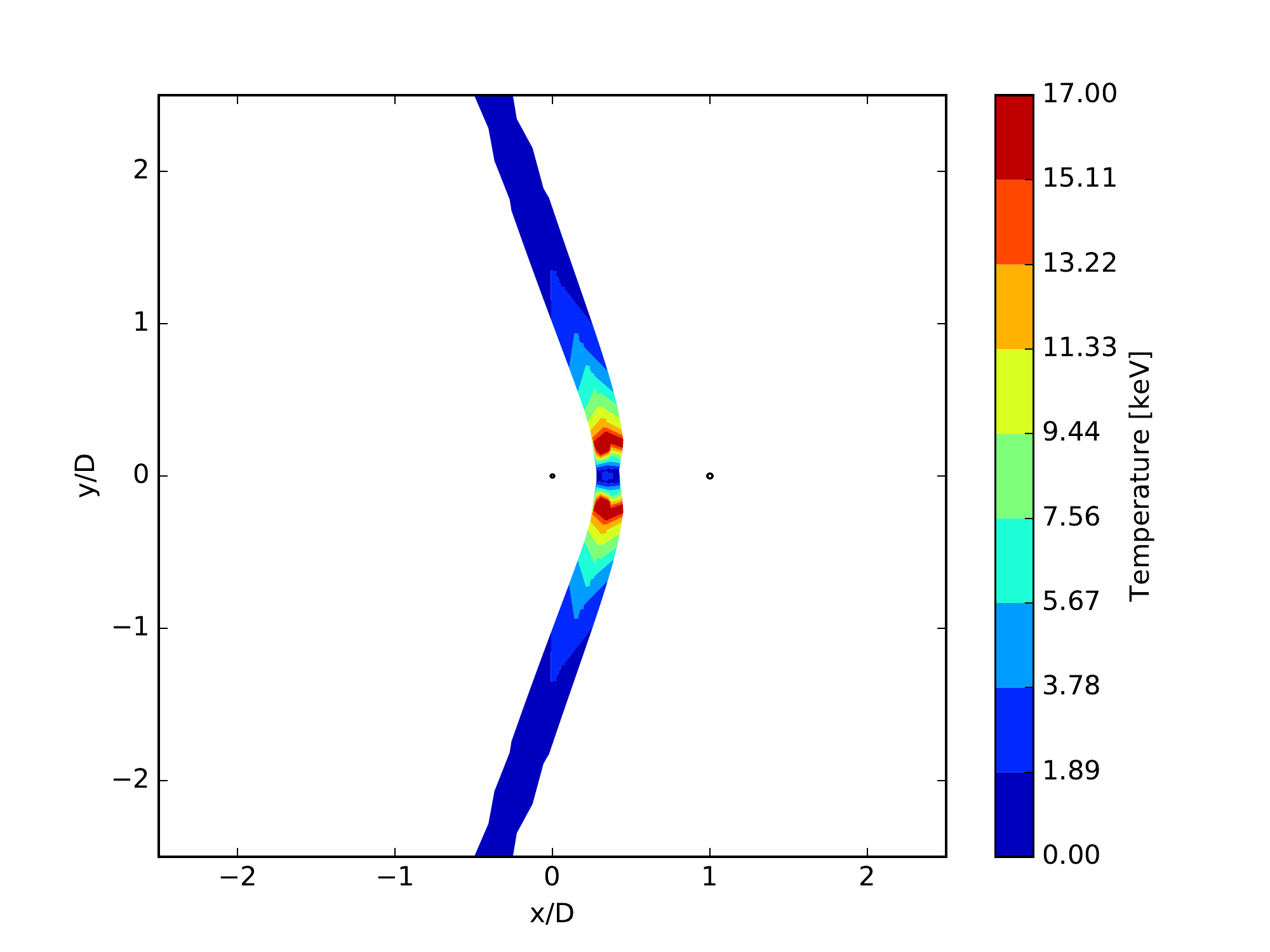}}
\put(400,300){\includegraphics[trim= 0cm 0cm 0cm 1.2cm,clip, width=5.2cm]{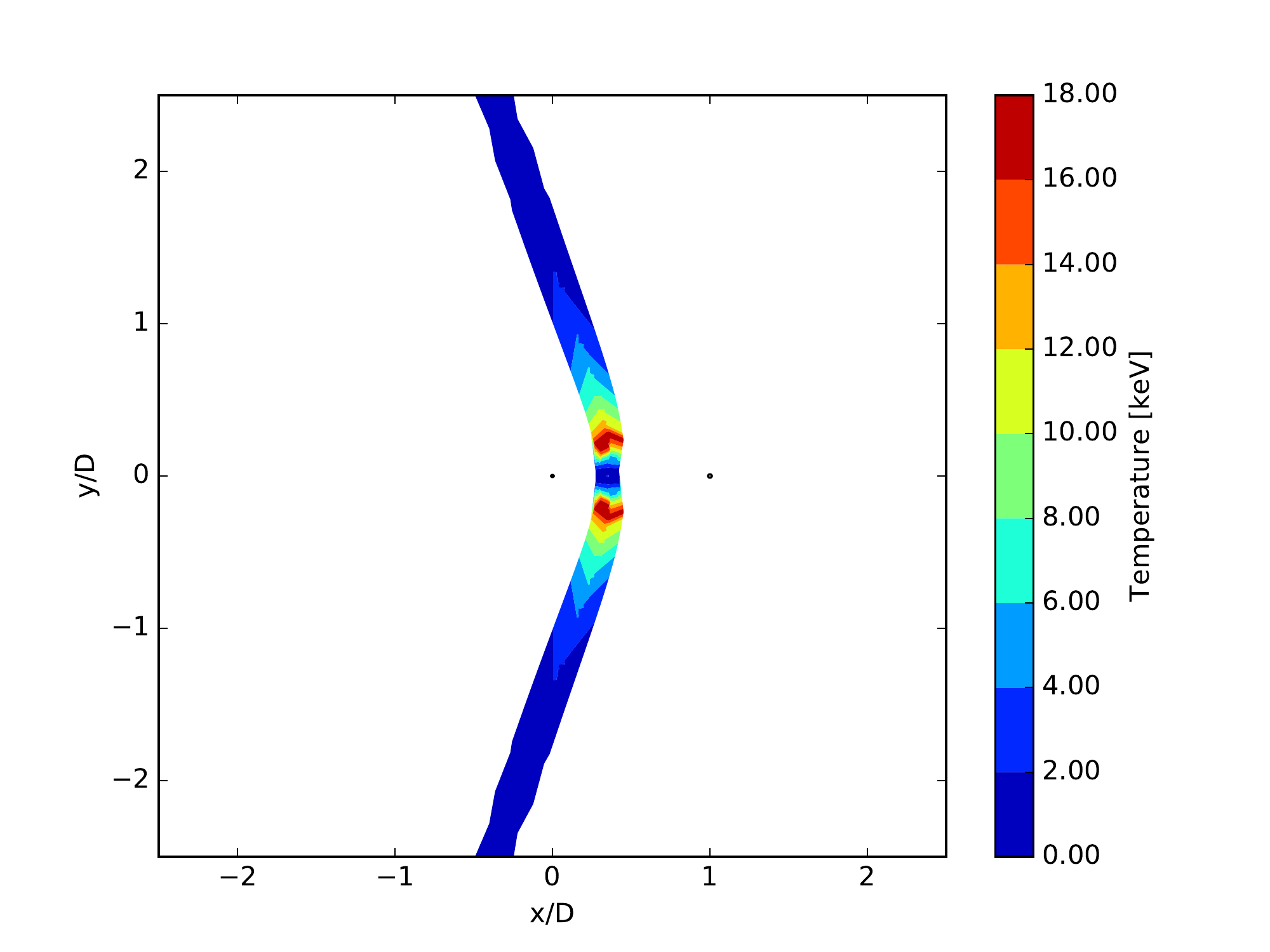}}

\put(0,200){\includegraphics[trim= 0cm 0cm 0cm 1.2cm,clip, width=5.2cm]{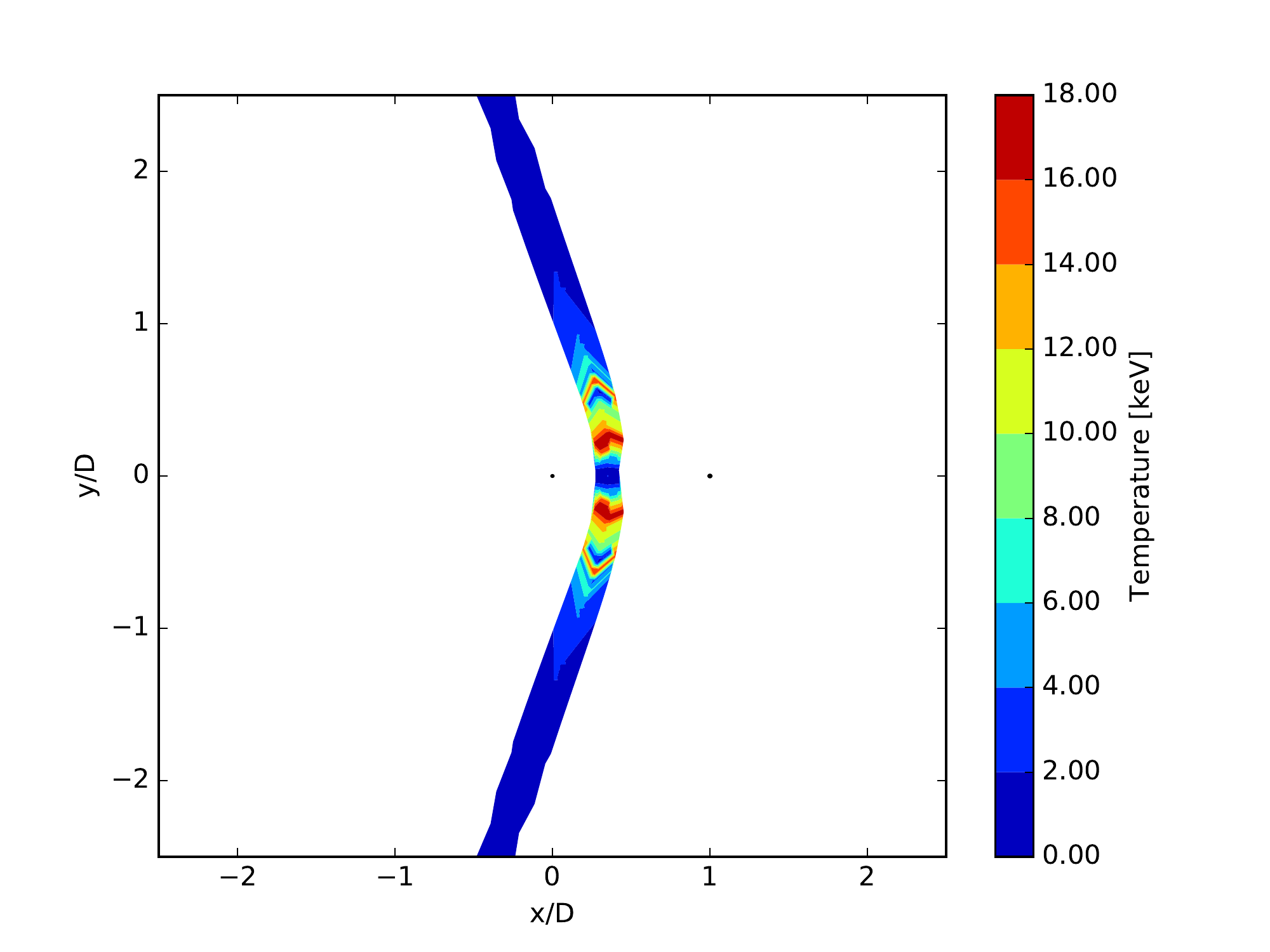}}
\put(133,200){\includegraphics[trim= 0cm 0cm 0cm 1.2cm,clip, width=5.2cm]{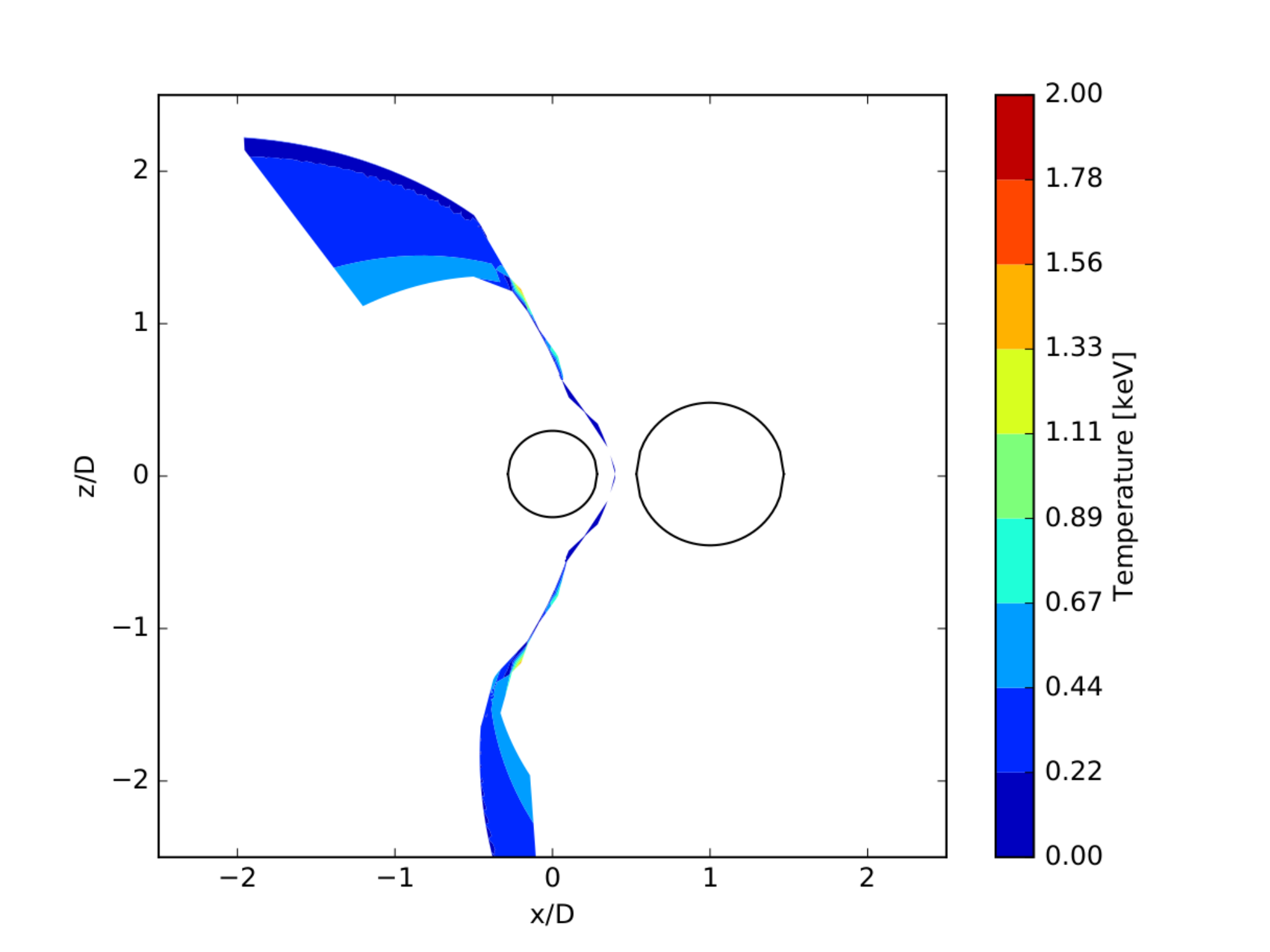}}
\put(266,200){\includegraphics[trim= 0cm 0cm 0cm 1.2cm,clip, width=5.2cm]{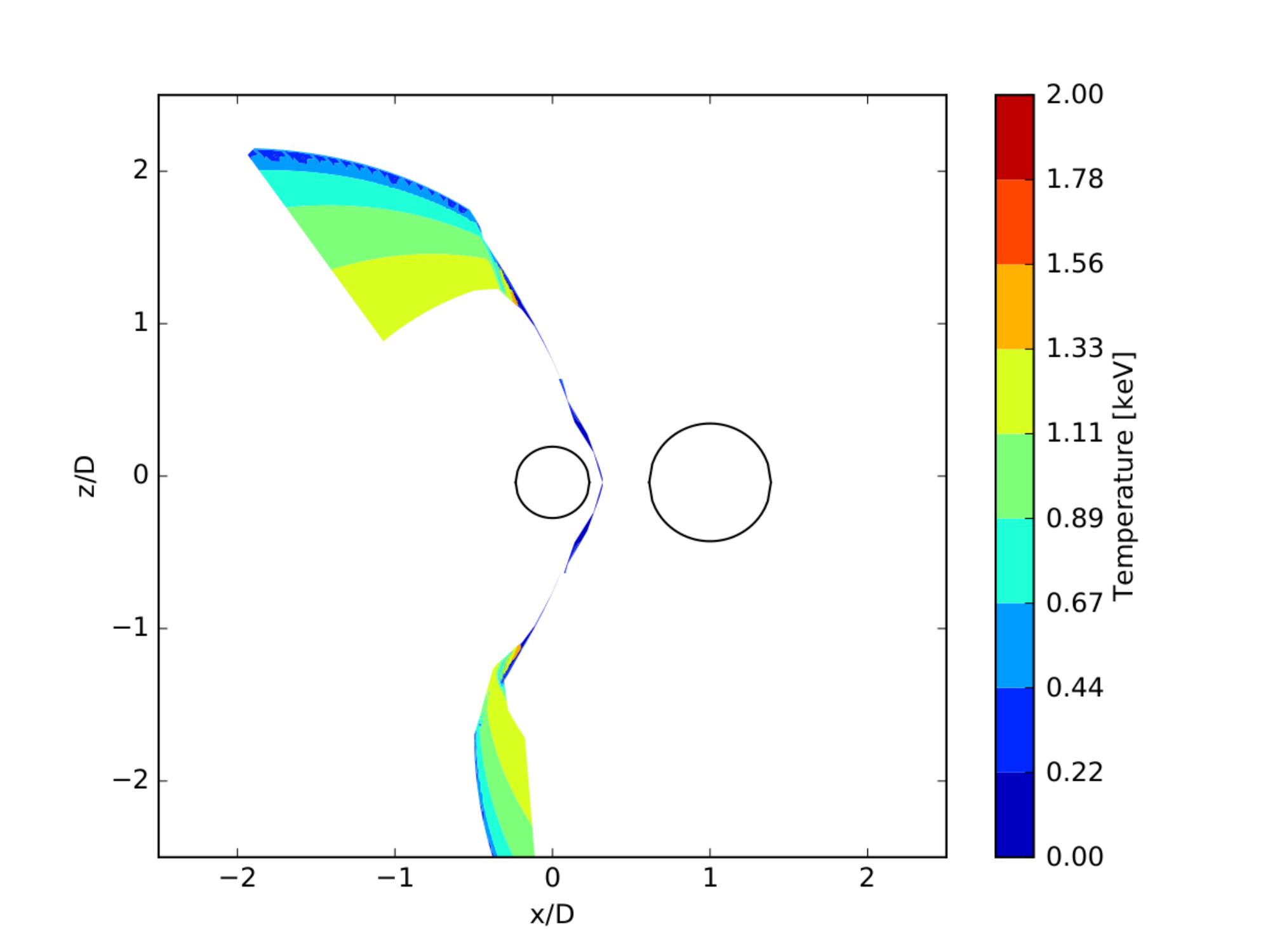}}
\put(400,200){\includegraphics[trim= 0cm 0cm 0cm 1.2cm,clip, width=5.2cm]{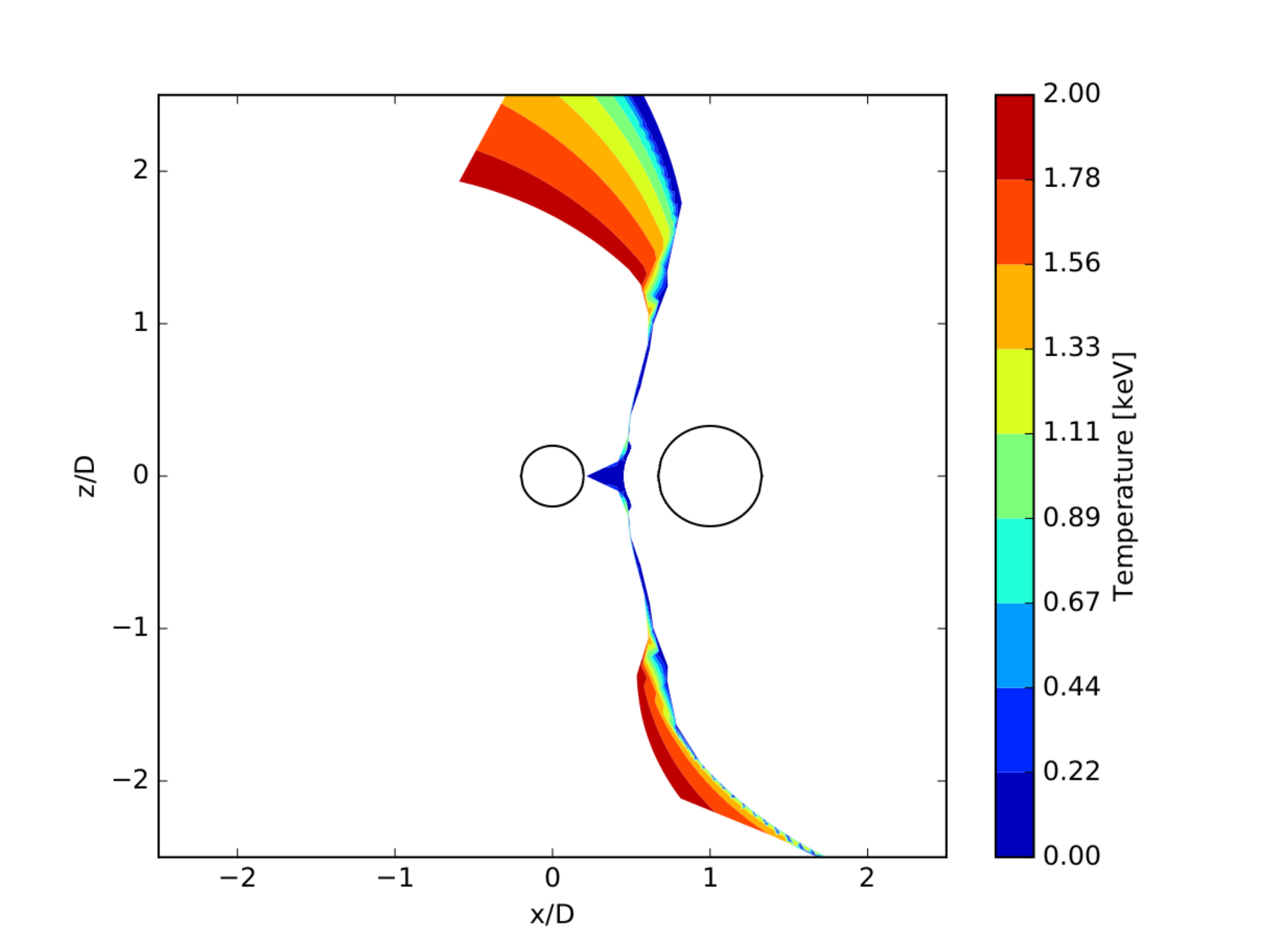}}

\put(0,100){\includegraphics[trim= 0cm 0cm 0cm 1.2cm,clip, width=5.2cm]{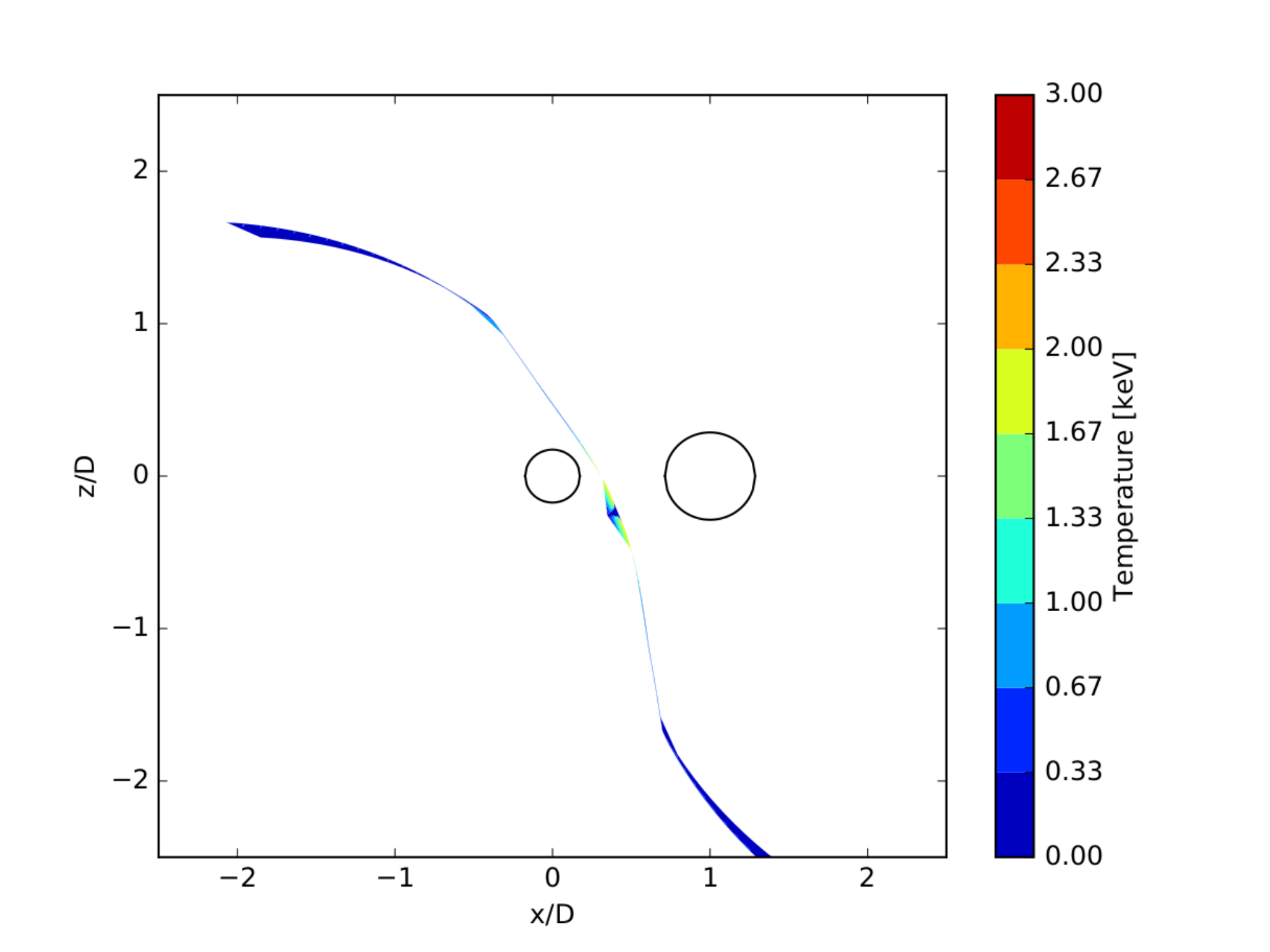}}
\put(133,100){\includegraphics[trim= 0cm 0cm 0cm 1.2cm,clip, width=5.2cm]{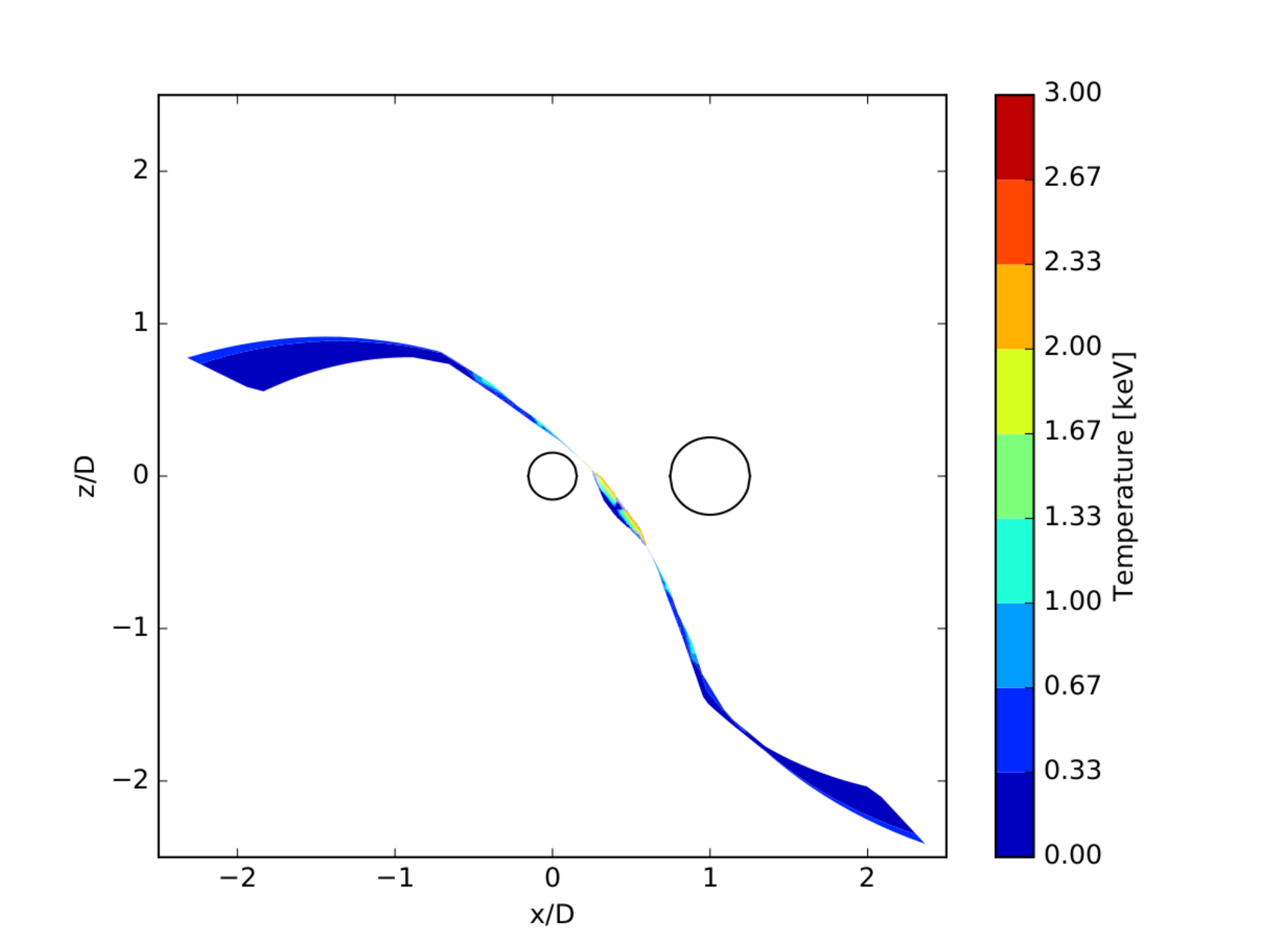}}
\put(266,100){\includegraphics[trim= 0cm 0cm 0cm 1.2cm,clip, width=5.2cm]{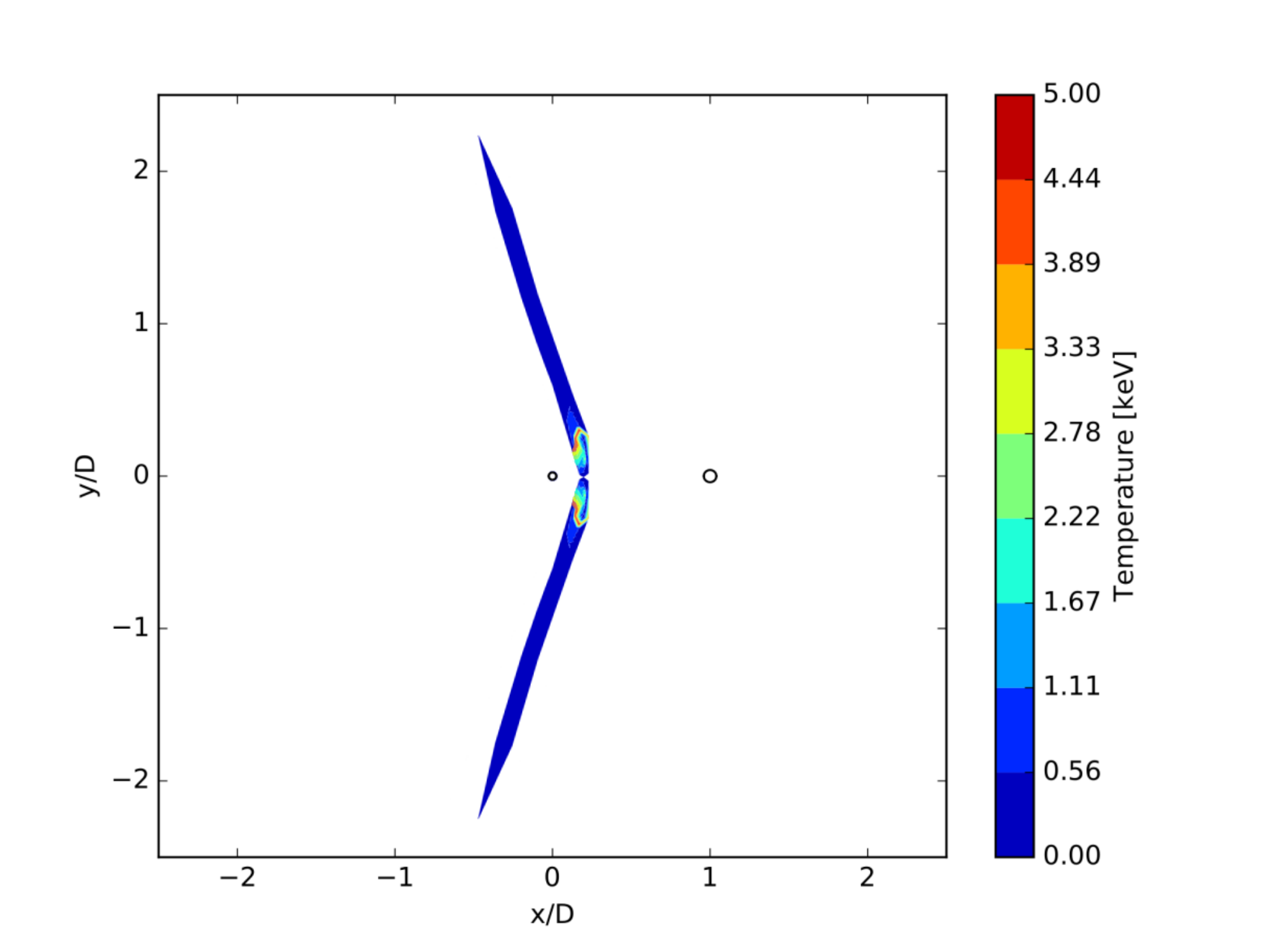}}
\put(400,100){\includegraphics[trim= 0cm 0cm 0cm 1.2cm,clip, width=5.2cm]{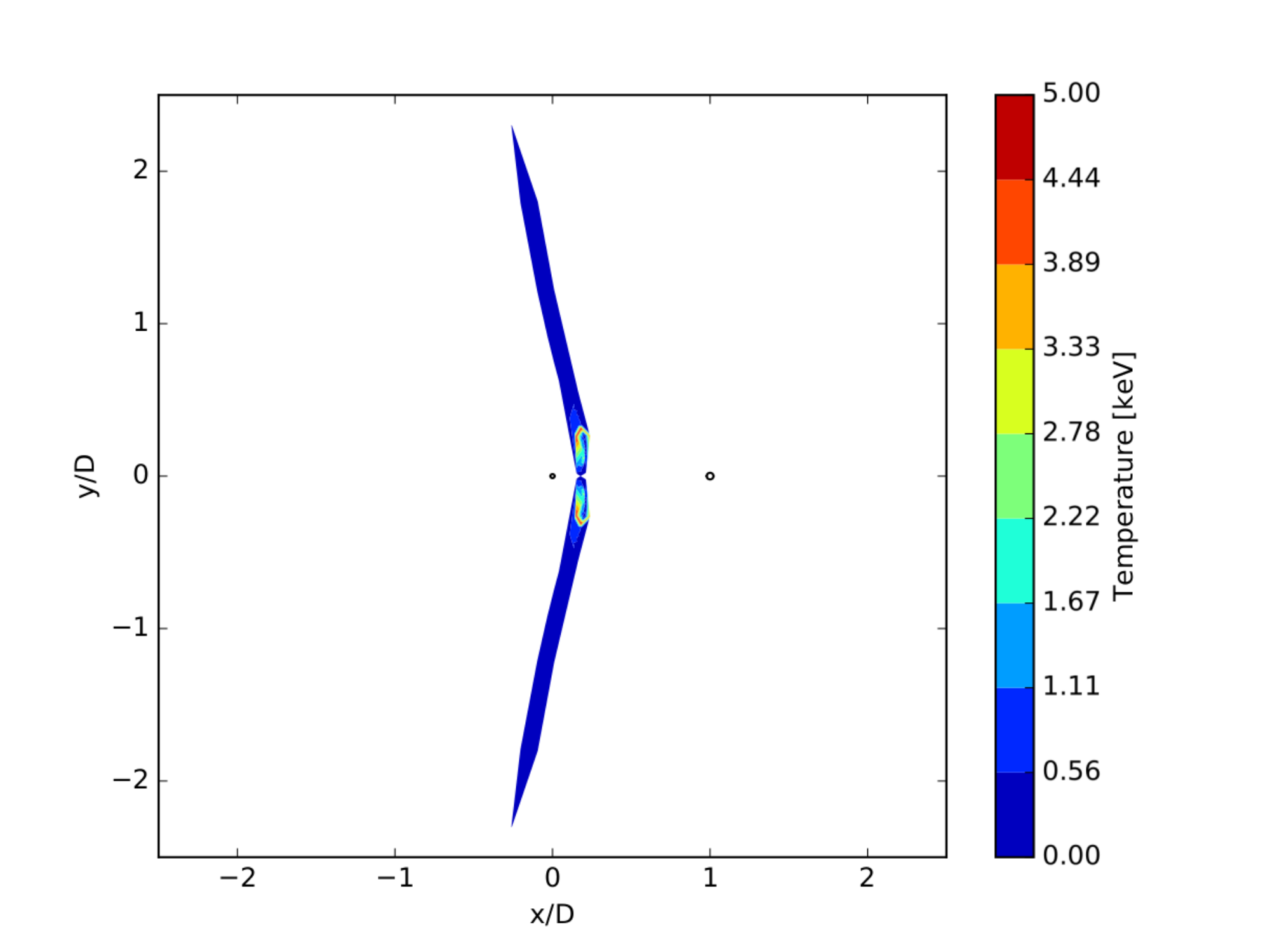}}

\put(0,0){\includegraphics[trim= 0cm 0cm 0cm 1.2cm,clip, width=5.2cm]{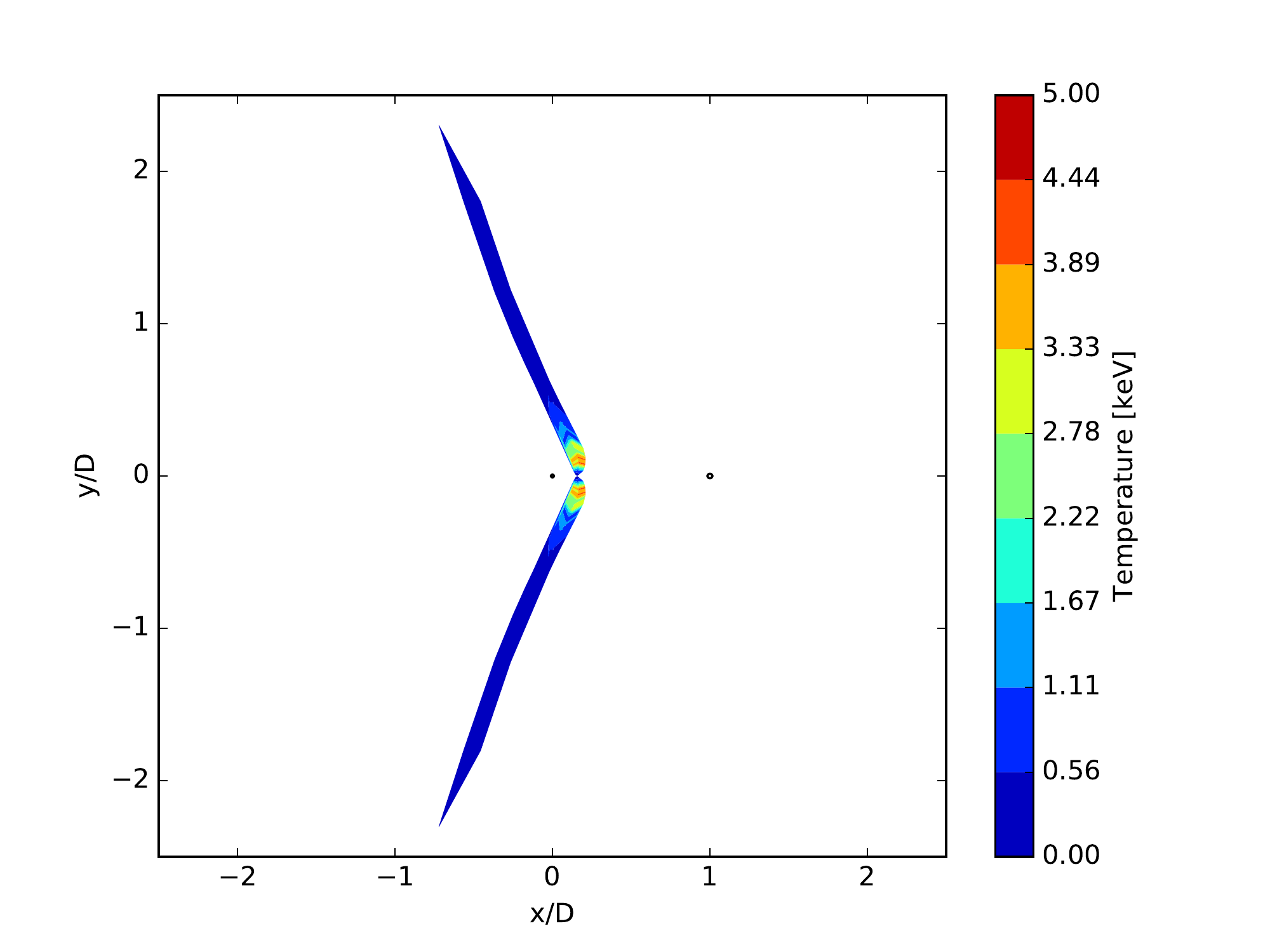}}
\put(133,0){\includegraphics[trim= 0cm 0cm 0cm 1.2cm,clip, width=5.2cm]{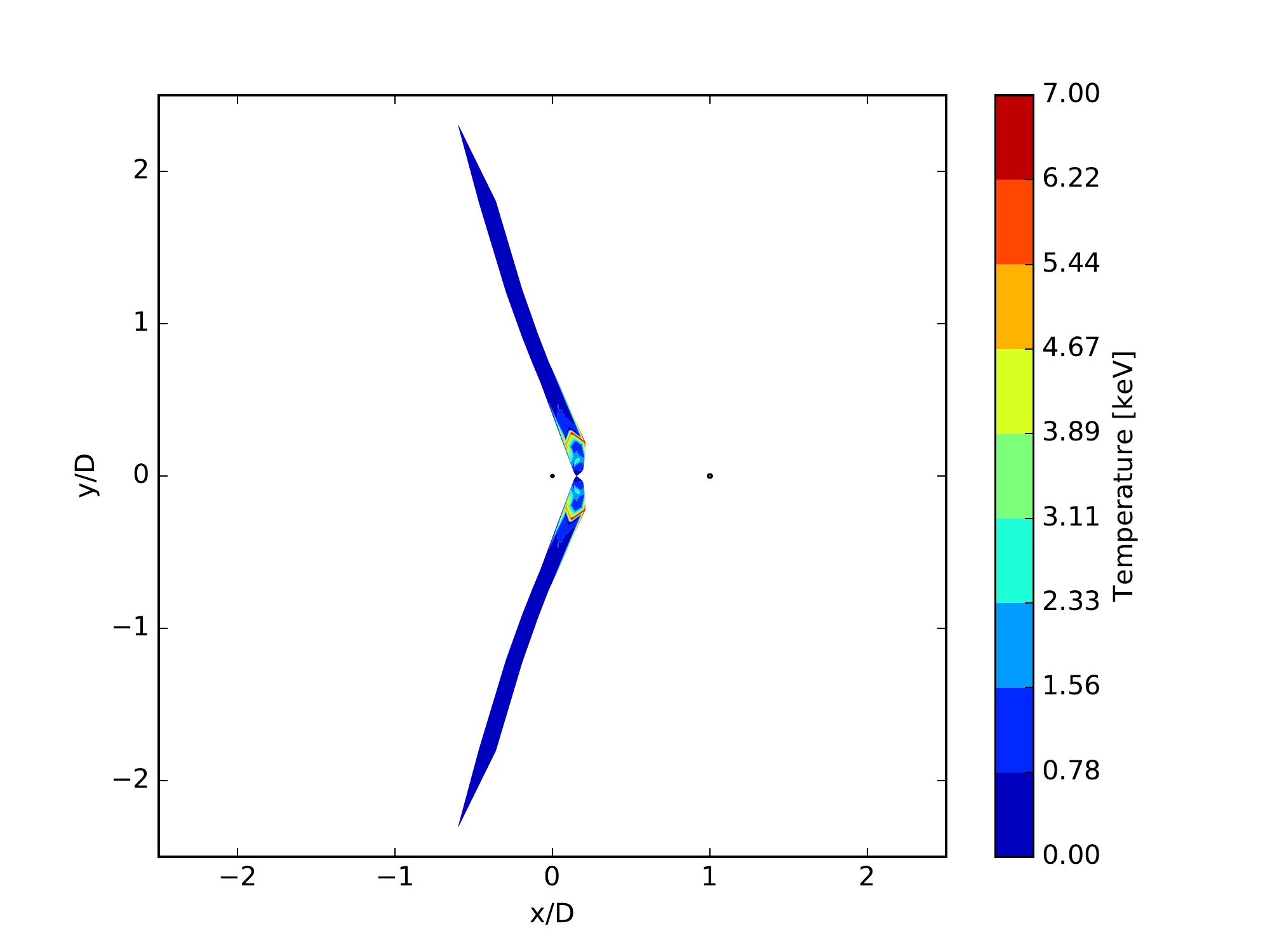}}

\put(22,690){\scriptsize{O7V+O7V}}
\put(22,682){\scriptsize{$d=57.6\,\rsun{}$}}
\put(155,690){\scriptsize{O7V+O7V}}
\put(155,682){\scriptsize{$d=70.8\,\rsun{}$}}
\put(288,690){\scriptsize{O7V+O7V}}
\put(288,682){\scriptsize{$d=84.1\,\rsun{}$}}
\put(422,690){\scriptsize{O7V+O7V}}
\put(422,682){\scriptsize{$d=97.3\,\rsun{}$}}

\put(22,590){\scriptsize{O7V+O7V}}
\put(22,582){\scriptsize{$d=110.5\,\rsun{}$}}
\put(155,590){\scriptsize{O7V+O7V}}
\put(155,582){\scriptsize{$d=582.9\,\rsun{}$}}
\put(288,590){\scriptsize{O7V+O7V}}
\put(288,582){\scriptsize{$d=1055.2\,\rsun{}$}}
\put(422,590){\scriptsize{O7V+O7V}}
\put(422,582){\scriptsize{$d=1527.6\,\rsun{}$}}

\put(22,490){\scriptsize{O7V+O7V}}
\put(22,482){\scriptsize{$d=2000.0\,\rsun{}$}}
\put(155,490){\scriptsize{O5I+O3III}}
\put(155,482){\scriptsize{$d=46.4\,\rsun{}$}}
\put(288,490){\scriptsize{O5I+O3III}}
\put(288,482){\scriptsize{$d=55.1\,\rsun{}$}}
\put(422,490){\scriptsize{O5I+O3III}}
\put(422,482){\scriptsize{$d=63.8\,\rsun{}$}}

\put(22,390){\scriptsize{O5I+O3III}}
\put(22,382){\scriptsize{$d=72.5\,\rsun{}$}}
\put(155,390){\scriptsize{O5I+O3III}}
\put(155,382){\scriptsize{$d=554.4\,\rsun{}$}}
\put(288,390){\scriptsize{O5I+O3III}}
\put(288,382){\scriptsize{$d=1036.2\,\rsun{}$}}
\put(422,390){\scriptsize{O5I+O3III}}
\put(422,382){\scriptsize{$d=1518.1\,\rsun{}$}}

\put(22,290){\scriptsize{O5I+O3III}}
\put(22,282){\scriptsize{$d=2000.0\,\rsun{}$}}
\put(155,290){\scriptsize{O9III+O9V}}
\put(155,282){\scriptsize{$d=48.3\,\rsun{}$}}
\put(288,290){\scriptsize{O9III+O9V}}
\put(288,282){\scriptsize{$d=58.5\,\rsun{}$}}
\put(422,290){\scriptsize{O9III+O9V}}
\put(422,282){\scriptsize{$d=68.7\,\rsun{}$}}

\put(22,190){\scriptsize{O9III+O9V}}
\put(22,182){\scriptsize{$d=78.9\,\rsun{}$}}
\put(155,190){\scriptsize{O9III+O9V}}
\put(155,182){\scriptsize{$d=89.1\,\rsun{}$}}
\put(288,190){\scriptsize{O9III+O9V}}
\put(288,182){\scriptsize{$d=566.8\,\rsun{}$}}
\put(422,190){\scriptsize{O9III+O9V}}
\put(422,182){\scriptsize{$d=1044.5\,\rsun{}$}}

\put(22,90){\scriptsize{O9III+O9V}}
\put(22,82){\scriptsize{$d=1522.3\,\rsun{}$}}
\put(155,90){\scriptsize{O9III+O9V}}
\put(155,82){\scriptsize{$d=2000.0\,\rsun{}$}}
\end{picture} 
\caption{Temperature distribution inside the wind shock region for the O5I+O3III, O7V+O7V, and O9III+O9V systems.}
\label{temperature}
\end{figure*}

Figures~\ref{density} and \ref{temperature} represent the associated density and temperature distributions inside the wind shock region.
One can notice the strong effect of the Coriolis deflection on the shape of the wind interaction zone for the six shortest orbital separations for each system.
Comparing the characteristics of the shock in adiabatic and radiative regime, the maximum temperature of the plasma in adiabatic regime is higher than in the radiative regime, as expected.
This is due to the impact of radiative cooling and the lower separation between the stars for systems in the radiative regime, which prevents the winds from reaching their maximum velocity before they collide.
However, the highest densities in the interaction zone are reached in radiative regime because density scales with $1/r^2$. 
We note that the predicted fluxes of the Fe~K resonance line are between five and ten orders of magnitude higher for radiative cases compared to adiabatic situations. 
This mainly reflects the difference of the density inside the emitting cells as the emission scales with density squared and is much less sensitive to temperature (in the temperature regime considered here). 
Yet, this result must be taken with great caution as the formalism of \citet{antokhin04} was found to predict X-ray luminosities that are far too high when compared to observations \citep{debecker04b} and the real emission of a radiative wind interaction region is likely reduced by the thin-shell instabilities \citep{kee14}.

\subsubsection{O7V+O7V system}
We now consider the first binary: the O7V+O7V system (top panel of Fig.~\ref{line_prof}).
First, we observe that the system does not emit the Fe~K lines if the stars are separated by less than about 70$\,\rsun{}$.
This is due to the temperature of the shocked plasma that is too low to emit this line.
For wider separations, the stellar winds have more time to accelerate before colliding, leading to a post-shock plasma at temperatures above 1.08\,keV (where the Fe~K lines start to be emitted, although at a low level at first).
The highest temperatures are reached far from the line of centres.
Indeed, close to the line of centres, the radiative inhibition from the companion star is very strong and nearly cancels the acceleration of the stellar wind.
At the shock position, the winds have thus reached only several hundreds of kilometres per second before colliding, which is not enough to efficiently heat the shocked plasma.
However, far from the line of centres, the pre-shock wind velocity is higher than $1000\,\mathrm{km\,s^{-1}}$ since the effect of the radiative inhibition is lower at these distances and off-axis angles.
Considering only the component of the velocity normal to the shock, the available energy is high enough to heat up the plasma.

In the radiative configurations, the wind interaction region between the stars is very narrow.
Because the emission is directly proportional to the volume of the emission region, we expect a small contribution to the emission to be produced in this region.

At a separation of $d = 70.8\,\rsun{}$, the plasma temperature is very close to the lower limit for the emission of the Fe~K lines.
This leads to a noisy line profile without a well-defined shape (numerical noise).

At $d = 84.1\,\rsun{}$, the line profiles are very narrow and have a very similar shape for each inclination.
This effect is explained by the size of the emission region.
Indeed, the emission of the resonance line is maximum for plasma temperatures of about 5\,keV and rapidly drops for lower values of the temperature.
The maximum temperature of the shocked plasma for this orbital separation is about 2\,keV. 
The volume of the plasma at the highest temperature is maximum close to the orbital plane of the stars and for the positive part of the z-axis (in the direction of the orbital motion).
The emission thus mostly arises in a narrow shell in the Coriolis-deflected part of the shock region located in the direction of the orbital motion. 
This narrow emission region leads to a narrow range in radial velocities of the emitting gas. 

At $d = 110.5\,\rsun{}$, the maximum temperature is higher (close to 3\,keV).
Moreover, the shock width is higher between the stars and, at this distance, the stellar winds have enough time to accelerate despite the radiative inhibition by the companion star.
We thus expect the highest contribution to the emission to be produced in the hot, wide, and dense region between the stars.
The shape of the shock in this region encompasses a wider range of Doppler shifts for the emitted photons even for low inclinations.
The line profiles are thus wider for this orbital separation.
But, considering the highest inclinations (72 or $90^\circ$), the line profiles at phase 0 and 0.4 become narrower.
This is because of the position of the emitting region. 
Indeed, at these inclinations, the emitting region is nearly perpendicular to the observer at phases close to the conjunction ($\phi=0$ and 0.5).
The radial velocities of emitted photons are thus close to zero leading to narrow line profiles centred on the energy of the lines at rest.

Considering the adiabatic configurations, the wind shock region is wide (about $0.25\,d$) leading to the volume of the shocked plasma having a temperature exceeding 1\,keV to be very high.
But the increase in volume does not compensate for the decay of the density, leading to a predicted line flux that is lower than in the radiative regime.
Because of the absence of the Coriolis effect, the shock is totally symmetric around the line of centres.
Due to this symmetry, the line profiles at phases 0.2 and 0.8  and phases 0.4 and 0.6 have exactly the same shape.
Finally, there is a small impact of the separation on the temperature because the winds have nearly reached their terminal velocities above the line of centres.
As orbital separation increases, we are able to distinguish the double peak feature.
This double peak feature and its dependence on inclination and orbital phase are well explained by the direction of the observer.
Indeed, at low inclinations, the wind shock region is nearly parallel to the direction of the observer at all phases.
Most of the emitted photons thus have a high negative or positive Doppler shift (about 1000 kilometres per second).
As the inclination increases, the shocks become more and more normal to the direction of the observer at conjunction phases and the emitted photons thus have a very small Doppler shift, leading to a narrow, single-peaked, profile at phase 0.
At other phases, the double peak feature is still observed because the shock is more parallel to the direction of the observer.

This system can be used for the comparison with the O6.5V+O7V binary HD~159176 characterised by an orbital period of about $3.367$\,days \citep{penny16}.
These authors determined an inclination of $43.5^\circ\pm4.5^\circ$ leading to a semi-major axis of $42.4\,\rsun{}$ implying a radiative cooling of the wind shock.
Comparing with the theoretical results, the orbital separation of the system is probably too short to emit the Fe~K triplet and the wind shock region is probably crashing onto the stellar surface.
Moreover, the masses derived by \citet{penny16} are higher than those we used in our simulations for this system.
This leads to a radiative inhibition of the winds that is higher than computed here and thus a shocked plasma temperature that is even lower than in our model.
This strengthens the conclusion of the absence of Fe~K triplet emission.
The \textit{XMM-Newton} observation of this system showed indeed that no X-ray emission is detected above 5\,keV \citep{debecker04b}, which is consistent with our simulations.

\subsubsection{O5I+O3III system}
We now analyse the O5I+O3III system (middle panel of Fig.~\ref{line_prof}).
At a separation of $46.4\,\rsun{}$, the maximum temperature of the shock is lower than 2\,keV.
The emission of the Fe~K resonance line is thus not very strong.
Moreover, the emitting region is wide compared to the O7V+O7V system.
The line profiles are thus spread over a wider range in radial velocity leading to a broad and poorly-defined profile of the resonance line.
For the other lines of the Fe~K triplet, the temperature is not sufficient to allow distinguishable emission leading to a very noisy profile (numerical noise).
Any detection of feature in the profile shape is thus very difficult (see Fig.~\ref{fig_sat_fe}).

At $d = 55.1\,\rsun{}$, the resonance line is stronger and its shape is better defined thanks to shock regions that have a higher temperature.
The double peak feature is now clearly detectable.
For an inclination of $90^\circ$, we clearly observe that the resonance line at phases 0.4 and 0.2 is blueshifted while at phases 0.6 and 0.8 it appears redshifted.

At $d = 63.8\,\rsun{}$, once again, the higher the inclination, the higher the shift of the lines, especially at phases 0 and 0.4.
This is also explained by the skewing angle that rotates the shock around the centre of mass of the system because of the Coriolis effect.

At $d = 72.5\,\rsun{}$, the shock has two thin regions at the end of the shock tail where the temperature increases up to 5\,keV, that it to say, close to the maximum of the  resonance line emissivity.
These spots thus contribute significantly to the emission of the lines.
The general behaviour is the same as for a separation of $63.8\,\rsun{}$: the higher the inclination, the higher the shift of the lines.
The difference is in the distribution of the Doppler shifts of the emitted photons: The emitting regions being smaller at $d=72.5\,\rsun{}$, the velocity range of the emitted photons is narrower leading to narrower profiles.

In the adiabatic regime, the shock is symmetric around the line of centres.
The line profiles at phases 0.2 and 0.8 and phases 0.4 and 0.6 thus have exactly the same shape.
Contrary to the O7V+O7V system, the maximum temperature inside the shock is very high (up to 18\,keV) leading to a stronger emission of the resonance line.
Moreover, the double peak feature is less obvious because of the curvature of the shock.
Finally, at an inclination of $90^\circ$ and phase 0, the lines are less clearly defined than for the O7V+O7V system because the main emitting regions (where the temperature is close to 6\,keV) are distributed over several locations along the shock up to an angle of about $50^\circ$ above the line of centres. 
This leads to several groups of radial velocities for the emitting gas and thus several energies for each observed line.

The observed system that can be compared to our synthetic system is the O5-5.5I+O3-4III binary Cyg~OB2~\#9 characterised by an orbital period of $858.4\pm1.5$\,days and an eccentricity of $0.713\pm0.016$ \citep{naze12}.
These authors determined a minimum semi-major axis of $a\sin{i}=1525.3\pm59.5\,\rsun{}$ leading to an adiabatic cooling of the shocked wind material.
Comparing with the masses listed by \citet{martins05}, \citet{naze12} inferred an inclination of $62^\circ$ for this system.
Given the high eccentricity, the range of orbital separations is $[386.5-2307.0]\,\rsun{}$.
This system can be compared to the O5I+O3III system for which we computed the line profiles in adiabatic regime for these orbital separations: 554.4, 1036.2, 1518.1, and 2000.0$\,\rsun{}$.
Given the argument of periastron $\omega=192.1^\circ$ \citep{naze12}, these distances correspond to phases of 0.58, 0.65, 0.72, and 0.85 for Cyg~OB2~\#9 (phase equal to zero at the conjunction with the star displaying the most powerful wind in front).
The line profile that would be observed at these phases for this system would thus be between those computed for the O5I+O3III system with an inclination of $54^{\circ}$ and $72^\circ$ in an adiabatic regime for phase $\phi=0.6$ and $\phi=0.8$.
A line at about 6.7\,keV is well observed in the X-ray spectra of Cyg~OB2~\#9 taken with \textit{Swift} and \textit{XMM-Newton} \citep{naze12} but the resolution of the current generation of X-ray satellite is not high enough to precisely analyse the shape of the line profile.

\subsubsection{O9III+O9V system}
We now analyse the O9III+O9V system (bottom panel of Fig.~\ref{line_prof}).
At $d = 48.3\,\rsun{}$, only two very small regions at the end of the shock cap have a temperature higher than 1\,keV.
This leads to a double peak feature of the resonance line.
However, as for the previous system, these regions are too small and too cool to strongly emit the other lines of the triplet leading to an overall profile that lacks any distinctive feature (see Fig.~\ref{fig_sat_fe}).

At $d = 58.5\,\rsun{}$, the temperature of the shock is higher with the hottest regions (about 1.5\,keV) located at the end of the shock cap.
Wider emission regions of lower temperature (about 1.2\,keV) are located in the Coriolis-deflected shock tail.
For the lowest inclinations, the profiles have the same shape with narrow lines. 
But, as the inclination increases, the line profiles at phase 0 are narrower than those computed at other phases.
This indicates that the highest contribution to the line profiles comes from the small regions of highest temperature.
Indeed, in these regions, the range of Doppler shifts of the emitted photons is small leading to narrow lines.
At an inclination of $72^\circ$ and phase 0, the radial velocities are nearly normal to the direction of the observer while at phases 0.8 and 0.2, part of the radial velocities are more parallel to the direction of the observer leading a broader profile and a small asymmetry in the line shape.

At $d = 68.7\,\rsun{}$, the temperature is even higher and the hottest regions are now located in the Coriolis tail.
The highest component of the emission is thus created at the shock on the side of star 1 (with the less powerful wind).
The general behaviour of the lines is the same as for the previous value of the separation.

For the latter two orbital separations in the radiative configuration, we observe that the lines get narrower with increasing orbital separation.
This is because the region that contributes most of the emission is located in the shock cap and its extent away from the line of centres decreases with increasing orbital separation, thereby leading to a narrower range of Doppler shifts.

In the adiabatic regime, we clearly observe the redshift of the lines for phase 0 and a small blueshift for phases 0.4 and 0.6, which are superimposed due to the symmetry of the shock.
This effect is better observed as the orbital separation increases.
The temperature distributions in the shock at all four orbital separations are very similar leading to very similar shapes of the line profiles.

This system can be used for the comparison with the O9III+O9.7V binary HD~152247 located at a distance of 1.52\,kpc \citep{sana12}.
This is a wide binary characterised by an orbital period of $581.71\pm0.70$\,days, an eccentricity of $0.593\pm0.015$, and an inclination of $76.8\pm2.8^\circ$ \citep{lebouquin16}.
\citet{lebouquin16} determined a semi-major axis of $940.0\,\rsun{}$ leading to a shocked wind in the adiabatic cooling regime.
Given the eccentricity, the range of orbital separations is $[382.6-1497.5]\,\rsun{}$.
For the O9III+O9V system, we computed the line profiles in adiabatic regime for orbital separations of 566.8, 1044.5, 1522.3, and 2000.0$\,\rsun{}$.
Given the argument of periastron $\omega=144.6^\circ$ \citep{lebouquin16}, the distances lying in the range of orbital separations of HD~152247 correspond to phases of 0.10 and 0.23 (phase equal to zero at the conjunction with the star displaying the most powerful wind in front).
The line profile that would be observed at these two phases for this system would thus be close to those computed for the O9III+O9V system with an inclination of $72^\circ$ and a separation of 566.8 ($\phi=0$) and 1044.5$\,\rsun{}$ ($\phi=0.2$).
The integrated line fluxes predicted for the 566.8 and 1044.5$\,\rsun{}$ orbital separations are low, making a detection of the line at the distance of the system very unlikely. 
Indeed, the \textit{XMM-Newton} observation of this binary did not reveal any strong emission near 6.7\,keV \citep{sana06}.

\subsection{Perspectives for forthcoming X-ray observatories}
Current-generation X-ray satellites have a low efficiency at high energies (above 5\,keV).
Although an emission line at about 6.7\,keV was observed in a few bright binary systems (e.g. WR\,25 \citealt{raasen03,arora19}, $\eta$\,Car \citealt{tsuboi97,corcoran01,pittard02,viotti02,leutenegger03} or WR\,140 \citealt{maeda99,rauw16b}), the accurate analysis of the shape of this line remains very difficult.

Fortunately, in the future, two X-ray satellites will allow a high-sensitivity and high-resolution coverage of the energy domain around 6--7\,keV. The Japanese X-Ray Spectroscopy and Imaging Mission \textit{XRISM} will carry a high-resolution micro-calorimeter spectrometer called \textit{Resolve} \citep{tashiro18}. \textit{Athena} \citep{nandra13}, ESA's next generation X-ray observatory, will carry a high-sensitivity high-resolution bolometric spectrograph (the X-ray Integral Field Unit, \textit{X-IFU}, \citealt{barret13, ravera14,barret18}). 
We convolved the line profiles of the O9III+O9V system studied in Sect.~\ref{three_sys} with the synthetic instrumental response of the \textit{Athena/X-IFU} and the \textit{XRISM/Resolve} to predict what should be observed with the two instruments.
This O9III+O9V system was chosen because it shows wide variations in the shape of the line profiles along the orbital separation and viewing angle.
For the \textit{Athena/X-IFU} instrument, we used the newest version of the theoretical response matrix file (RMF) and ancillary response file (ARF) computed assuming an active mirror aperture radius of $259-1183\,$mm, a 2.3\,mm rib spacing, an on-axis observation and a thick optical blocking filter (mostly used for the O-type binaries, which are often very bright in optical).
The RMF is  XIFU\_cc\_baselineconf\_thickfilter\_2018\_10\_10.rmf and the ARF is XIFU\_cc\_baselineconf\_thickfilter\_2018\_10\_10.arf.
For the \textit{XRISM/Resolve} instrument, we used the RMF computed for the high-resolution grade xarm\_res\_h5ev\_20170818.rmf.
The ARF of \textit{XRISM/Resolve}, xarm\_res\_flt\_pa\_20170818.arf, includes the quantum efficiency and the optical blocking filter transmission of the \textit{Hitomi SXS} detector. 

The emission at energy $E$ of the convolved line profile $LP_\mathrm{conv}$ is:
\begin{equation}
 LP_\mathrm{conv}(E)=\int^{\infty}_0 R(E,E')\,A(E')\,LP_\mathrm{th}(E')\,dE'\,,
\end{equation}
where $LP_\mathrm{th}(E')$ is the emission of the theoretical line profile at an energy $E'$, $R(E,E')$ is the probability that a photon of energy $E'$ is recorded with an energy $E$, and $A(E')$ is the effective area at the energy $E'$.

\begin{figure*}
\centering
\includegraphics[trim= 0cm 1cm 0cm 1cm,clip, width=13.5cm]{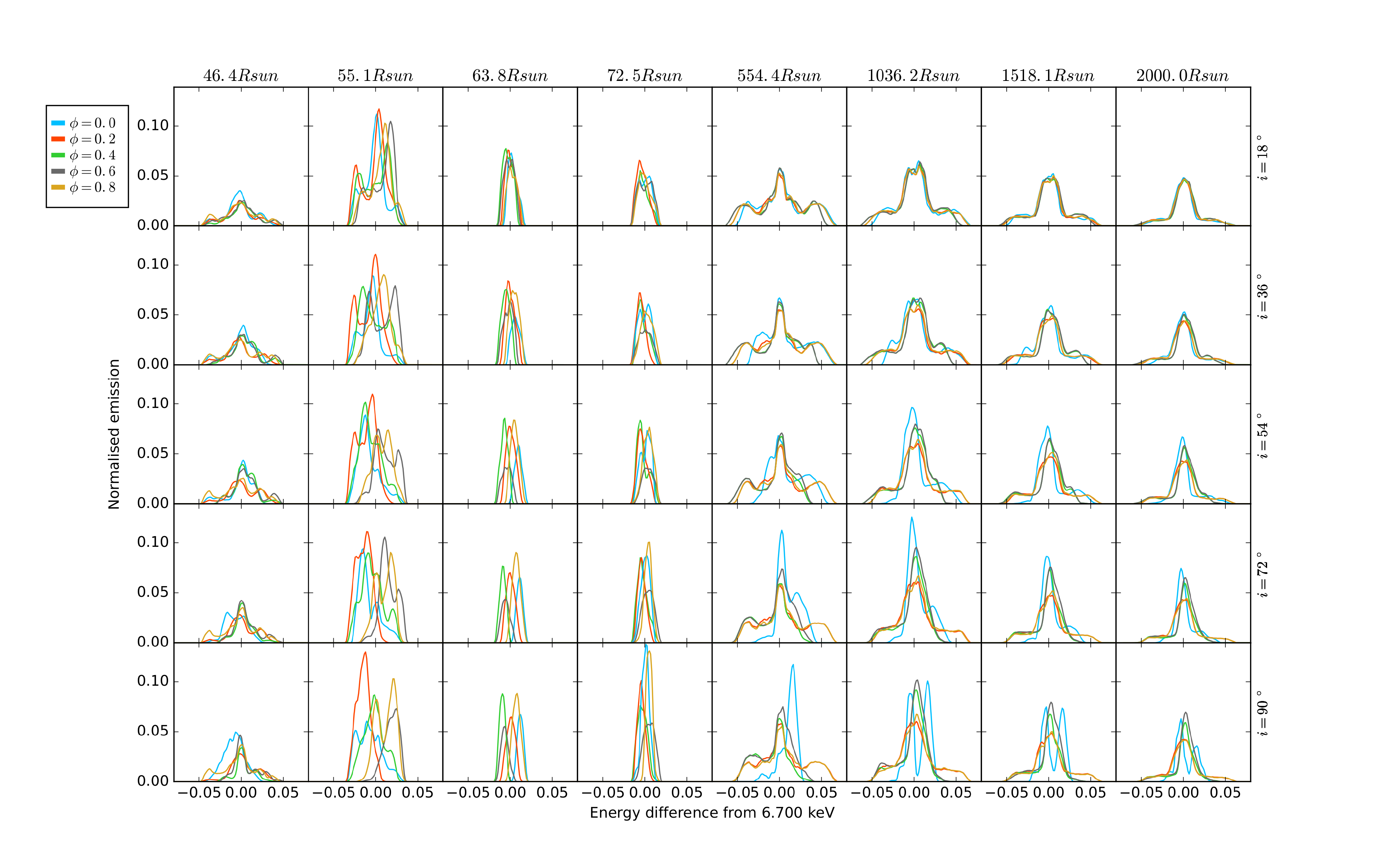}\\
\includegraphics[trim= 0cm 1cm 0cm 1cm,clip, width=13.5cm]{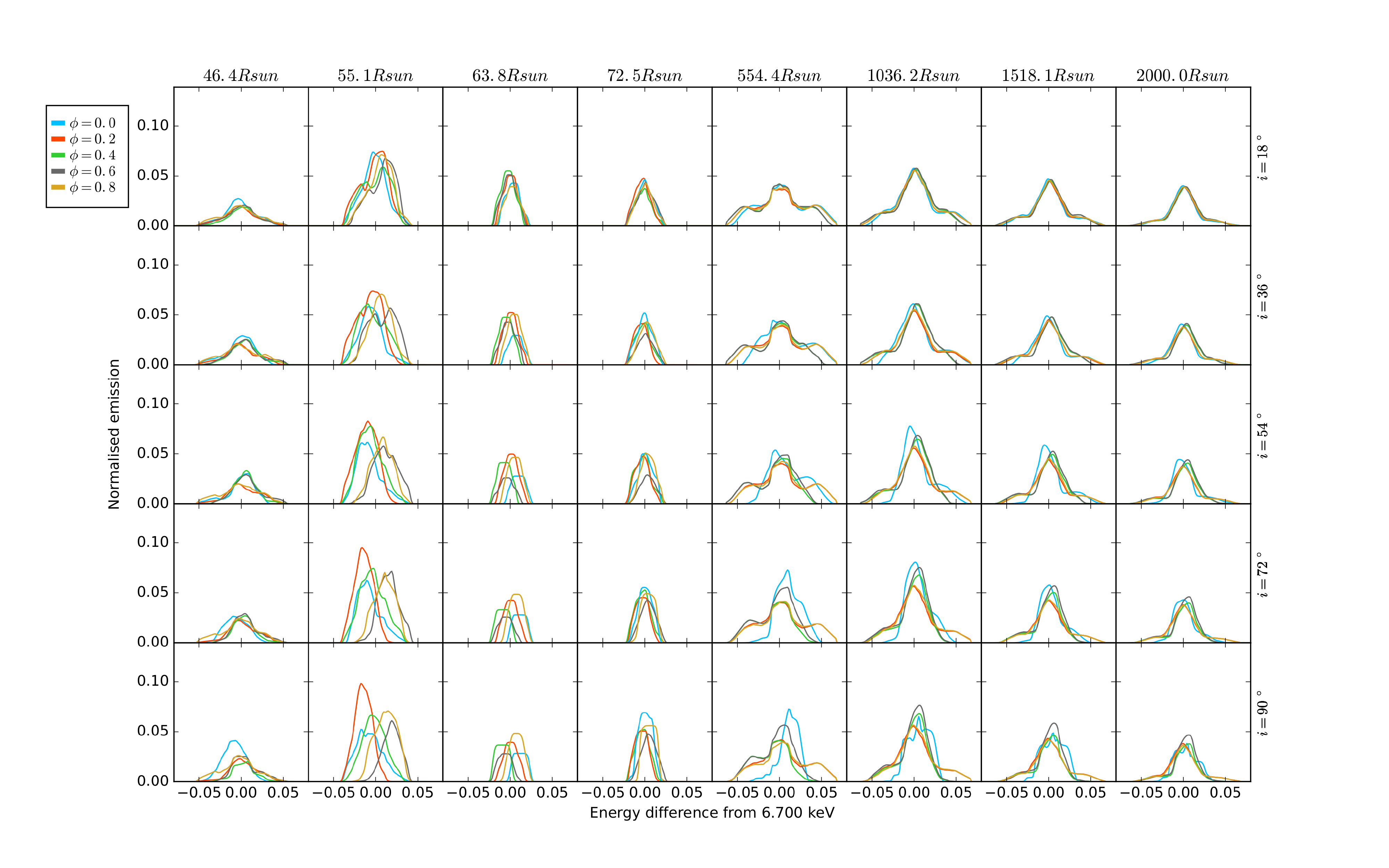}
\caption{Line profiles of the O9III+O9V system convolved with the response of \textit{Athena/X-IFU} (top panel) and \textit{XRISM/Resolve} (bottom panel).}
\label{line_prof_conv}
\end{figure*}

Figure~\ref{line_prof_conv} shows the line profiles convolved with the response of the \textit{Athena/X-IFU} and the \textit{XRISM/Resolve} instruments.
The improvement of the spectral resolution of both instruments over that of the current X-ray satellites will allow the first detailed analyses of the morphology of the Fe K line profiles, enabling a very precise comparison between the observed and theoretical line profiles. Whilst \textit{XRISM/Resolve} will allow a good description of the overall shape of the line profiles, the finest structures of the profiles, such as close double peak features, will only be resolved with \textit{Athena/X-IFU}.

Knowing the orbital parameters of a system, we will be able to find the line profiles whose shape is the closest to the observed one and from there retrieve the physical parameters of the wind interaction zone.
These parameters can then be compared to the stellar wind properties deduced from optical, UV and IR observations of the system to infer further information about the physics of winds in massive star binaries.

\section{Obtaining and using \progname{}}
\label{prog}
\progname{} is an interactive program written in \texttt{Python} (v2.7) and located on GitHub\,\footnote{\href{https://github.com/emossoux/LIFELINE/}{https://github.com/emossoux/LIFELINE/}}.
It can be redistributed and/or modified under the terms of the GNU General Public License as published by the Free Software Foundation, either version 3 of the License, or any later version.
GitHub allows a collaborative use of the program: If the user computes the line profile for a new binary, a new inclination or phase, or a new ion, he/she can increase the database for the future users.
The manual provided with the program contains a complete explanation on how to download and use \progname{} but the main points are provided hereafter.

In addition to the \progname{} program, the \texttt{Cloudy} free program \citep{ferland17} is needed to construct the cooling function of a specific atomic species.
To compute the characteristics of radiative shocks, the program also needs the \texttt{Python} interface of ATOMDB (pyatomdb\,\footnote{\href{https://pypi.org/project/pyatomdb}{pypi.org/project/pyatomdb}}) to compute the evolution of the temperature inside the shock and the ionisation fraction.

\progname{} can be used in three modes: Perform the overall computation, that is to say, the velocity distribution, the shock characteristics, and the line profile; compute the shock characteristics and the line profile using a pre-computed velocity distribution; or compute only the line profile using pre-computed shock characteristics and velocity distributions.

To use \progname{}, in addition to the code parameters (defining the directories, the requested ion, the mode of computation, etc.), the user must provide a file containing the stellar parameters of the studied binaries.
According to these parameters, the program will determine the cooling mechanism occurring inside the shock based on the criterion of \citet{stevens92} and whether or not the Coriolis deflection must be accounted for in the wind shock construction. 
In \progname{}, the Coriolis effect is included whenever $v_\mathrm{orb}>0.1\,v$ with $v_\mathrm{orb}$ the orbital velocity defined in Sect.~\ref{eq:cor} and $v$ the pre-shock velocity at the stagnation point.

\section{Summary and conclusions}
\label{summary}
Stars in massive binary systems are characterised by very powerful winds.
Between the stars, these winds collide thereby forming an interaction region, limited by two hydrodynamical shocks separated by a contact discontinuity surface.
If the preshock wind velocity is wide enough, the shocked plasma can be heated up to high temperatures and emits in the X-ray domain, mostly as X-ray lines.
The morphology of these lines strongly depends on the characteristics of the shock region and thus on the characteristics of the stellar winds.
The comparison between observed and theoretical line profiles should thus allow us to retrieve the characteristics of the stellar winds.

In this paper, we presented the \progname{} program for the simulation of the X-ray LIne proFiles in massivE coLliding wInd biNariEs.
This is a self-consistent program allowing the computation of the distribution of the wind velocity, the characterisation of the wind shock region, and the computation of the line profile.
Using this program, we generated line profiles of 780 binary systems for 25 couples of phases and inclinations.
We first computed, for each couple of stars, the 2D wind velocity distribution accounting for the radiative acceleration by the star, the radiative braking from the companion star, and the gravitational forces from both stars.
We then assumed a symmetry of the wind around the line of centres to determine the 3D velocity field of each wind.
From these winds, we finally computed the shape, the position and the physical characteristics of the resulting wind interaction region.
For each couple of stars, we defined ten orbital separations, allowing us to simulate six interaction zones in the radiative cooling regime and four cases in the adiabatic regime.
Whenever necessary, the Coriolis skewing and deflection of the shock was accounted for.
Finally, the line profiles were computed considering the characteristics of each emitting cell in the interaction zone.
The emitted photons were followed along their path towards the observer and the absorption by the shocked material and by the cool stellar winds was computed.
The histograms of the Doppler shifts of the observed photons were computed to create the line profiles.

In this paper, the line profiles of three systems were detailed.
The systems were chosen to sample three different values of the wind momentum ratio and different stellar parameters.
As expected for adiabatic shocks, the lower the wind momentum ratio, the more curved the shock around the star with the less powerful wind.
The line profiles are thus more asymmetric.
For interaction regions in the radiative regime, the orbital separation (and thus the winds distribution) and the Coriolis effects imply a more complex shape of the profiles.
Nevertheless, as illustrated by our three case studies, the profiles are sufficiently different to allow us to distinguish between profiles created in systems with different stellar characteristics.

We finally convolved the line profiles with the instrumental response matrices of the forthcoming \textit{Athena/X-IFU} and \textit{XRISM/Resolve} instruments and showed that the anticipated performances of these instruments will allow a detailed comparison between the theoretical and observed line profiles.

\begin{acknowledgements}
  This work was supported by the Fonds National de la Recherche Scientifique - FNRS, notably under grant n$^\mathrm{o}$ T.0192.19.
  We thank the European Space Agency (ESA) and the Belgian Federal Science Policy Office (BELSPO) for their support in the framework of the PRODEX Programme (contracts XMaS and HErMeS). 
\end{acknowledgements}

\begin{appendix}
\onecolumn
\section{Equations of the radiative inhibition of the stellar winds}
\label{NS}
\subsection{Distribution of the wind velocity}
We assume that the wind is axisymmetric about the line of centres.
We thus do not work with the azimuthal component of the Navier-Stokes equation in spherical coordinates.
The norm of the wind velocity is thus $v^2=\ur^2+\ut^2$.
We also assume a zero viscosity.
The steady state Navier-Stokes equation thus reduces to:
\begin{equation}
\left( \begin{array}{c} \ur\frac{\partial\ur}{\partial r}+\frac{\ut}{r_1}\frac{\partial \ur}{\partial\theta}-\frac{\ut^2}{r_1} \\   
\ur\frac{\partial\ut}{\partial r}+\frac{\ut}{r_1}\frac{\partial \ut}{\partial\theta}+\frac{\ur\,\ut}{r_1}\end{array} \right)
=-\frac{1}{\rho}\left( \begin{array}{c} \frac{\partial P}{\partial r}\\
   \frac{1}{r_1}\frac{\partial P}{\partial\theta}\end{array} \right)
+\left( \begin{array}{c} F_\mathrm{r}\\
   F_\mathrm{\theta}\end{array} \right)\,.
   \label{NS_us}
\end{equation}

The pressure of an ideal gas is $P=a^2\rho/\gamma$ with $\gamma$ the adiabatic index and $a^2=\gamma\,k_\mathrm{B}T/(\mu m_\mathrm{H})$ the isothermal sound speed with $T$ the temperature of the winds and $\mu$ the mean molecular weight.
Considering an isothermal gas ($\partial a/\partial r=0$), we thus have:
\begin{equation}
\begin{split}
\frac{\partial P}{\partial r}&=\frac{a^2}{\gamma}\frac{\partial \rho}{\partial r}\,\mathrm{and}\\
\frac{\partial P}{\partial \theta}&=\frac{a^2}{\gamma}\frac{\partial \rho}{\partial \theta}\,.
\end{split}
\label{deriv_p}
\end{equation}

Considering star 1, the mass-loss rate is:
\begin{equation}
\dot{M}=4\pi\,\rho\,r_1^2\,v = 4\pi\,\rho\,r_1^2\sqrt{\ur^2+\ut^2}
\end{equation}
Considering a uniform and constant mass loss rate, its derivatives are:
\begin{equation}
\begin{split}
\frac{\partial \dot{M}}{\partial r}&=0=4\pi\left(\rho\,v\,2r_1+\frac{\rho\,r_1^2}{2v}\frac{\partial (\ur^2+\ut^2)}{\partial r} + r_1^2v\frac{\partial \rho}{\partial r}\right)\,\mathrm{and}\\
\frac{\partial \dot{M}}{\partial \theta}&=0=4\pi\left(\frac{\rho\,r_1^2}{2v}\frac{\partial (\ur^2+\ut^2)}{\partial \theta} + r_1^2v\frac{\partial \rho}{\partial \theta}\right)\,.
\end{split}
\end{equation}
We thus have:
\begin{equation}
\begin{split}
-\pmb{\frac{a^2}{\gamma\rho}}\frac{\partial \rho}{\partial r}&=\pmb{\frac{a^2}{\gamma\rho}}\frac{2\rho}{r_1}+\pmb{\frac{a^2}{\gamma\rho}}\frac{\rho}{v^2}\left(\ur\frac{\partial\ur}{\partial r}+\ut\frac{\partial\ut}{\partial r}\right)\,\mathrm{and}\\
-\pmb{\frac{a^2}{\gamma\rho r_1}}\frac{\partial \rho}{\partial \theta}&=\pmb{\frac{a^2}{\gamma\rho r_1}}\frac{\rho}{v^2}\left(\ur\frac{\partial\ur}{\partial \theta}+\ut\frac{\partial\ut}{\partial \theta}\right)\,.
\end{split}
\label{deriv_rho}
\end{equation}

Equation~\ref{deriv_rho} can be injected in Eq.~\ref{deriv_p}.
The components of Eq.~\ref{NS_us} are thus:
\begin{equation}
\begin{split}
\label{equations}
&\left(1-\frac{a^2}{\gamma v^2}\right)\ur\frac{\partial \ur}{\partial r}+\frac{\ut}{r_1}\frac{\partial \ur}{\partial \theta}-\frac{a^2}{\gamma v^2}\ut\frac{\partial \ut}{\partial r}-\frac{\ut^2}{r_1}-\frac{2a^2}{\gamma r_1}-F_\mathrm{r}=0\,\mathrm{and}\\
&\left(1-\frac{a^2}{\gamma v^2}\right)\ut\frac{\partial \ut}{\partial \theta}+r_1\ur\frac{\partial \ut}{\partial r}-\frac{a^2}{\gamma v^2}\ur\frac{\partial \ur}{\partial \theta}+\ur\ut-r_1\,F_\mathrm{\theta}=0\,.
\end{split}
\end{equation}

The last missing terms are the components of the forces $(F_\mathrm{r}, F_\mathrm{\theta})$, which are the sum of the gravity ($F_\mathrm{grav}$) and the radiative acceleration ($F_\mathrm{R}$) from both stars (see Fig.~\ref{forces}).
\begin{equation}
\begin{split}
F_\mathrm{r}&=F_\mathrm{R,1}-F_\mathrm{grav,1}+F_\mathrm{grav,2}\cos{(\theta_1+\theta_2)}-F_\mathrm{R,2}\cos{(\theta_1+\theta_2)}\,,\\
F_\mathrm{\theta}&=F_\mathrm{grav,2}\sin{(\theta_1+\theta_2)}-F_\mathrm{R,2}\sin{(\theta_1+\theta_2)}\,.
\end{split}
\end{equation}
Points located at $\theta_1>\pi/2$ and $\tan\theta_2<R_1/d$ are not influenced by the radiation pressure from star 2.
Once the forces are known, we can resolve the equations and determine the velocity components.

\subsection{Solving the equations}
\begin{figure}
\centering
\includegraphics[width=8cm]{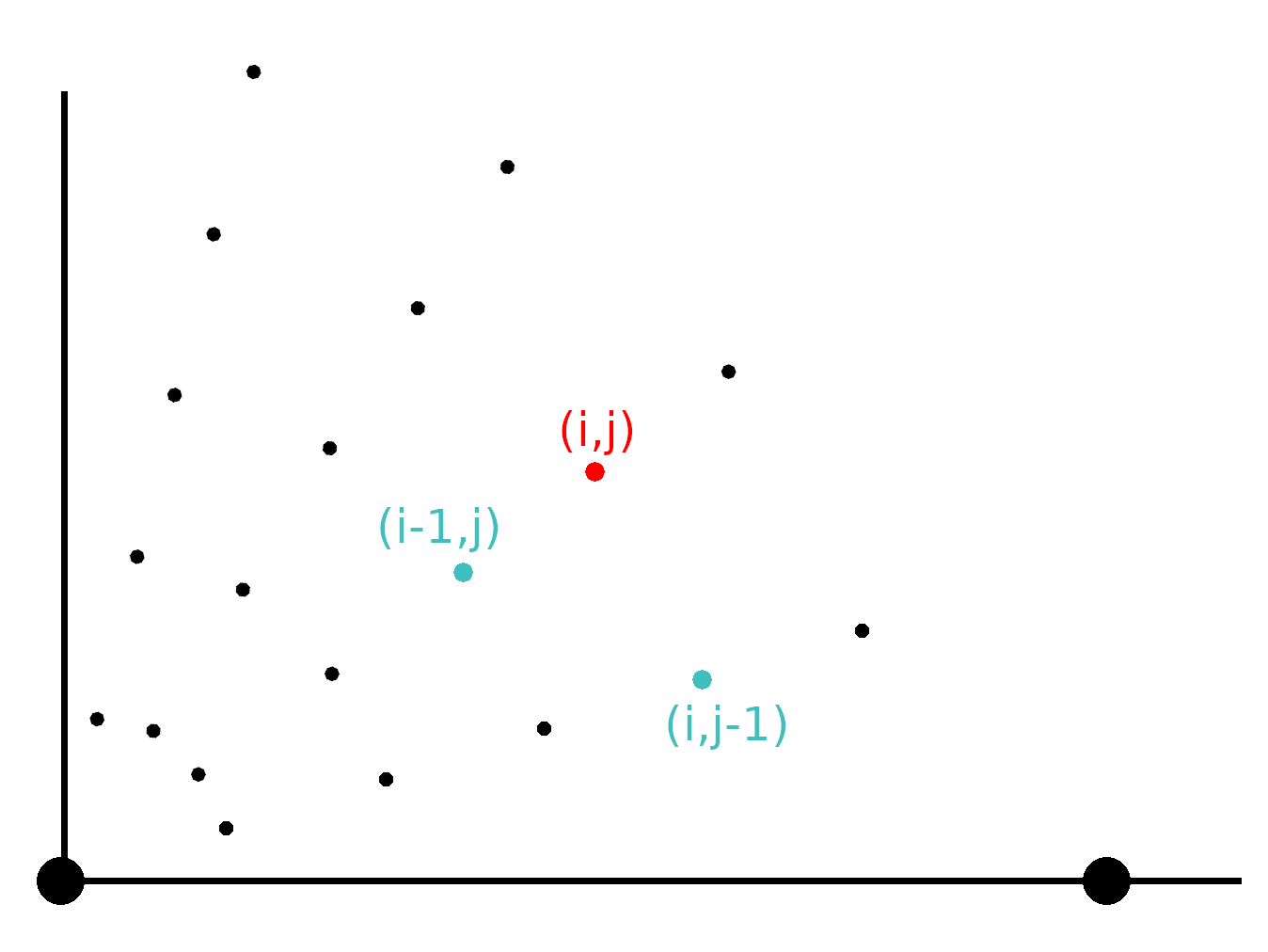}
\caption{Representation of the polar grid used to compute the stellar wind distribution.}
\label{fig:grid}
\end{figure}

We solve equations \ref{equations} using the finite-differences method defining, at a step $m$, $\partial f/\partial z=(f_\mathrm{m}-f_\mathrm{m-1})/\Delta z$.
In 2D, a point can be defined by its position $(i,j)$ on the grid with $i$ the step along the first direction and $j$ the step along the second direction.
In polar coordinates, let $\ur$ and $\ut$ be the radial and angular component of the wind velocity at the grid point $(i,j)$ (see Fig.~\ref{fig:grid}).
We name $\urr$ and $\utr$ the radial and angular component at the point $(i-1,j)$, and $\urt$ and $\utt$ the radial and angular components at the point $(i,j-1)$.
The equations are thus:
\begin{equation}
\begin{split}
eq_1=&\left(1-\frac{a^2}{\gamma v^2}\right)\,\ur\,\frac{\ur-\urr}{\Delta r} - \frac{\ut^2}{r_1} + \frac{\ut}{r_1}\,\frac{\ur-\urt}{\Delta \theta} - \frac{a^2\,\ut}{\gamma v^2}\frac{\ut-\utr}{\Delta r} - \frac{G\,M_2\,(1-\Gamma_2)\,\cos{(\theta_1+\theta_2)}}{r_2^2}  \\
&+ \frac{G\,M_1\,(1-\Gamma_1)}{r_1^2} - g_\mathrm{rad}\,(F_1\,K_1-F_2\,K_2\,\cos{(\theta_1+\theta_2)}) - 2\frac{a^2}{\gamma\,r_1}\,\mathrm{and}\\
eq_2=&\left(1-\frac{a^2}{\gamma v^2}\right)\frac{\ut}{r_1}\,\frac{\ut- \utt}{\Delta \theta} + \frac{\ut\,\ur}{r_1} + \ur\frac{\ut- \utr}{\Delta r} - \frac{a^2\,\ur}{\gamma v^2}\frac{\ur-\urt}{r_1\,\Delta \theta} - \frac{G\,M_2\,(1-\Gamma_2)\,\sin{(\theta_1+\theta_2)}}{r_2^2}\,, \\
&+ g_\mathrm{rad}\,F_2\,K_2\,\sin{(\theta_1+\theta_2)}
\end{split}
\label{eq12}
\end{equation}
where $\rho=(d\dot{M}/d\Omega)/(4\pi\,v\,r_1^2)$ and $g_\mathrm{rad}=\frac{\sigma_\mathrm{e}^{1-\alpha}\,k}{c\,(V_\mathrm{th}\,\rho\,v)^{\alpha}}\,\left(\left| \ur\frac{\ur-\urr}{\Delta r}+\ut\frac{\ut- \utr}{\Delta r}\right|\right)^{\alpha}$.
The velocity components $\ur$ and $\ut$ are computed by resolving $eq_1=0$ and $eq_2=0$.
Since these are non-linear equations that lack an analytical solution, they must be solved numerically. 
For this purpose, we have chosen to use the Broyden-Fletcher-Goldfarb-Shanno (BFGS) algorithm that minimises the quantity $(eq_1^2 + eq_2^2)^{0.5}$, which is equivalent to solving simultaneously the two equations $eq_1 = 0$ and $eq_2 = 0$ to within the maximal tolerance on the residual that we set to $0.002\,\mathrm{cm\,s^{-2}}$.
To refine the values of $\ur$ and $\ut$, we run the BFGS algorithm with a decreasing tolerance until the convergence.
At each execution, the values of $\ur$ and $\ut$ are initialised to the values determined at the previous execution, and the tolerance is divided by 50 compared to its previous value.

The derivative of these equations are
\scriptsize
\begin{equation}
\begin{aligned}
\frac{\partial eq_1}{\partial \ur}=&\frac{(2\ur-\urr)(\gamma\ut^2-a^2)+\gamma \ur^2(4\ur-3\urr)}{\Delta r\,\gamma v^2}&+&\frac{2\ur(a^2\ut(\ut-\utr)-\ur(\ur-\urr)(v^2\gamma-a^2))}{\Delta r\,\gamma v^4}&+&\frac{\ut}{r_1\,\Delta\theta}&-&C(F_1\,K_1-F_2\,K_2\,\delta_1)(2\ur-\urr)\,\mathrm{,}\\
\frac{\partial eq_1}{\partial \ut}=&\frac{-a^2(2\ut-\utr)+2\gamma\ur\ut(\ur-\urr)}{\Delta r\,\gamma v^2}&+&\frac{2\ut(a^2\ut(\ut-\utr)-\ur(\ur-\urr)(v^2\gamma-a^2))}{\Delta r\,\gamma v^4}&-&\frac{2\ut}{r_1}+\frac{\ur-\urr}{r_1\Delta \theta}&-&C(F_1\,K_1-F_2\,K_2\,\delta_1)(2\ut-\utr)\,\mathrm{,}\\
\frac{\partial eq_2}{\partial \ur}=&\frac{-a^2(2\ur-\urt)+2\gamma\ur\ut(\ut-\utt)}{r_1\,\Delta \theta\,\gamma v^2}&+&\frac{2\ur(a^2\ur(\ur-\urt)-\ut(\ut-\utt)(v^2\gamma-a^2))}{r_1\Delta \theta\,\gamma v^4}&+&\frac{\ut}{r_1}+\frac{\ut-\utr}{\Delta r}&+&C\,F_2\,K_2\,\delta_2(2\ur-\urr)\,\mathrm{, and}\\
\frac{\partial eq_2}{\partial \ut}=&\frac{(2\ut-\utt)(\gamma\ur^2-a^2)+\gamma \ut^2(4\ut-3\utt)}{r_1\Delta \theta\,\gamma v^2}&+&\frac{2\ut(a^2\ur(\ur-\urt)-\ut(\ut-\utt)(v^2\gamma-a^2))}{r_1\Delta \theta\,\gamma v^4}&+&\ur\left(\frac{1}{\Delta r}+\frac{1}{r_1}\right)&+&C\,F_2\,K_2\,\delta_2(2\ut-\utr)\,,\\
\end{aligned}
\end{equation}
\normalsize
with $C=\frac{\alpha\,\sigma_\mathrm{e}\,k}{c}\left(\frac{4\pi r_1^2}{\sigma_\mathrm{e}\,V_\mathrm{th}\,d\dot{M}/d\Omega\,\Delta r}\right)^{\alpha}\,\left|\ur(\ur-\urr)+\ut(\ut-\utr)\right|^{\alpha-1}$.
The Jacobian of $(eq_1^2+eq_2^2)^{0.5}$ is thus
\begin{equation}
   J =(eq_1^2+eq_2^2)^{-0.5}\left( \begin{array}{c} eq_1\frac{\partial eq_1}{\partial \ur}+eq_2\frac{\partial eq_2}{\partial \ur}\\
   eq_1\frac{\partial eq_1}{\partial \ut}+eq_2\frac{\partial eq_2}{\partial \ut}\end{array} \right)\,.
\end{equation}

\section{Total line profiles of the Fe~K triplet}
\label{total_lp}
\begin{figure*}
\centering
\includegraphics[trim= 2cm 1cm 2cm 1cm,clip, width=13.7cm]{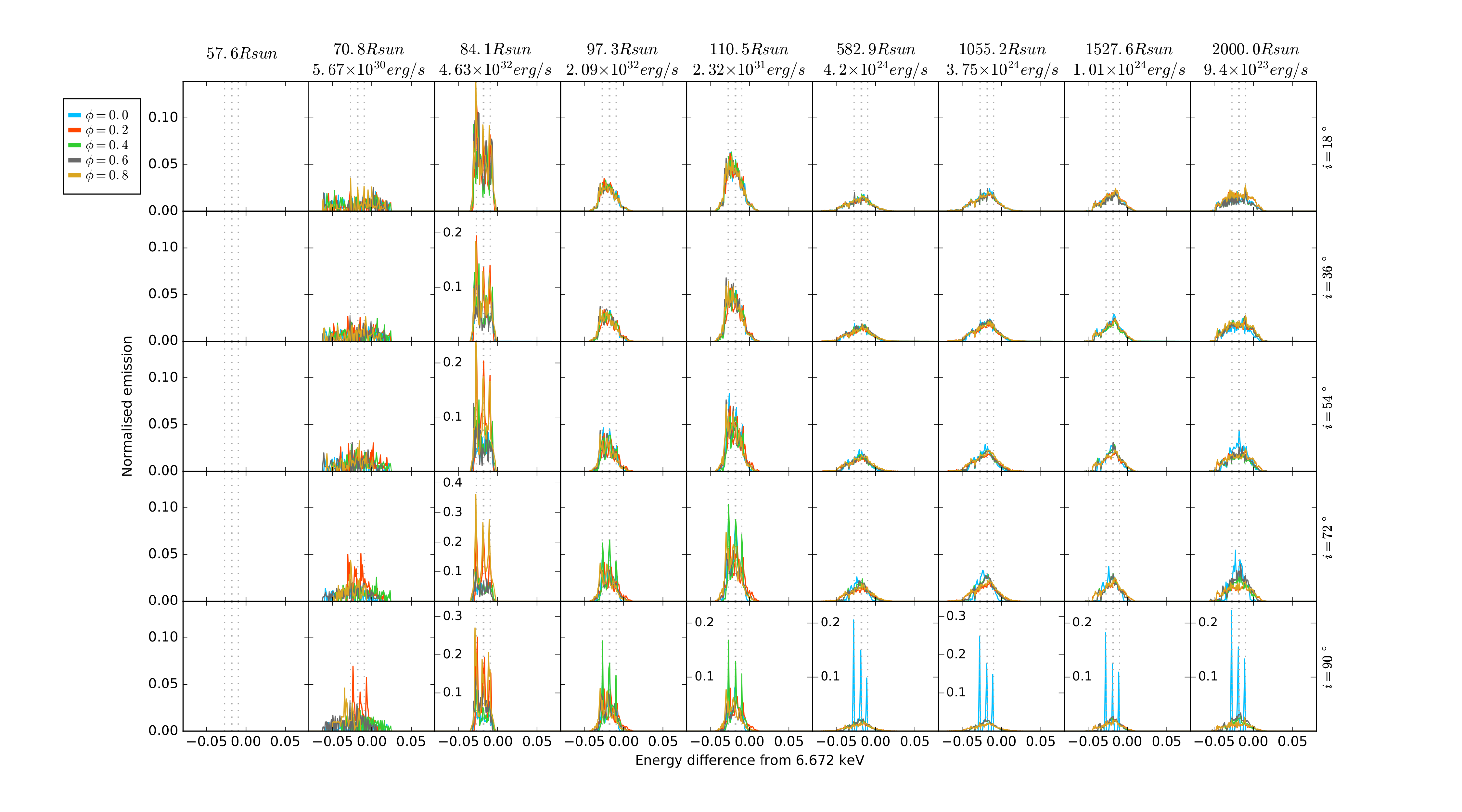}\\
\includegraphics[trim= 1cm 1cm 2cm 1cm,clip, width=13.7cm]{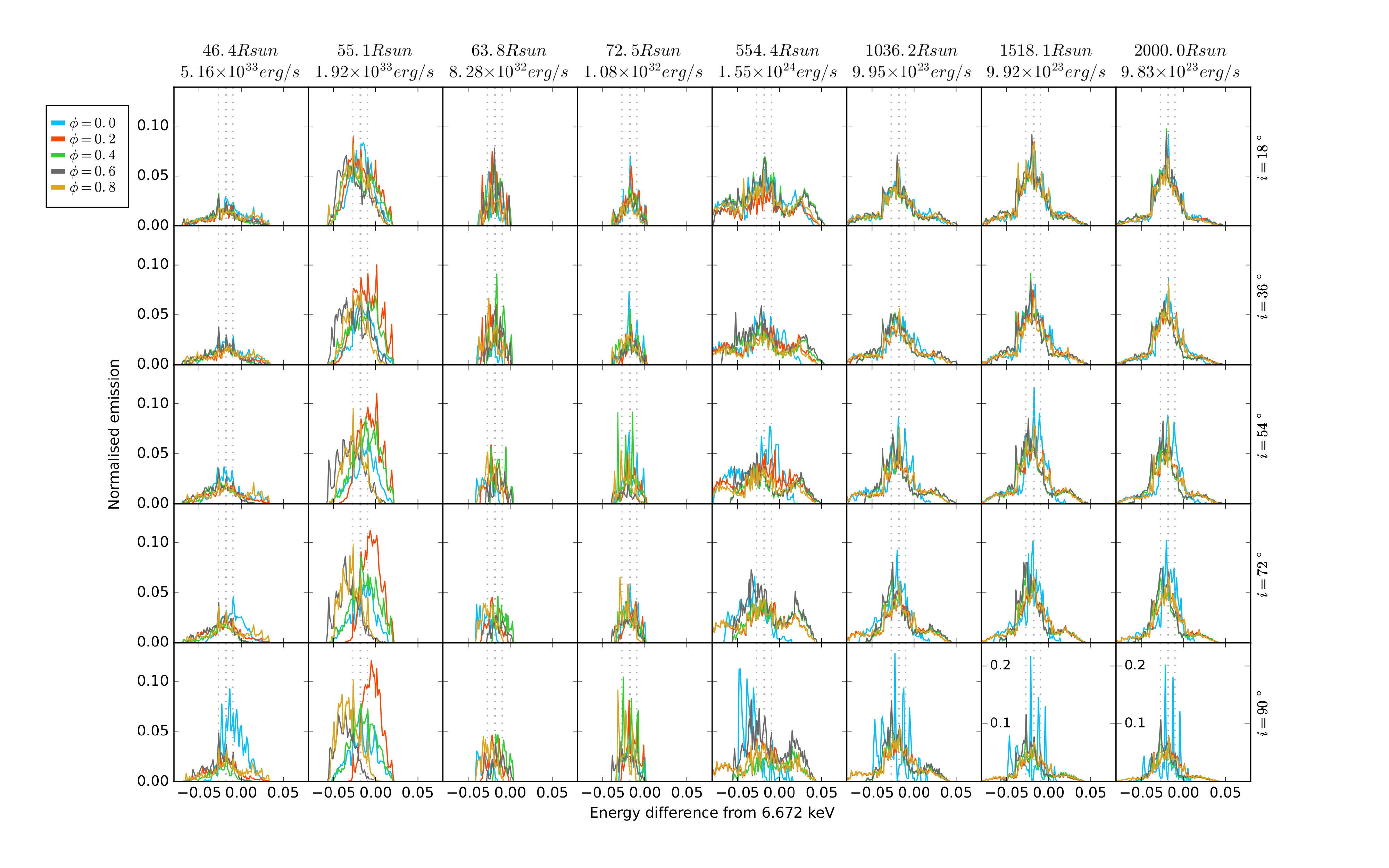}\\
\includegraphics[trim= 2cm 1cm 2cm 1cm,clip, width=13.7cm]{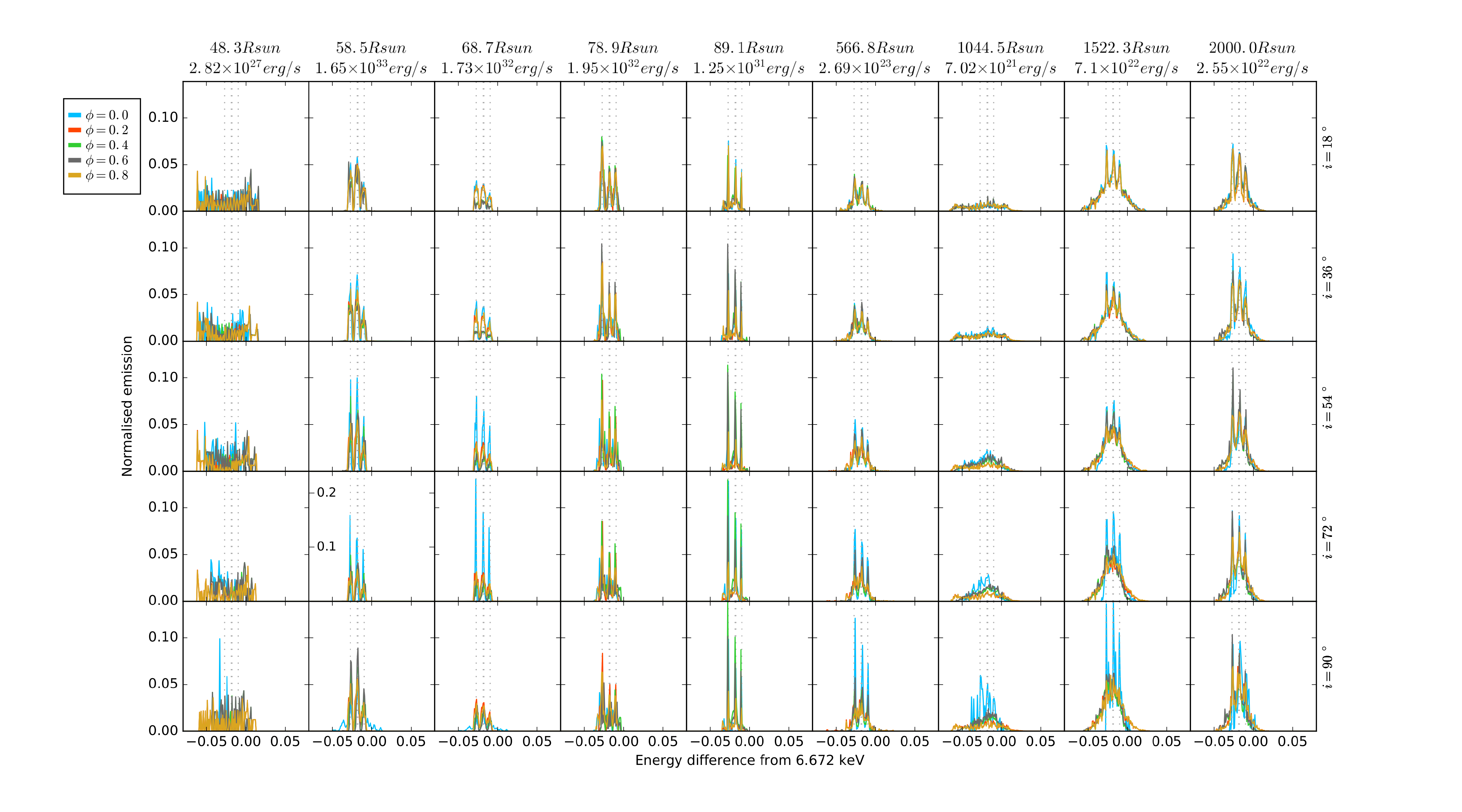}
\caption{The line profiles of the Fe~{\sc xxiv} satellite lines computed for three systems: O7V+O7V (top panel), O5I+O3III (middle panel), and O9III+O9V (bottom panel).}
\label{fig_sat}
\end{figure*}

As explained in Sect.~\ref{iron_line}, four satellite lines from Fe~{\sc xxiv} are expected to emit at energies in the energy range of the Fe~K triplet.
Figure~\ref{fig_sat} shows the line profiles of the satellite lines for the three stellar systems described in Sect.~\ref{three_sys}.
The vertical dotted lines show the energy of the four satellite lines at rest.

\begin{figure*}
\centering
\includegraphics[trim= 2cm 1cm 2cm 1cm,clip, width=14cm]{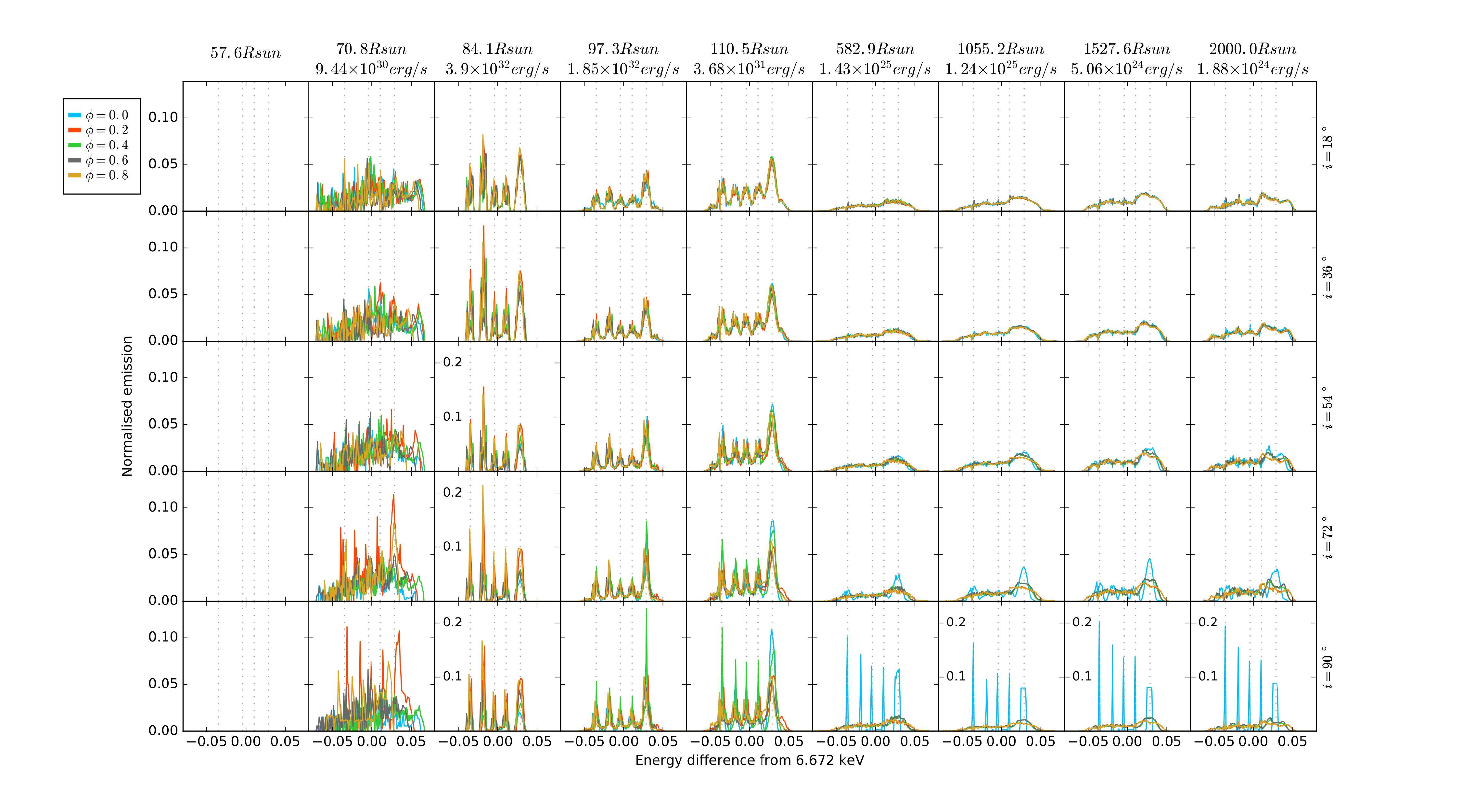}\\
\includegraphics[trim= 1cm 1cm 2cm 1cm,clip, width=14cm]{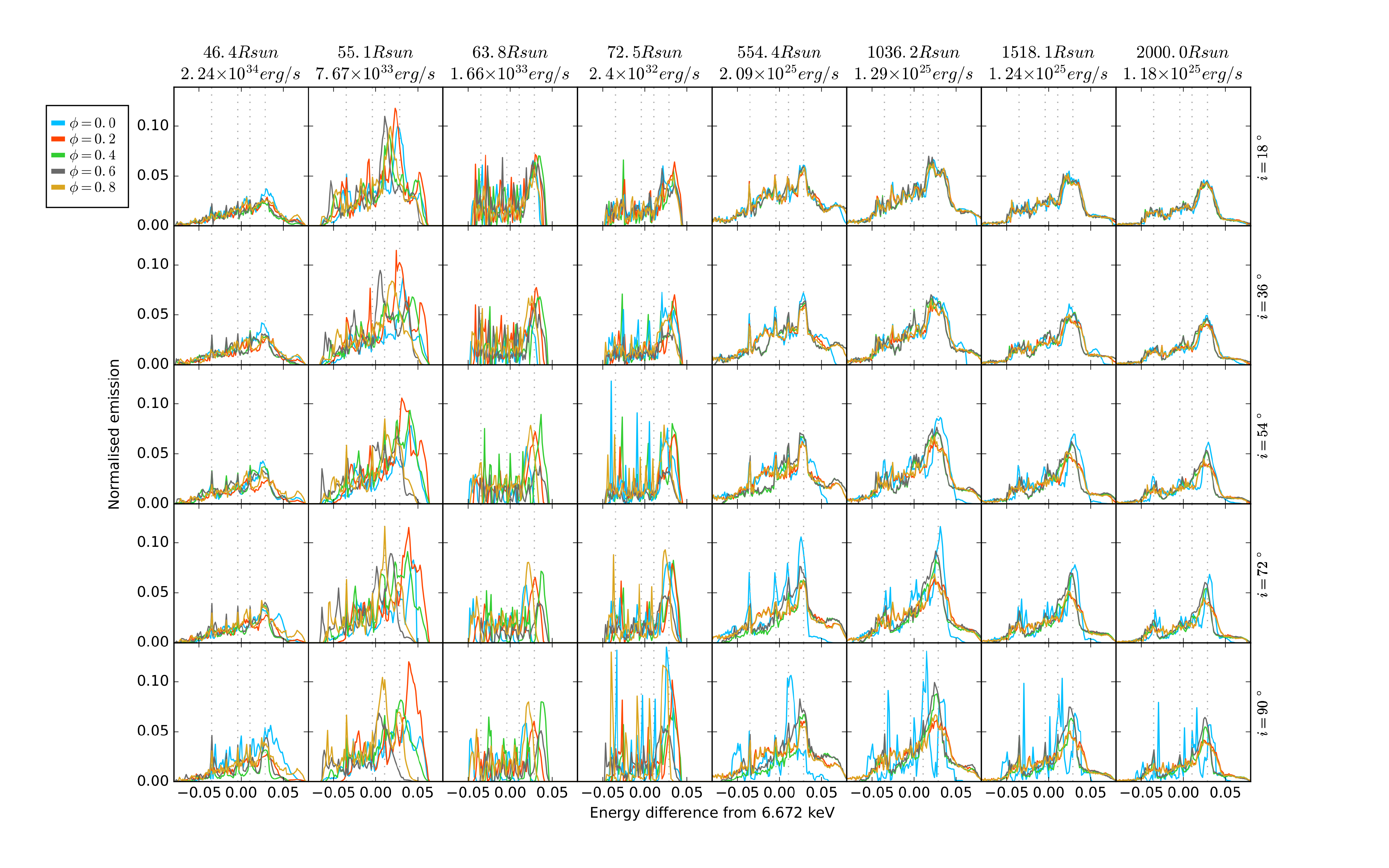}\\
\includegraphics[trim= 2cm 1cm 2cm 1cm,clip, width=14cm]{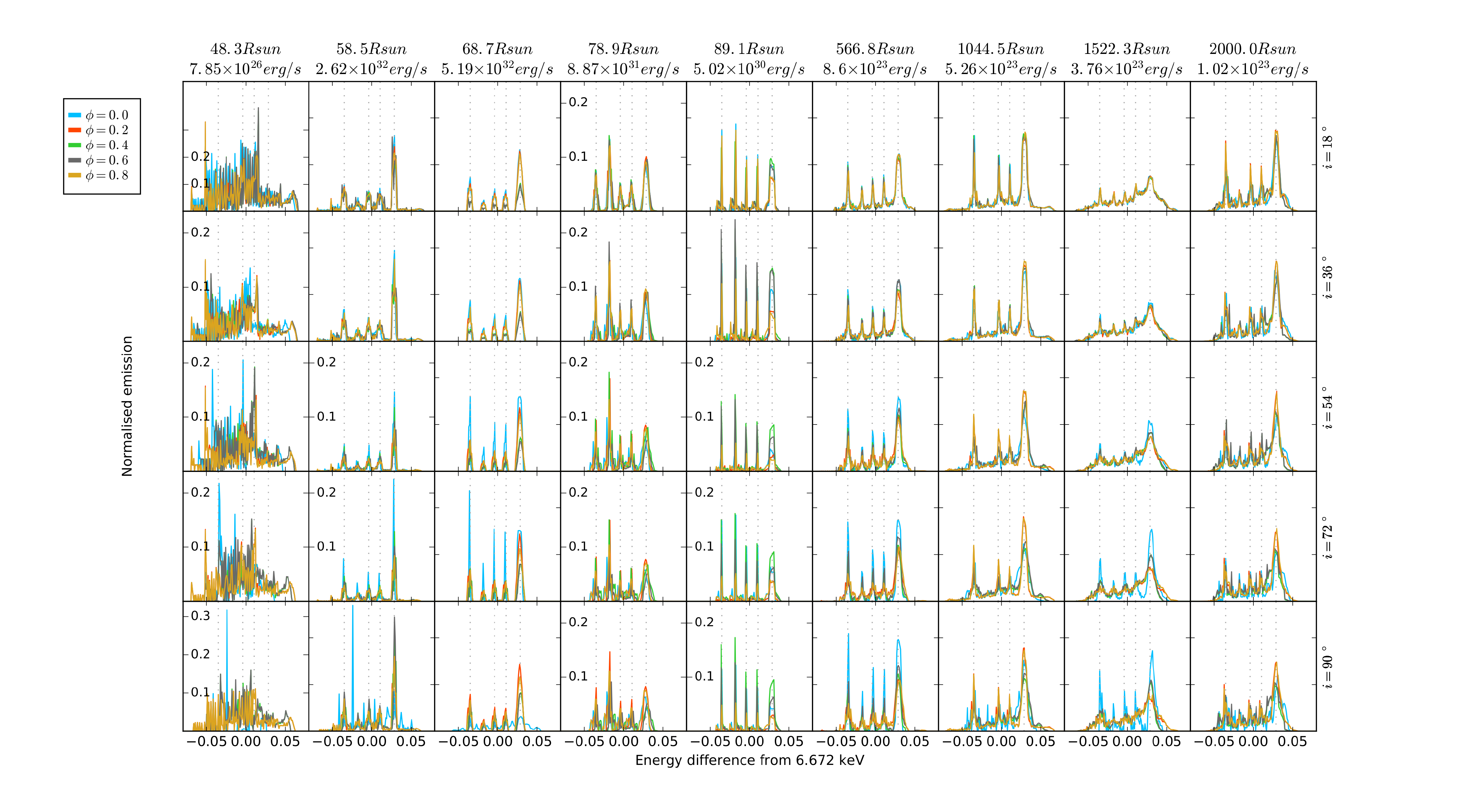}
\caption{The line profiles of the Fe~K triplet plus the Fe~{\sc xxiv} satellite lines computed for three systems: O7V+O7V (top panel), O5I+O3III (middle panel), and O9III+O9V (bottom panel).}
\label{fig_sat_fe}
\end{figure*}

These satellite lines will thus be blended with the Fe~{\sc xxv} triplet.
Figure~\ref{fig_sat_fe} shows the resulting complex morphology of the Fe~K line region for the three stellar systems. The vertical dotted lines show the energy of the four lines of the Fe~K triplet. 

\end{appendix}

\bibliographystyle{aa}
\bibliography{biblio_lifeline}

\end{document}